\documentclass[
preprint, 
onecolumn,
superscriptaddress,
nofootinbib,
]{revtex4-2}

\usepackage{amsmath}
\usepackage{amssymb}
\usepackage{graphicx} 
\graphicspath{ {figures/} } 
\usepackage{tikz}
\usepackage{courier}
\usepackage{hyperref}
\usepackage{xcolor}
\hypersetup{
	colorlinks,
	linkcolor={blue!50!black},
	citecolor={blue!50!black},
	urlcolor={blue!80!black}
}\usepackage{changepage}        
\usepackage{booktabs}
\usepackage{pythonhighlight}

\newcommand{\pbs}{Dept. of Psychological \& Brain Sciences, Indiana University, Bloomington, IN, 47405}
\newcommand{\sice}{School of Informatics, Computing \& Engineering, Indiana University, Bloomington, IN, 47405}

\begin{document}
	
	\title{Information Theory for Complex Systems Scientists: \\ What, Why, \& How?}
	
	\author{Thomas F. Varley}
	\email{tfvarley@uvm.edu}
	\affiliation{\pbs} 
	\affiliation{\sice} 
	
	\begin{abstract}
		In the 21st century, many of the crucial scientific and technical issues facing humanity can be understood as problems associated with understanding, modeling, and ultimately controlling \textit{complex systems}: systems comprised of a large number of non-trivially interacting components whose collective behavior can be difficult to predict. Information theory, a branch of mathematics historically associated with questions about encoding and decoding messages, has emerged as something of a \textit{lingua franca} for those studying complex systems, far exceeding its original narrow domain of communication systems engineering. In the context of complexity science, information theory provides a set of tools which allow researchers to describe a variety of dependencies, including interactions between the component parts of a system, interactions between a system and it's environment, and the mereological interaction between the parts and the ``whole". 
		
		In this review, we aim to provide an accessible introduction to the core of modern information theory, aimed specifically at aspiring (and established) complex systems scientists. This includes standard measures, such as Shannon entropy, relative entropy, and mutual information, before building to more advanced topics, including: information dynamics, measures of statistical complexity, information decomposition, and effective network inference. In addition to detailing the formal definitions, we also make an effort to discuss how information theory can be \textit{interpreted} and to develop the intuitions behind abstract concepts like ``entropy". We aim to enable interested readers to understand what information is, and how it is used, at a fundamental level to better further their own research and education. 
	\end{abstract}
	
	\maketitle
	\date{\today}
	
	\newpage
	
	\tableofcontents
	
	\newpage
	
	\section{What Are Complex Systems?}
	\label{sec:complex_systems}
	If the first decades of the 21$^{\text{th}}$ century have provided any insight into what the scientific challenges of tomorrow will be, it is that many of the core problems are tied together by a common thread: complexity emerging from highly-interconnected systems. From climate change, to pandemics, from disinformation to economic inequality, a defining characteristic of many issues facing the modern world is that they deal with the problem of modeling, predicting, and controlling \textit{emergent} properties of systems comprised of thousands, or millions of interacting elements. In the case of pandemics, those elements are individual humans and animals, milling and seething across the globe. Animals and humans interact, and in doing so, spread evolving pathogens; they fall ill, thy recover, and occasionally, they die. In the context of ``fake news", the elements form an ecosystem of humans and ever-more human-like artificial intelligences, all navigating online environments controlled by increasingly opaque algorithms. Climate change threatens the stability of a huge number of interconnected, complex systems, from coral reef ecosystems (which bleach in the face of sustained high temperatures \cite{orlowski_chasing_2017}), to agriculture and food-production systems (possibly leading to food shocks, and in extreme cases, famine \cite{von_braun_climate_2020}).
	
	These problems of complexity are interconnected as well: far from being restricted to the traditional disciplinary silos of Academia (physics, chemistry, biology, etc), complex systems can involve many different areas of science simultaneously, and require seeing connections across domains. For example, it takes tremendous amounts of energy to train the artificial intelligences that increasingly dominate our online lives. In an economy run on fossil fuel, this means that the systems driving one phenomenon (``fake news", social media bubbles, ``infodemics" \cite{cinelli_covid-19_2020}) are simultaneously exacerbating another one (climate change) with their large carbon footprints and outrageous water demands \cite{strubell_energy_2019,henderson_towards_nodate,li_making_2023}. These simple examples highlight some of the fundamental challenges of modern complex systems science:
	
	\begin{enumerate}
		\item The systems that make up our world are fundamentally \textit{interconnected} in ways that defy the typical breakdown of academic science disciplines. It is not often that climatologists collaborate with machine learning engineers (although as awareness of the carbon costs of computation increases this is starting to change \cite{strubell_energy_2019,li_making_2023}).
		\item All of the systems described above are highly non-linear \cite{strogatz_nonlinear_2018}, exhibiting complex phenomena such as critical phase changes (or ``tipping points") \cite{scheffer_early-warning_2009}, cascading failures \cite{dobson_complex_2007}, extreme ranges of intensities \cite{newman_power_2005}, and self-organization \cite{ashby_principles_1991} (to name but a few of the interesting properties complex systems can display).  
		\item These systems are \textit{multiscale:} at the fundamental level, they involve millions, or even billions of moving parts: modeling the entire structure would require astronomical amounts of data, computing power, and very detailed mathematical and/or algorithmic models of behavior. Consequently, the ability to think critically about the distinction between ``wholes" and ``parts" is key \cite{gutknecht_bits_2020}. So is understanding when it is possible to reduce the dimensionality via processes such as coarse-graining \cite{hoel_when_2017}. Furthermore, ``higher-order" or ``synergistic" effects present only at the macro-scale can complicate analyses when systems don't neatly decompose into sums of parts \cite{rosas_disentangling_2022}.
		\item The processes that make up the world are \textit{dynamic:} the past informs on the future and individual moments cannot be assumed to be decoupled from the history that lead up to that point \cite{crutchfield_calculi_1994}. This can take the form of chaotic sensitive dependence on initial conditions, dynamic hysteresis, or long-term autocorrelations (also called ``memory"), all of which can describe the influence of the past on the future. 
	\end{enumerate}
	
	What might be called ``classical" statistical modeling, which makes assumptions of linearity, reductionism, and independence can struggle when attempting to disentangle complex systems. Warren Weaver, a pioneer of early complex system science described a schema with which systems can be categorized: 
	
	\begin{itemize}
		\item[] \textit{``Simple" systems}, which are comprised of a small number of interacting components that can be easily described using mathematical models. For example, Newtonian physics does an excellent job describing the collision between two objects moving through space. 
		\item[] \textit{``Disorganized" complex systems}, which are comprised of a large number of largely non-interacting components and are amenable to aggregation and linear modelling. For example, there are billions of atoms in a gas, but their interactions are trivial and the logic of statistical mechanics describes their (macro-scale) behaviour well.
		\item[] \textit{``Organized" complex systems}, which make up almost all of the really interesting systems in nature. These systems are composed of large numbers of parts (as in the disorganized systems), but the interactions are non-trivial, with complex correlations and patterns spanning multiple scales.  
	\end{itemize}
	
	Organized complex systems, in some sense, combine the worst (or best) of both worlds. The wicked combination of a large number of elements, coupled by non-trivial interactions results in systems that are not particularly amenable to the simple and elegant analytic models of the past. Instead, the only way forward is to tackle the complexity head on, using computer simulations, structure learning, and large models \cite{weaver_science_1948}.
	
	What, then, is the way forward? Science in the 21$^{st}$ century needs a toolkit that can be applied to as many complex systems as possible, both to understand their inner workings, and facilitate interdisciplinary interactions between fields that may never have had much reason to come into contact with each-other, and therefore lack a common language. Furthermore, rather than running from the nonlinear, hard-to-model nature of complex data, the science of tomorrow must be able to tackle these systems without throwing out important information or making over-simplifying assumptions of linearity, independence, or equilibrium conditions. Increasingly, the emerging answer to this problem is ``information theory," which is poised to become the \textit{lingua franca} of our increasingly complexity-aware scientific enterprise. 
	
	\newpage
	
	\section{What Is Information?}
	
	Information theory was first developed by Claude Shannon, an engineer working at Bell Labs on the problem of how to reliably transmit a signal from a sender to a receiver over some potentially noisy channel (such as a telegraph line). In a now-famous moment of remarkable genius, Shannon produced almost the entirety of classical information theory in a single monograph: \textit{A Mathematical Theory of Communication}, first published in 1948 \cite{shannon_mathematical_1948}. In this highly practical context, ``information" was understood as referring to signals being propagated through a channel. We will describe the idea of information in formal detail below, but intuitively, it can be understood in terms of how our uncertainly is resolved by the data made available to us. If someone speaks into a telephone receiver, but all the person at the other end hears is static white-noise, the ``information" has clearly been lost: the listener has no way of inferring what the speaker said. In contrast, if the voice comes through loud and clear, then the listener has no uncertainty about what the speaker said, and information has been transmitted. If the channel is noisy, but not totally compromised (perhaps the sound quality is poor), the listener may not be able to perfectly catch every word, but can still reconstruct the gist of the message based on the parts that they could hear. Shannon provided a mathematically rigorous formal language with which one might quantify how much information was transmitted (or lost), and what the fundamental limits on transmission are in noisy environments.\footnote{It is not hyperbolic to say that this work formed one of the fundamental foundations on which our modern, digital world was built (on par with the development of electricity, or the internal combustion engine in terms of impact).}
	
	The significance of this development was not lost on Shannon's fellow cyberneticians, and information theory quickly received interest from a broad and interdisciplinary group of scientists, mathematicians, and even philosophers. The question of what information theory's ``natural domain" is has been a contentious one ever since. Despite Shannon's initial narrow, and eminently practical focus, during the cybernetic revolution of the mid-20$^{\text{th}}$ information theory was quickly applied to a wide range of fields, from sociology to biology \cite{ashby_introduction_1956,gleick_information_2011}. Shannon himself, alarmed at what he saw as a the hasty misapplication of his theories to disciplines where they didn't belong, pushed back, claiming that, by the mid-1950s that information theory had become a ``bandwagon" and was been ``oversold" by overeager researchers \cite{shannon_bandwagon_1956}. Despite Shannon's apparent misgivings, however, in the modern day, information theory has retained its popularity, finding applications in diverse fields such as neuroscience \cite{timme_tutorial_2018,fagerholm_primer_2023}, ecology \cite{ulanowicz_information_2001,harte_maximum_2014}, artificial intelligence \cite{ruiz_information_2009,mackay_information_2003}, sociology \cite{varley_untangling_2022}, climatology \cite{greene_information-theoretic_2013,majda_quantifying_2010,goodwell_temporal_2017,hlinka_non-linear_2014}, and deep connections have been found between information theoretic formalisms and fundamental physical models such as thermodynamics (forming the basis of ``stochastic thermodynamics" \cite{peliti_stochastic_2021}). Finally, information theory has become a key foundation on which a variety of formal theories of ``emergence" have been constructed \cite{seth_measuring_2010,hoel_quantifying_2013,barnett_dynamical_2021,hoel_can_2016,mediano_greater_2022,varley_emergence_2022,varley_flickering_2023,rosas_software_2024}. Emergence, the phenomenon where a system collectively displays properties not trivially reducible to some property of its component parts is arguably the defining feature of complex systems, and the links between multivariate information theory and emergence solidify the deep link between the fields. 
	
	It is interesting to consider the question of how it came to be that Shannon's theory, developed within the practical confines of communications engineering, has been able to find such widespread success far beyond its original home. Has the bandwagon simply continued to run amok, heedless of Shannon's 1956 warning? I would (perhaps unsurprisingly) argue ``no" (it's hard to imagine why I would have written this tutorial review otherwise). Rather, I claim that information theory's native domain is \textit{not just} the field of communication engineering where it was born, but rather \textbf{information theory is better understood as being fundamentally about the general mathematics of \textit{inference} \cite{mackay_information_2003}, particularly, \textit{inference under conditions of uncertainty.}} The original applications of information theory (how to reliably transmit a message from a sender to a receiver along a noisy channel) is itself a particular kind of inference problem; a special case of a more general framework. When the receiver receives a message transmitted in the presence of noise, the problem of decoding that message is one of making inferences about the most probable contents of the original message (``what was the sender likely saying"). Similarly, when engineering a usable communication channel, the goal is to give the receiver the best possible chance of correctly inferring the senders message. \textit{It is the logic of inference (encoding, decoding, probability and statistics) that informs the development of communications channels, not the other way around.} 
	
	This perspective of ``information theory as inference" resolves another long-standing tension in the field. Information theory is agnostic to the question of the ``meaning" of a message: it never mattered to Shannon what message was being sent over a noisy channel, only the statistics of the symbols that make it up. This distinction has been referred to as the difference between the \textit{syntactic} information (which information theory is concerned with), and the \textit{semantic} information (which it is not) \cite{kolchinsky_semantic_2018}. As long as information theory has been understood as a ``mathematical theory of communication", the question of meaning (semantic content) is a very natural one: everyone cares more about the meaning of a message than the statistics of the symbols used to transmit it \footnote{Except engineers, of course.} and information theory's failure to address this has lead some to doubt its appropriateness as a theory of communication. 
	
	This paper is written with the perspective that focusing on the ``communication" angle is a mistake, or at least, insufficiently general. When considering ``information theory as inference", it is clear that the question of semantic information is largely irrelevant: we are not interested in what a message might mean, but rather, ensuring that we have inferred the correct message from some space of possible messages. The same mathematics, however can apply to any issue where we are attempting to increase our degree of certainty about the state of some as-yet-unknown variable. Whether it is a message coming down a pipe, the future state of an evolving system, or the degree of mutual coupling between two variables. These are the core questions that form the basic problem of modeling complex systems in any domain. 
	
	\subsubsection*{A Note to the Mathematically Nervous}
	
	It is impossible to write a comprehensive review of information theory (a field of mathematics in its own right) in complex systems (a very quantitative meta-field) without relying on a certain amount of mathematics. However, throughout this review I have done my best to bolster the formulae with discussions about the intuitions behind the math, and how the measures might be interpreted. Complex systems draws interest from many fields, and many (especially students just discovering the interest for the first time) may not have a comprehensive mathematics background. To paraphrase Albert Einstein, we have tried to ``make everything as simple as possible, but no simpler". For readers who wish to refresh on the basics of probability theory (on which information theory is fundamentally based), see the Supplementary Information). 
	
	With that caveat in mind, we now turn to the most fundamental building block of information theory: the entropy.
	
	\subsection{Entropy}
	\label{sec:entropy}
	
	The fundamental questions information theory equips us to ask (and answer) are:
	
	\begin{enumerate}
		\item How uncertain am I about the state of some thing I could observe?
		\item How much ``work" (for want of a better word) do I need to do to become \textit{more certain} than I am now?
	\end{enumerate} 
	
	It is important to note that these two core questions are not (on the surface) actually about information, but rather, they are about \textit{uncertainty}, and how uncertainty changes. This forms the core of Shannon's great insight: information is the reduction of uncertainty associated with observing data. To even begin to mathematically understand information, we must first understand uncertainty. This is done with the ``entropy", a mathematical function that quantifies how uncertain we are about the state of some variable given its statistics. 
	
	There is a (likely apocryphal) story that, upon developing his theory, Claude Shannon approached the famous physicist and fellow cybernetician John von Neumann and asked what he should call his new measure. Von Neumann (allegedly) replied: \textit{``you should call it entropy...since no one really knows what entropy is, you'll always have the upper hand in a debate."} (Paraphrased). While tongue-in-cheek, this story highlights the reality that, despite Shannon's originally narrow focus, the structure he developed is remarkably general and lends itself to many interpretations (and more than a few misinterpretations as well). 
	
	Many intuitive understandings of information entropy have been proposed, and many authors have their own ways of introducing it. Here I provide two interpretations that make sense to me, but I encourage interested readers to go find as many explanations of entropy as they can. 
	
	\subsubsection{Entropy as Expected Surprise}
	\label{sec:entropy:quantify}
	
	To answer Question 1 ("how uncertain am I about the state of something in the world"), we need a measure of uncertainty; a way to quantify exactly how certain we are about whatever it is we believe. To understand what we mean by quantifying ``uncertainty", consider the case of a fair die, where the probability of any face coming up is 1/6. If you were gambling on the die, you would have no reason to pick any face over another and consequently \textit{no outcome would be more surprising than any other.} You would be no more shocked to roll a 2 than a 5 and are, in some sense, maximally uncertain about the outcome of a roll. This notion of ``surprise" as being related to ``uncertainty" becomes clearer if we instead imagine that you are a cheater and, at some point, managed to switch the fair die for a loaded one of your own. This tricky die is balanced in such a way that 2 comes up 2/3 of the time, and all other faces come up with probability 1/15. Knowing this probability distribution ahead of time, you will be \textit{less surprised} to see a 2 come up than you would be to see a 2 come up on the fair die. Similarly, you would be \textit{more surprised} to see any other number turn up on the loaded die than you would to see that same number on the fair die. In this case, you are more certain about the outcome of a roll with the tricky die than you would be in the case of the fair one.  
	
	Everyone has an intuitive notion that some events are more surprising than others, and that the amount of surprise you experience is directly related to how likely you were to make a correct prediction. You'd be more surprised to receive a spaceship for Christmas than a pair of socks, because socks are a much more likely gift than spaceships! This formed the basis of Shannon's derivation of the entropy, and is the first, fundamental mathematical measure we will introduce. Shannon \cite{shannon_mathematical_1948} realized that a quantitative measure of surprise (noted as $h()$) should satisfy a set of four basic criteria. For a random variable $X$ that draws states from the support set $\mathcal{X}$ according to the probability distribution $P(x)$:
	
	\begin{enumerate}
		\item The surprise associated with a particular outcome $X = x$ should be inversely proportional to the probability $P(x)$: the more probable an event, the less surprising it is.
		\item Surprise should never be negative. Formally:\newline $h(x) \ge 0\text{ for all }x \in \mathcal{X}$.
		\item Events that are guaranteed to happen are totally unsurprising. Formally: \newline If $P(x) = 1$ than $h(x) = 0$.
		\item Surprise should be additive: our surprise at the outcome of two simultaneous, independent coin flips should be equal to our surprise at the outcome of the same coin flipped two different times. Formally:\newline $h(x_1,x_2) = h(x_1) + h(x_2) \iff X_1 \bot X_2$.
	\end{enumerate}
	
	He went on to show that the function uniquely specified by these desiderata is:
	
	\begin{equation}
	h(x) = \log\bigg(\frac{1}{P(x)} \bigg)
	\end{equation}
	
	This value is often referred to as the surprise, Shannon information content, or the local entropy of a specific outcome of random variable $X$. The less probable an outcome, the more surprising it is. We can now ask ``on average, how surprised are we as we watch this variable over time?" Rare events are very surprising but we aren't that surprised very often. In contrast, frequent events are unsurprising, so we spend a lot of time not being very surprised. This can be quantified by the \textit{expected surprise} of a distribution $P(X)$:
	
	\begin{align}
	H(X) &:= \mathbb{E}_X[h(x)] \\ &= -\sum_{x \in \mathcal{X}} P(x) \log(P(x))
	\label{eq:entropy}
	\end{align}
	
	This is the now-famous equation of Shannon entropy: a function that takes in a discrete probability distribution and returns a measure of, on average, how uncertain are we about the state of the associated random variable. $H(X)$ is maximal when all outcomes are equiprobable, which matches the intuition laid out in the fair dice example above: when all outcomes have the same chance of occurring, we are maximally uncertain about the outcome of the next roll. Furthermore, by convention we say that $0\log(0) = 0$, so impossible events have no impact on our uncertainty. Knowing that a fair die can never roll a 100 doesn't decrease our uncertainty about which of the six faces will come up next. 
	
	\subsubsection{Entropy as Required Information}
	\label{sec:entropy:required}
	
	Another, related, way of thinking about entropy is based on Question 2: ``How much work do I need to do to become more certain than I am now?". Forget about surprise for a moment, and ask ``on average, what is the minimum number of yes/no questions would I need to determine the number rolled on each dice (the fair and the weighted)?" 
	
	In the case of the fair die, you have no reason to suspect any particular face to be more likely than another, so a guess of ``it is $x$" will only be right one time in six. A more efficient approach might be to start whittling down the space of possibilities with clever questions like ``is it an odd number." Assuming that you are very good good at selecting questions, on average, it will take you $\approx 2.59$ questions to uniquely determine the state of the die. In the case of the loaded die, however, if you start with ``is it 2", you'll be right two times in three. On average, it will only take $\approx 1.7$ guesses to determine the outcome of a roll of the loaded die. 
	
	Every time you ask a question, you \textit{rule out} possible outcomes, reducing your uncertainty and \textit{gaining information about the state of the die} (for a fascinating discussion of information as probability mass exclusions, see \cite{finn_probability_2018}). In the case of the fair die, you need to ask more questions (that is to say, you need more information) to uniquely specify the state of the die, while in the case of the loaded one, since you already know that it's more likely to be two, you need less information. You come into the game with information already ``in your pocket", as it were.
	
	Remarkably, for a random variable $X$, the average number of yes/no questions required to uniquely specify the state of $X$ is given by the entropy function $H(X)$, when the base of the logarithm in Eq. \ref{eq:entropy} is two. 
	
	The link between uncertainty and information is deep (as we will see below). Broadly, we can think of the entropy of a random variable as a measure of the capacity of that variable to represent information, although for those just being exposed to information theory for the first time, the link between the two concepts can often be confusing. For didactic purposes, we recommend thinking of entropy as measuring uncertainty, and later, information as measuring a reduction in uncertainty. 
	
	\subsubsection{Joint Entropy}
	
	The classic Shannon entropy is formulated to describe the uncertainty associated with a single variable (Equation \ref{eq:entropy}). A defining feature of complex systems, however, is the presence of many interacting elements, and a core use of information theory when researching complex systems is understanding how elements constrain and disclose information about each-other. The simplest generalization of entropy is the joint entropy, which treats sets of variables as a coarse-grained ``macro"-variable and gives a measure of the uncertainty about the ``macro"-state of the whole system. 
	
	The joint entropy of two variables $H(X_1, X_2)$ is given by:
	
	\begin{align}
	H(X_1, X_2) &= \mathbb{E}_{X_1,X_2}[h(x_1, x_2)] \\  
	&= -\sum_{\substack{x_1 \in \mathcal{X}_1 \\ x_2 \in \mathcal{X}_2}} P(x_1, x_2) \log (P(x_1, x_2)) 
	\end{align}
	
	The generalization to more than two variables is obvious. If and only if variables $X_1$ and $X_2$ are all statistically independent, then the joint entropy is equal to the sum of the marginal entropies:
	
	\begin{equation}
	H(X_1, X_2) = H(X_1) + H(X_2) \iff X_1 \bot X_2
	\end{equation}
	
	In the case of complex systems, this independence condition is rarely satisfied and the joint and marginal entropies are generally related by the inequality:
	
	\begin{equation}
	H(X_1, X_2) \le H(X_1) + H(X_2)
	\label{eq:cond_ent_ineq}
	\end{equation}
	
	Equation \ref{eq:cond_ent_ineq} tells us that we can never be \textit{more} uncertain about the states of $X_1$ and $X_2$ when considering them together than if we add our uncertainty about each variable individually. We note this by saying that the entropy function is fundamentally ``subadditive." As we will see in Sec. \ref{sec:mutual_information}, if $H(X_1,X_2) \le H(X_1) + H(X_2)$, then $X_1$ and $X_2$ have a non-zero mutual information and knowing the state of one necessarily reduces our uncertainty about the state of the other. 
	
	The joint entropy formula highlights a key fact about information theory: variables can be "coarse-grained", and every joint distribution can be cast as a single-variable distribution by aggregating multiple ``micro"-scale variables into a single ``macro"-scale variable. For example, imagine $X_1$ and $X_2$ are two fair, independent coins. The joint states are given by \{00, 01, 10, 11\}, however we could easily describe both variables as a single 	``macro"-variable \textbf{X} with states \{A,B,C,D\}. In this case:
	
	\begin{equation}
	H(X_1,X_2) = H(\textbf{X})
	\end{equation}
	
	More complex configurations are also allowable. For example, if we had four micro-state variables $X_1,X_2,X_3,X_4$, they could either be coarse-grained into \textbf{X}, or we could map $X_1$ and $X_2$ to a macro $\textbf{X}_\alpha$ and $X_3$ and $X_4$ to macro $\textbf{X}_\beta$. Once again:
	
	\begin{equation}
	H(X_1,X_2,X_3,X_4) = H(\textbf{X}_\alpha, \textbf{X}_\beta)
	\end{equation}
	
	This ability to move easily between analysis of ``wholes" (eg, \textbf{X, X}$_\alpha$, and \textbf{X}$_\beta$), and constituent ``parts" (eg. $X_1,X_2,X_3$, and $X_4$) makes information theory very useful when assessing the structure of complex systems, which are typically multi-scale, displaying non-trivial dynamics and structure at a variety of levels in a hierarchy. For example, a group of interacting neurons can be considered at the level of the whole set (e.g. a ganglion on brain region), or at the level of each neuron and how subsets of the group interact with each-other. For an example of this, see Sec. \ref{sec:mi_tc}.
	
	\subsubsection{Conditional Entropy}
	
	The conditional entropy quantifies how uncertain we are about the state of variable \textit{given that we have knowledge of another variable}. As with many information theoretic measures, it is best understood in a Bayesian context: our uncertainty about a variable will change depending on what information we have about the context around it. \textit{The entropy of a variable describes \textbf{your} uncertainty about the state of that variable, and is not a value intrinsic to the real world in the way something like mass is.} Our level of certainty changes depending on what information we have.  
	
	For two variables, the conditional entropy $H(X_1 | X_2)$ gives us a measure of how uncertain we are about the state of $X_1$ after we account for information from $X_2$. It is defined by:
	
	\begin{align}
	H(X_1 | X_2) &= \mathbb{E}_{X_1,X_2}[h(x_1 | x_2)] \\ 
	&= -\sum_{\substack{x_1 \in \mathcal{X}_1 \\ x_2 \in \mathcal{X}_2}} P(x_1, x_2) \log (P(x_1 | x_2)) 
	\label{eq:cond_ent_main}
	\end{align}
	
	Note that the expectation is taken with respect to the joint probability $P(X_1,X_2)$ rather than the conditional probability $P(X_1|X_2)$. This is because of the presence of two interacting variables. If we wanted to calculate the conditional entropy of $X_1$, given that we see $X_2$ in some particular, fixed state $x_2$, then the entropy formula takes on it's usual pattern:
	
	\begin{align}
	H(X_1 | X_2=x_2) &= \mathbb{E}_{X_1|X_2=x_2}[h(x_1 | X_2=x_2)] \\ &= -\sum_{x_1 \in \mathcal{X}_1}P(x_1 | X_2=x_2)\log(P(x_1 | X_2=x_2))
	\end{align}
	
	However, if we want to know the effect of \textit{all} possible $x_2 \in \mathcal{X}_2$, we need to sum the conditional entropy of each $x_2$, and weight the resulting contribution of each by the probability that we see that particular state of $X_2 = x_2$. 
	
	\begin{equation}
	H(X_1 | X_2) = -\sum_{x_2 \in X_2}P(x_2)H(X_1 | X_2 = x_2)
	\end{equation}
	
	This results in a ``nested expectation", which is equivalent to Eq. \ref{eq:cond_ent_main}. The conditional entropy can also be written in terms of the joint and marginal entropies:
	
	\begin{equation}
	H(X_1 | X_2) = H(X_1, X_2) - H(X_2)
	\end{equation}
	
	The intuition is that the uncertainty about $X_1$ that is left over when uncertainty about $X_2$ is removed from the joint state of both. As with the joint entropy, if and only if all the variables are independent then:
	
	\begin{equation}
	H(X_1 | X_2) = H(X_1) \iff X_1 \bot X_2
	\end{equation}
	
	In other words, if knowing the state of $X_2$ does nothing to reduce our uncertainty about the state of $X_1$, then they must be independent. This is, in some sense, the most fundamental definition of independence possible. In general:
	
	\begin{equation}
	H(X_1 | X_2) \le H(X_1)
	\end{equation}
	
	As with the joint entropy inequalities, this is understood as indicating that knowing the state of other variables can only ever \textit{decrease} our uncertainty (or in the case of independent variables, leave it unchanged). There is no context in which knowing information about $X_2$ could ``obfuscate" our knowledge about $X_1$ and make us more uncertain.  
	
	It is crucial to understand that, like all entropy measures, the conditional entropy returns a measure of uncertainty about a variable. There is a tendency among students to read $H(X_1 | X_2)$ as ``the amount of information $X_2$ provides about $X_1$", which is an incorrect interpretation. This confuses the conditional entropy with the mutual information.
	
	\subsection{Relative Entropy}
	\label{sec:dkl}
	
	Also known as the Kullback-Liebler Divergence ($D_{KL}$), the relative entropy provides a measure of how our certainty about the state a variable (or set of variables) changes depending on our beliefs (in this case, various probability distributions $P$ and $Q$ we could use to model a random variable $X$). Formally:
	
	\begin{align}
	D_{KL}(P || Q) &= \mathbb{E}_{P(X)}\bigg[\log\bigg(\frac{P(X)}{Q(X)}\bigg)\bigg] \\ &= \sum_{x \in \mathcal{X}}P(x) \log \bigg(\frac{P(x)}{Q(x)} \bigg)
	\end{align}
	
	The KL-divergence has a very natural Bayesian interpretation: if $Q(X)$ is our prior distribution on $X$, and $P(X)$ is our posterior distribution inferred given some data, than the KL-Divergence is the information \textit{gained} when we revise our prior beliefs $Q$ to the updated posterior beliefs $P$. Conversely, the KL-Divergence can be thought of as the amount of information \textit{lost} when $Q$ is used to approximate $P$.
	
	\textbf{Note}: The term $D_{KL}(P || Q)$ is typically read as "the KL-Divergence from $Q$ to $P$," despite the fact that $P$ is written first in the argument. It can alternately be read as "the relative entropy of $P$ with respect to $Q$."
	
	The KL-divergence between two distributions can be (naively) thought of as giving a ``distance" between the two distributions, although it is important to realize that it is not a true metric. It cannot be assumed that $D_{KL}(P || Q) = D_{KL}(Q||P)$, nor can it be assumed that it satisfies the triangle inequality (i.e. for three distribution $P$, $Q$, and $W$, $D_{KL}(P||W) \not\le D_{KL}(P||Q) + D_{KL}(Q || W)$). A ``corrected" KL-divergence, called the Jensen-Shannon divergence is symmetric however \cite{cover_elements_2012}:
	
	\begin{align}
		D_{JS}(P||Q) := \frac{D_{KL}(P||M) + D_{KL}(Q||M)}{2}
	\end{align}

	Where $M$ is the mixture distribution constructed from $P$ and $Q$:
	
	\begin{align}
		M := \frac{P+Q}{2}
	\end{align}

	While the $D_{JS}$ is symmetric, it is still not a true metric, however the square root is: $\sqrt{D_{JS}(P||Q)}$ is often called the Jensen-Shannon distance from $P$ to $Q$ \cite{endres_new_2003}. 
	
	\subsubsection{Local Relative Entropy}
	
	As with the Shannon entropy, the relative entropy measure is an expectation value averaged over every value of $x \in \mathcal{X}$. This expectation can be ``unrolled" to get the information associated with every specific value $X$ can take on. 
	
	\begin{align}
	d_{kl}(P(x)||Q(x)) &= \log\bigg(\frac{P(x)}{Q(x)}\bigg) \\ 
	&= h_Q(x) - h_P(x)
	\end{align}
	
	We can see that $d_{kl}(P(x)||Q(x))$ tells us how much more surprised we are to see $X=x$ when we are basing our predictions on the prior distribution of $Q(X)$ compared to when we use the posterior distribution $P(X)$. The full KL-divergence, then, can be thought of as the average, element-wise difference in the surprise for every state $x$ in the two distributions $P$ and $Q$, weighted by the posterior distribution. The full relative entropy is always greater than zero, however, for specific local values, it can be less than zero, indicating that, for these values, we would be less surprised to see them if we were basing our predictions on the prior, rather than the posterior. 
	
	\subsection{Mutual Information}
	\label{sec:mutual_information}

	\begin{figure}
		\begin{center}
			
			\def\firstcircle{(0,0) circle (3cm)}
			\def\secondcircle{(3,0) circle (3cm)}
			
			\begin{tikzpicture}
			
			\draw \firstcircle node[] at (-1.5,0) {$H(X|Y)$};
			\draw \secondcircle node[] at (4.5,0) {$H(Y|X)$};
			
			\node[] at (1.5,0) {$I(X;Y)$};
			\node[] at (1.5,3.5) {$\underline{H(X,Y)}$};
			\node[] at (5.5,2.5) {$\underline{H(Y)}$};
			\node[] at (-2.5,2.5) {$\underline{H(X)}$};
			
			\end{tikzpicture}
		\end{center}
		\label{fig:mi_venn}
		\caption{An entropy diagram showing how the marginal, conditional, and joint entropies are related the mutual information for two interacting variables $X$ and $Y$. The area of each circle corresponds to the \textit{amount of uncertainty} we have about the state of each variable \textbf{not} the amount of information we have (this is a common misinterpretation). Each circle corresponds to our total uncertainty about $X$ or $Y$ respectively. The intersection of the Venn diagram is that uncertainty that is common to both variables. If we were to resolve \textit{all} of our uncertainty about $Y$, we would also be resolving some uncertainty about $X$: this overlap is the mutual information $I(X;Y)$. We can also see the conditional entropy is that uncertainty specific to one variable that is left over after resolving the uncertainty specific to the other. Finally, the joint entropy is given by the union of both marginal entropies. From this diagram we can get a visual intuition for the various definitions of mutual information: it's clear that $I(X;Y) = H(X) + H(Y) - H(X,Y) = H(X,Y) - H(Y|X) - H(X|Y) $.}	
	\end{figure}
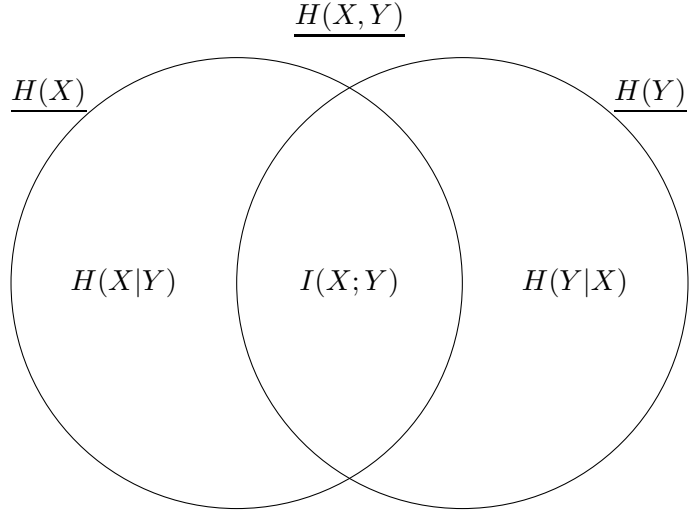

	Like the Shannon entropy, the mutual information between two variables $X_1$ and $X_2$ is one of the foundational measures of information theory.\textit{ Mutual information provides a measure of the statistical dependency between two random variables: how much does knowing the state of $X_1$ reduce our uncertainty about the state of $X_2$ (and vice versa)?} Recall from Sec. \ref{sec:entropy:required} that ``information" is equivalent to ``reduction in uncertainty", so mutual information provides the first real measure of information we have encountered so far. For two random variables, it has a number of equivalent forms:
	
	\begin{align}
	\label{eq:mi_cond_1}
	I(X_1;X_2) &:= H(X_1) - H(X_1|X_2) \\
	&= H(X_2) - H(X_2|X_1) \label{eq:mi_cond_2}\\
	&= H(X_1)+H(X_2)-H(X_1,X_2) \label{eq:mi_1}\\
	&= H(X_1,X_2) - H(X_1|X_2) - H(X_2|X_1) \label{eq:mi_2}
	\end{align}
	
	It can be helpful to break Eq. \ref{eq:mi_cond_1} down into it's component parts. $H(X_1)$ quantifies our initial uncertainty about the state of $X_1$. The conditional entropy $H(X_1|X_2)$ quantifies how much uncertainty remains \textit{after} learning the state of $X_2$. The difference between these terms is the uncertainty about the state of $X_1$ that is \textit{resolved} by learning the state of $X_2$: the information $X_2$ discloses about $X_1$. Note that while generally $H(X_1) \not= H(X_2)$ and $H(X_1|X_2)\not=H(X_2|X_1)$, the mutual information is symmetric in its arguments (i.e. $I(X_1;X_2)=I(X_2;X_1)$). The other formulations can be derived via simple algebra. See Fig. \ref{fig:mi_venn} for a visual aid into the relationships between Eqs. \ref{eq:mi_cond_1}, \ref{eq:mi_cond_2}, \ref{eq:mi_1}, and \ref{eq:mi_2}.  
	
	The mutual information can also be expressed in terms of the Kullback-Leibler divergence:
	
	\begin{equation}
	I(X_1, X_2) = D_{KL}(P(X_1, X_2) || P(X_1)\times P(X_2))
	\label{eq:mi_dkl}
	\end{equation}
	
	This means that the measure has a natural Bayesian interpretation: in this case, the prior distribution is the product of the marginals: how we would expect the pair of variables to behave if they were independent, with no statistical coupling. The posterior is the true joint probability distribution. The mutual information gives us the \textit{increase in information} we gain when we update our prior (the two variables are uncoupled) to the posterior (the variables are coupled and disclose information about each-other). If $X_1$ and $X_2$ truly are coupled, than knowing the state of one should reduce our uncertainty about the state of the other and vice-versa. 
	
	As with the entropy, the mutual information can also be understood as an average measure, formulated as an expectation over the joint state $P(X_1,X_2)$ as seen in Eq. \ref{eq:mi_main}:
	
	\begin{align}
	I(X_1 ; X_2) &= \mathbb{E}_{X_1,X_2}\bigg[\log \bigg( \frac{P(x_1| x_2)}{P(x_1)} \bigg)\bigg] \\ 
	&= 
	\sum_{\substack{x_1 \in \mathcal{X}_1 \\ x_2 \in \mathcal{X}_2}} P(x_1, x_2) \log \bigg ( \frac{P(x_1| x_2)}{P(x_1)} \bigg)
	\label{eq:mi_main}
	\end{align}
	
	The expectation formulation provides another window into the Bayesian interpretation of mutual information. The denominator, $P(X_1)$ can be thought of as our prior beliefs about the distribution of $X_1$. The numerator $P(X_1|X_2)$ is the posterior: our revised beliefs about the distribution of $X_1$ after we have factored in knowledge about the state of $X_2$. 
	
	\subsubsection{Local Mutual Information}
	
	Like all the other information theoretic measures we have seen so far, mutual information is an expected value, which can be ``unrolled", to get the information associated with every possible combination of $X_1$ and $X_2$ values. 
	
	\begin{align}
	i(x_1, x_2) &= \log\bigg(\frac{P(x_1|x_2)}{P(x_1)}\bigg) \\
	&= \log\bigg(\frac{P(x_1,x_2)}{P(x_1)\times P(x_2)}\bigg)
	\end{align}
	
	As with local relative entropy, this measure can be positive or negative. If $P(x_1|x_2) > P(x_1)$, then $i(x_1,x_2) > 0$. Conversely, if $P(x_1|x_2) < P(x_1)$, then $i(x_1,x_2) < 0$. This second case can be thought of as ``misinformation:" it appears that the particular joint state we are observing would be \textit{more likely} if $X_1$ and $X_2$ are independent than if they are coupled. Conveniently, the local mutual information relates to the local entropies in all the same ways as the expected mutual information relates to the expected entropies:
	
	\begin{eqnarray}
	i(x_1, x_2) &=& h(x_1) + h(x_2) - h(x_1, x_2) \\
	&=& h(x_1,x_2) - h(x_1|x_2) - h(x_2|x_1) \\
	&=& h(x_1) - h(x_1 | x_2)
	\end{eqnarray}
	
	The local mutual information is not as well-known as the global expected value, however it has been used in a number of different contexts, such as computational chemistry \cite{cmelo_profiling_2021}, natural language processing \cite{church_word_1990}, network science \cite{luo_highly-accurate_2021,varley_flickering_2023}, computational neuroscience \cite{pope_time-varying_2024}, and image segmentation \cite{gueguen_local_2014}. In timeseries analysis, the local mutual information can be used to construct an ``edge-timeseries" \cite{faskowitz_edge-centric_2020, sporns_dynamic_2021}, which provides a measure of instantaneous coupling between two processes (for example, when are two processes behaving ``is if" they were coupled vs. ``is if" they were independent).
	
	\subsubsection{Conditional mutual information}
	\label{sec:cond_mi}
	
	As with entropy, the amount of information one element of a system discloses about another can change depending on what information is available from a third (or more) variables. The conditional mutual information between $X_1$ and $X_2$, given $X_3$ is given as:
	
	\begin{eqnarray}
	I(X_1 ; X_2 | X_3) &=& H(X_1 | X_3) + H(X_2 | X_3) - H(X_1, X_2 | X_3)
	\label{eq:cond_mi_1} \\
	&=& H(X_1 | X_3) - H(X_1 | X_2, X_3) \label{eq:cond_mi_2} \\
	&=& H(X_1, X_3) + H(X_2, X_3) - H(X_1, X_2, X_3) - H(X_3) \label{eq:cond_mi_3}
	\end{eqnarray}
	
	Equation \ref{eq:cond_mi_1} shows that the conditional mutual information can be thought of as a measure of the degree to which knowing the state of $X_1$ resolves uncertainty about $X_2$ when information from $X_3$ has been accounted for. This implies that information is distributed over three (or more) variables in a way that constrains how the relationship between $X_1$ and $X_2$ is interpreted. 
	As with the standard, bivariate mutual information, the local conditional mutual information can also be written out in terms of local joint and conditional entropies in a manner analogous to Eqs. \ref{eq:cond_mi_1}, \ref{eq:cond_mi_2}, and \ref{eq:cond_mi_3}.
	
	There are two contexts in which this kind of three-way interaction can occur: when information is \textit{redundantly} shared over multiple variables and when information is \textit{synergistically} present in the joint states of multiple variables. In both cases, examining only the pairwise relationships between elements provides an inaccurate picture of the system under study. Generally, redundant information will result in over-estimates of the extent to which elements are interacting, while synergistic information will result in under-estimates of th same. As we will discuss in Sections \ref{sec:fc_nets} and \ref{sec:eff_nets}, this can significantly complicate the problem of constructing network models of complex systems from data. 
	
	\begin{flushleft}
		\small{\textbf{Redundant (Shared) Information}}
	\end{flushleft}
	
	To understand redundant information, consider the case where $X_1$ and $X_2$ both ``listen to" $X_3$, but have no coupling between each-other (no information passes from $X_1$ to $X_2$, but they both receive information from $X_3$). 
	
	\begin{align}
		X_1 \leftarrow X_3 \rightarrow X_2 \nonumber
	\end{align}
	
	Here we have a "common driver effect": even though there is no coupling between $X_1$ and $X_2$, we find that $I(X_1 ; X_2) > 0$ bit because both are evolving based on a shared upstream input. If we were trying to determine whether $X_1$ was correlated with $X_2$, we would erroneously infer a link between the pair. However, if we condition the mutual information between $X_1$ and $X_2$ on $X_3$, we find that: $I(X_1;X_2|X_3) = 0$ bit. 
	
	Why does this happen? 
	
	Consider Eq. \ref{eq:cond_mi_1}: all the entropy terms have become conditional: if $X_1$ and $X_2$ are both "listening" to the upstream element $X_3$, then knowing the state of $X_3$ resolves uncertainty about the states of both $X_1$ and $X_2$: $H(X_1 | X_3) \rightarrow 0$ bit and $H(X_2 | X_3) \rightarrow 0$ bit. Furthermore, because $X_3$ is a common driver of both $X_1$ and $X_2$, conditioning the joint states of both downstream elements on $X_3$ results in $H(X_1, X_2 | X_3) \rightarrow 0$ because $X_3$ is providing the \textit{same} information to both child elements. Whatever intrinsic uncertainty is left in $X_1$ after all the information from $X_3$ is accounted for won't be resolved by knowledge about $X_2$: at this point, the redundant information has been conditioned out, and since there's no connection between the two, there won't be any statistical dependency.
	
	Here, the process of conditioning the mutual information between two variables and a third is very similar to the process of "regressing out" or "controlling for" a variable in more standard statistical analyses, although as usual, there are no assumptions about linearity or Gaussianity being made here.  
	
	\begin{flushleft}
		\small{\textbf{Synergistic Information}}
	\end{flushleft}
	
	Synergistic information is a little bit more complicated than redundant information, although toy models can help build intuition. Defined generally, synergy referrers to information that is only present in the joint-states of two or more elements of the system and cannot be extracted from single variables considered alone. 
	
	Consider three variables, $X_1$, $X_2$, and $X_3$, where $X_1$ and $X_2$ are both random, maximally entropic binary variables, and $X_3 = XOR(X_1, X_2)$ (for the lookup table for the logical XOR gate, see Appendix). 
	
	\begin{align}
		X_1 \rightarrow X_3 \leftarrow X_2 \nonumber
	\end{align}

	Despite the fact that $X_3$ is a deterministic function of both $X_1$ and $X_2$, both $I(X_1 ; X_3) = 0$ bit and $I(X_2 ; X_3) = 0$ bit. Due to the synergistic nature of the XOR function, the relationship between all three variables only comes into view when information from all three are accounted for: $I(X_1 ; X_3 | X_2) = 1$ bit. This occurs because the output of the XOR function cannot be known from a single argument: knowledge about both $X_1$ and $X_2$ are necessary to determine whether $XOR(X_1, X_2)=0$ or 1. 
	
	The somewhat counter-intuitive result is that attempts at pairwise inference of connections between $X_1$, $X_2$, and $X_3$ will all return values of 0, however by conditioning we can find that a sort of hyper-edge exists between $X_1,X_2$ and $X_3$: it is only when both $X_1$ and $X_2$ are considered together that we can gain information about $X_3$. Individually, they are completely uninformative:
	
	\begin{align}
		I(X_1;X_3) = 0 \nonumber \\
		I(X_2;X_3) = 0 \nonumber \\ 
		I(X_1,X_2;X_3) = 1 \nonumber 
	\end{align}

	\subsection{Multivariate Generalization of Mutual Information}
	\label{sec:multi_mi}

	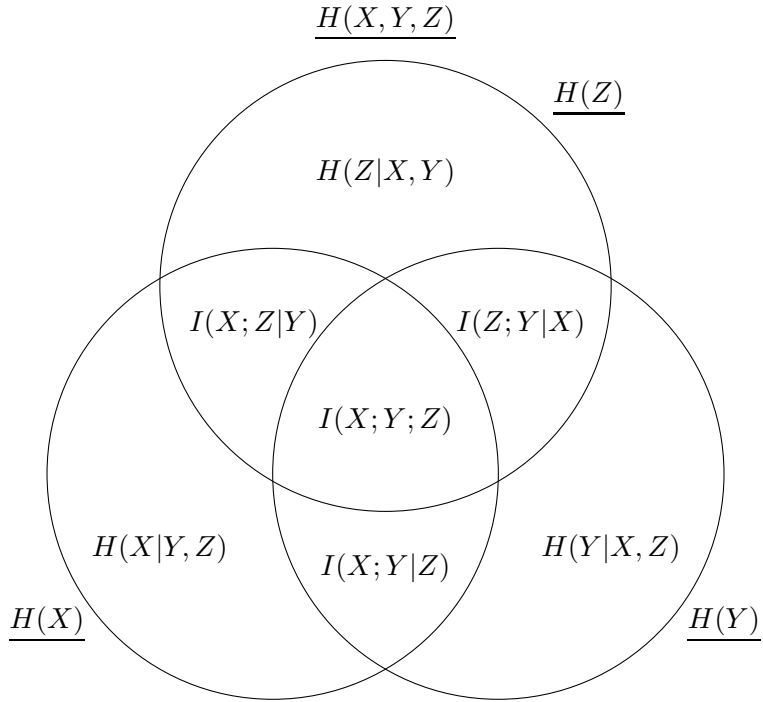
\begin{figure}
		\begin{center}
			
			\def\firstcircle{(0,0) circle (3.5cm)}
			\def\secondcircle{(3,0) circle (3.5cm)}
			\def\thirdcircle{(1.5,2.5) circle (3.5cm)}
			
			\begin{tikzpicture}
			
			\draw \firstcircle node[] at (-1.75,-1) {$H(X_1|X_2,X_3)$};
			\draw \secondcircle node[] at (4.75, -1) {$H(X_2|X_1,X_3)$};
			\draw \thirdcircle node[] at (1.5, 4) {$H(X_3|X_1,X_2)$};
			
			\node[] at (1.5,0.7) {$I(X_1;X_2;X_3)$};
			\node[] at (1.5,-1.5) {$I(X_1;X_2|X_3)$};
			\node[] at (3.75,2.5) {$I(X_3;X_2|X_1)$};
			\node[] at (-0.75, 2.5) {$I(X_1;X_3|X_2)$};
			
			\node[] at (1.5,6.5) {$\underline{H(X_1,X_2,X_3)}$};
			\node[] at (4.75,5.25) {$\underline{H(X_3)}$};
			\node[] at (6.5,-2.5) {$\underline{H(X_2)}$};
			\node[] at (-3.5,-2.5) {$\underline{H(X_1)}$};
	
			\end{tikzpicture}
		\end{center}	
		\label{fig:multi_mi_venn}
		\caption{This Venn diagram generalizes the relationships introduced in Figure \ref{fig:mi_venn}. However, care should be taken when considering entropy diagrams for more than two variables: the innermost intersection ($I(X_1;X_2;X_3)$, corresponding to the co-information or interaction information) is \textit{not} strictly positive \cite{matsuda_physical_2000}. The sign refers to whether the trivariate relationship is redundancy or synergy dominated \cite{williams_nonnegative_2010}, discussed below.}
	\end{figure}

	A defining feature of complex systems, however, is that they are comprised of a large number of interacting elements, often numbering in the hundreds or thousands. This has lead to a natural interest in information theoretic tools that can assess the structure of many-element systems and how they can interact. Arguably the most straightforward attempts have been multivariate generalizations of mutual information, although this turns out to not be as straightforward as one might think.
	
	Several attempts have been made to generalize the notion of mutual information to more than two variables, although different intuitions can lead to different formalisms. For example, should multivariate mutual information refer to only that information which is redundantly present in \textit{all} $N$ elements, or should it refer to \textit{any} information that is present in at least two or more elements jointly? While these are identical concepts in the bivariate case, as more variables are added, the situation becomes more complex and different generalizations will quantify different ``notions" of shared information.
	
	\subsubsection{Total Correlation}
	\label{sec:mi_tc}
	The total correlation \cite{watanabe_information_1960}, also known as multiinformation \cite{studeny_multiinformation_1998}, integration \cite{tononi_measure_1994} is a multivariate mutual information that generalizes the notion of mutual information as the KL-Divergence between the joint and the independent distributions of all the elements. 
	
	\begin{equation}
	TC(\textbf{X}) = D_{KL}(P(X_1, X_2, ... X_N)||P(X_1) \times P(X_2)\times...\times P(X_N))
	\label{eq:tc_1}
	\end{equation}
	
	The total correlation is also the natural multivariate generalization of Eq. \ref{eq:mi_1}:
	
	\begin{equation}
	TC(\textbf{X}) = \bigg( \sum_{i=0}^{N}H(X_i) \bigg) - H(\textbf{X})
	\label{eq:tc_2}
	\end{equation}
	
	Examining Eq. \ref{eq:tc_2} can provide insights into the behaviour of the total correlation function. $TC(\textbf{X})$ will be high when all of the individual elements are maximally entropic (that is to say, all possible states appear equiprobably) but the joint state of the system has a low entropy. Said differently, total correlation is high when every individual element explores it's entire state space, but the joint state of the system is restricted to a small subset of the joint state-space.
	
	A good example of this would be synchronization (itself a core topic in complex systems research \cite{strogatz_sync_2012,okeeffe_oscillators_2017}): imagine a system comprised to 10 elements ($\textbf{X} = \{X_1, X_2,...X_{10}\}$), each of which can be in 8 states ($\{S_1, S_2, ... S_8\}$). If all the elements are synchronized and cycling through the sequence of states at the same time, then the entropy of each element will be 3 bit, and the sum of the entropies of all channels will be 30 bit. When considering the joint entropy of the system we find that, in contrast to the individual elements, each of which explores it's full state-space, the system as a whole is restricted to just 8 joint-states out of a possible $8^{10}$ configurations (since every observed joint state is just $S_i$ repeated 10 times). Consequently, the joint entropy is also 3 bit, giving a final total correlation of 27 bit. In contrast, the total correlation would be low when every element is independent of every other, in which case the sum of the individual entropies would be equal to the joint entropy. 
	
	Since the total correlation is an extension of the the KL-divergence between the joint distribution and the product of the marginals, it lends itself to the same intuitive Bayesian interpretation as bivariate mutual information. If our prior is that all elements of the system are disconnected, then we would expect $H(\textbf{X}) >> H(X_i)$, however, in the case of our totally synchronized elements, this would be incorrect and the \textit{increase} in our ability to predict the joint state of the system is indicated by the high total correlation value. That increase in predictive power is a result of the fact that the system explores a very small part of the joint state-space (8 configurations as opposed to $8^{10}$), and so we have far less uncertainty about the joint state at any given time. 
	
	\subsubsection{Dual Total Correlation}
	\label{sec:dtc}
	The dual total correlation \cite{han_linear_1975}, also known as the binding entropy \cite{rosas_quantifying_2019} is another generalization of mutual information to more than two interacting elements. It is based on a generalization of Eq. \ref{eq:mi_2}:
	
	\begin{equation}
	\label{eq:dtc}
	DTC(\textbf{X}) = H(\textbf{X}) - \sum_{i=0}^{N} H(X_i|\textbf{X}_{-i})
	\end{equation}
	
	Where $\textbf{X}_{-i}$ indicates the set of all elements \textit{excluding} the $ith$ element. The conditional entropy term $H(X_i | \textbf{X}_{-i})$ quantifies the uncertainty in the $i^{th}$ element that cannot be resolved with information from any other combination of elements in the system: the uncertainty that is intrinsic to that element in particular. This is often referred to as the residual entropy \cite{rosas_quantifying_2019}. The  dual total correlation gives the amount of information left over when the intrinsic information associated with each individual element is removed from the joint entropy of the whole: that is to say, all the information that is shared by two or more elements of the system. 
	
	Unlike the total correlation, which monotonically increases with the amount of dependency between the elements, the  dual total correlation has the interesting property of being low both when all the elements of the system are totally coupled (as in the synchronization example given above), and totally independent. In the case of the 10-element, totally synchronized system described above, the  dual total correlation is equivalent to the joint entropy (3 bit), since there is no residual entropy specific to any element (the state of any part of the system can be predicted with total accuracy with knowledge of any other element, since they're always all the same). Similarly, if all the elements are independent, then the joint entropy is high ($\approx \log(8^{10})$), but so is the sum of all the residual entropies (since, by definition, information about the $i^{th}$ element cannot be extracted from independent variables). 
	
	The  dual total correlation is high in an intermediate zone where the joint entropy of the whole system is high, but the residual entropy of every element is low (i.e. the state of any individual element can be predicted with a high degree of certainty with information present in some combination of other elements). The occurs when complex, informative dependencies exist between the elements in a way that doesn't overly-constrain the flexibility of the whole.
	
	\subsubsection{Co-Information}
	
	Also known as the interaction information \cite{mcgill_multivariate_1954}, the co-information is a multivariate generalization of mutual information that attempts to preserve the notion that mutual information is that information present in \textit{all} the interacting variables. Contrast this with the total correlation (which focuses on mutual information as the $D_{KL}$ between the joint and independent distributions) and the dual total correlation (which aggregates all the information not specific to a single variable) \cite{mackay_information_2003,cover_elements_2012}. 
	
	Based on the diagram in Fig. \ref{fig:multi_mi_venn}, we can see that, for three variables, the $C(X_1;X_2;X_3)$ is the information that is present in all three variables and can be calculated as: 
	
	\begin{equation}
	\begin{aligned}
	C(X;Y;Z) &=& I(X_1;X_2)-I(X_1;X_2|X_3) \label{eq:coinfo} \\ 
	&=& I(X_1;X_3)-I(X_1;X_3|X_2)  \\ 
	&=& I(X_2;X_3)-I(X_2;X_2|X_1)  
	\end{aligned}
	\end{equation}
	
	Equation \ref{eq:coinfo} shows that, for three elements, the co-information is the difference between the initial information that $X_1$ discloses about $X_2$ based on the joint probability distribution $P(X_1,X_2)$, and the information that $X_1$ discloses about $X_2$ after we have observed $X_3$ (i.e. based on the joint probability distribution $P(X_1,X_2,X_3)$). A similar analysis reveals that the co-information can alternatively be formalized using the intersection inclusion-exclusion criteria for overlapping sets :
	
	\begin{equation}
	\begin{aligned}
	C(X_1;X_2;X_3) &= H(X_1) + H(X_2) + H(X_3)& \\
	&- H(X_1,X_2) - H(X_1,X_3) - H(X_2,X_3)& \\
	&+ H(X_1,X_2,X_3)&
	\end{aligned}
	\end{equation}
	
	This sets up a general pattern for sets of $n$-interacting variables: the entropies of successively larger combinations of variables are added and subtracted based on the parity of the number of variables in the set:
	
	\begin{equation}
	C(\textbf{X}) = -\sum_{\xi \subseteq \textbf{X}} -1^{(|\textbf{X}| - |\xi|)}H(\xi) 
	\end{equation}
	
	Where $\xi$ indicates all possible subsets of $\textbf{X}$. Unfortunately, unlike the total and dual total correlations, which easily accommodate large numbers of variables without changing the interpretation, the co-information can be hard to interpret for more than 3 elements. A significant issue is that the co-information can be negative, which has historically limited its applicability outside of theoretical contexts. \cite{williams_nonnegative_2010} showed that, for three variables, the sign of co-information can be understood as indicating whether the statistical dependencies within the dataset are dominated by redundancy or synergy, although this pattern does not necessarily generalize (compare with the O-information, introduced in Sec. \ref{sec:complexity}). 
	
	Despite it's practical limitations, however, the co-information has served as a useful jumping off point for research into how mutual information may be further decomposed into redundant and synergistic components (for example, see \cite{ince_measuring_2017,ince_partial_2017,goodwell_temporal_2017}, for a discussion of partial information decomposition, see Sec. \ref{sec:pid}.
	
	\newpage 
	
	\section{Information Dynamics}
	\label{sec:info_dynamics}
	
	Information dynamics deals with the question of how complex systems ``compute" (note the scare quotes), and sits at the intersection of dynamical systems analysis, information theory, computer science, and information theory.
	
	To begin with, what does it mean for a system to ``compute"? This inevitably raises the complex, philosophical question ``what is computation", which we will sidestep entirely by operationalizing ``computation" in a complex system as the process by which the system decides its immediate next state \cite{varley_synergistic_2024}. This can happen at every scale: individual elements can be in multiple states and decide their immediate next state based on the interactions between their own internal dynamics and the states of all their neighbors. At the macro-scale, the dynamics of the joint-state of the whole system ``emerges" from the interactions of all it's constituent elements \cite{lizier_local_2013}. 
	
	Information dynamics are typically broken down into three components: 
	
	\begin{enumerate}
		\item Information storage \cite{lizier_local_2012}, which characterizes how the past of an element or system constrains its immediate future, 
		\item Information transfer \cite{lizier_local_2008}, which characterizes how the past of a source element constrains the immediate future of a target element, 
		\item Information modification \cite{lizier_information_2010,lizier_towards_2013}, which describes how multiple streams of information are combined to produce an output that is not trivially reducible to any input. 
	\end{enumerate}
	
	These three measures can be seen as \textit{very roughly} analogous to the fundamental processes at work in a digital computer: information can be actively stored in memory and retrieved, information can be relayed from one processor to another, and information can be integrated in the form of a ``computation," to produce some new information. When modeling a complex system, understanding the interplay between information storage, transmission, and modification can help elucidate how the behavior of individual components constrains the evolution of the system as a whole.  
	
	For this section, we will be focusing on the information dynamics of a single target (which is the most well-developed area of the field at present). Recently, the notion of information dynamics has been generalized to account for redundant and synergistic interactions between multiple variables: we will briefly discuss this in Sec. \ref{sec:phi_id}, and interested readers can consult \cite{mediano_beyond_2019,mediano_towards_2021}.
	
	\subsection{Conditional Entropy Rate}
	
	The simplest information dynamics measure is the conditional entropy rate \cite{crutchfield_regularities_2003,wibral_local_2014}, which measures how much uncertainty about the next state of $X$ remains after the past $k$ states are accounted for:
	
	\begin{equation}
	H_{\mu}(X) = \lim_{k\to\infty} H(X_{t} | X_{-k:t-1})
	\end{equation}
	
	As $k\to\infty$, the entropy rate gives a measure of unpredictability inherent in $X$ which can never be resolved simply by observing $X$ itself, no matter how long $X$ is watched. For continuous processes, $H_{\mu}$ is approximated by a number of measures, including sample entropy and approximate entropy \cite{wibral_local_2014}, and for discrete processes, can be approximated using the Lempel-Ziv complexity \cite{amigo_estimating_2004}.
	
	As with all entropy measures, it is possible to calculate the local conditional entropy rate associated with a particular configuration of $X$:
	
	\begin{equation}
	h_{\mu}(x) = \lim_{k\to\infty}h(x_t | x_{-k:t-1})
	\end{equation}

	where $k$ indicates the number of discrete past states to take into account. Following the usual formalism, $X_t$ refers to the state of $X$ at time $t$, while $X_{-k:t-1}$ indicates a vector of states, stretching from time $t-1$ all the way back to time $t-1-k$. This tells us how surprising any particular realization $x$ is, given that it is preceded by the particular sequence $x_{-k:t-1}$. When $h_{\mu}(x)$ is high, then \textit{X} has done something very unexpected and inconsistent with its past behavior, while when $h_{\mu}(x)$ is low, whatever state $X$ has just adopted would be unsurprising to anyone who had been watching it for the last $k$ timesteps.  The local conditional entropy rate has been used in thermodynamic interpretations of the transfer entropy measure \cite{prokopenko_thermodynamic_2013}.
	
	\subsection{Information Storage}
	
	The conditional entropy rate quantifies the ``left over" uncertainty intrinsic to $X$ after the past has been accounted for. Its dual measure is the active information storage \cite{lizier_local_2012,wibral_local_2014}, which quantifies how much information the \textit{past} of a variable discloses about its immediate next state. In this context, ``information storage" can be thought of as a measure of how dependent an element of a system is on it's own past. If every next state is chosen at random, then $H_{\mu}(X) = H(X)$ and $A(X) = 0$. Formally, active information storage is given by: 
	
	\begin{equation}
	A(X) = \lim_{k\to\infty}I(X_{-k:t-1};X_t)
	\end{equation}
	
	The active information storage and entropy rate are complementary: the uncertainty about the next state of $X$ ($H(X)$) can be decomposed into $A(X)$ and $H_{\mu}(X)$ \cite{lizier_local_2012}:
	
	\begin{equation}
	H(X) = A(X) + H_{\mu}(X)
	\end{equation}
	
	Intuitively, this shows us that some of our uncertainty about the next state of $X$ could be resolved by looking at $X$'s past, and a residual part that cannot be resolved. This is a key insight when attempting to model complex systems. For processes with a naturally high entropy rate, simply observing the past behavior will not help predict the future (and consequently, modeling is hard). Conversely, a highly predictable periodic process will have high active information storage and the future can be easily predicted from the past. Importantly, both of these measures (and indeed, all of the basic information dynamics) are purely observational; they do not provide any insight into the ``causal" structure of th system, and as such cannot be used to predict the effects of interventions or perturbations.
	
	Active information storage has been applied in a wide range of contexts, from understanding distributed processing in neural networks \cite{wibral_local_2014}, predictive coding models \cite{brodski-guerniero_information-theoretic_2017}, eye and gaze tracking \cite{wollstadt_quantifying_2021}, to understanding the dynamics of football matches \cite{hutchison_towards_2014}.
	
	The local active information storage describes how knowledge of the specific past states of $X$ informs on our uncertainty that $X$ will adopt the particular next state $x$ \cite{lizier_local_2012,lizier_local_2013}.
	
	\begin{equation}
	a(x_t) = i(x_{-k:t-1}; x_t)
	\end{equation}
	
	An example of local active information storage in biology might be a neuron: after firing an action potential, neurons typically go through a refractory period as the membrane potential is reset and for that duration, no subsequent firings can occur. In this context, knowing that our target neuron fired within the last $k$ milliseconds drastically reduces our uncertainty about what the next state of the neuron will be: it probably won't fire, because it's within the refractory window \cite{wibral_local_2014}. Local active information storage has also been used to understand swarming behaviours of crustaceans such as crabs \cite{x_rosalind_wang_measuring_2011, tomaru_information_2016}, and synchronization processes \cite{mutoh_synchronization_2015}.
	
	This allows a ``time-resolved" picture of how a single variable remembers and computes with its own past. When $a(x) > 0$, the system is behaving ``as we would expect", in that past states are reducing our uncertainty about future states in accordance with the overall statistics of the system. In contrast, when $a(x) < 0$, we are being ``misinformed" and the system appears to be transiently ``breaking" from it's own past to do something unexpected. 
	
	\subsection{Information Transfer}
	\label{sec:te}
	
	The second major type of information dynamics used to study complex systems is that of \textit{information transfer}. While active information storage examines a single variable's behavior through time, information transfer describes how the behavior of one (or more) elements in a system informs the behavior of another. The most well-established measure of information transfer is the transfer entropy \cite{schreiber_measuring_2000} (for a superb and extremely thorough discussion of transfer entropy in all its glory, see \cite{bossomaier_introduction_2016}).
	
	For two variables, $X$ and $Y$, the transfer entropy $TE(X\to Y)$ measures the amount of information knowing the \textit{past of $X$} gives us about the \textit{next state of $Y$}, in the form of a conditional mutual information:
	
	\begin{equation}
	\label{eq:te}
	TE(X \rightarrow Y) = I(X_{-k:t-1}; Y_t | Y_{-j:t-1})
	\end{equation}
	
	It is well worth the effort to spend some time unpacking this equation, as it contains a number of important, but subtle interpretations. $TE(X\to Y)$ can be thought of as \textit{the amount of information we get about the next state of Y by knowing the past states of X above and beyond what we would get simply knowing the past state of Y}. Phrased differently: how much information does $X$'s past provide about $Y$'s future that is not also provided about $Y$'s past?
	
	Transfer entropy is probably the most widely used information dynamic in the study of complex systems, having been used in studies of biological networks of neurons, \cite{timme_multiplex_2014,ito_large-scale_2014,antonello_self-organization_2022,shimono_functional_2015,nigam_rich-club_2016,varley_information-processing_2023}, cryptocurrencies \cite{dimpfl_group_2019,garcia-medina_network_2020,garcia-medina_transfer_2020,keskin_information-theoretic_2020}, precious metal economics \cite{luu_duc_huynh_effect_2020}, the interaction between social media sentiment and commodity prices \cite{keskin_information-theoretic_2020}, the directed relationship between atmospheric carbon dioxide levels and global warming \cite{stips_causal_2016}, climate models \cite{oh_time_2018,goodwell_temporal_2017,tongal_forecasting_2021} and more (for further citations, see \cite{bossomaier_introduction_2016}). Transfer entropy's nonparametric nature makes it particularly suited to analysis of complex systems where non-linearities are known to play a significant role in the dynamics. 
	
	Transfer entropy is a non-parametric generalization of the well-known Granger Causality: in the case of Gaussian random variables, the Granger Causality and the transfer entropy are equivalent \cite{barnett_granger_2009,razak_quantifying_2014}. Here, ``causality" refers to the statistical, rather than epistemic, definition, and care should be taken to avoid confusing transfer entropy, which is a measure of effective connectivity, with true causality \cite{ay_information_2008,lizier_differentiating_2010,razak_quantifying_2014}. The question of how to extract causal relationships from data is a rich, and rapidly growing field known as \textit{causal inference} (for an excellent primer, see \cite{pearl_causal_2016}), which is (regrettably) beyond the scope of this introduction.  
	
	As with local mutual information and local active information storage, local transfer entropy gives a high degree of spatial and/or temporal resolution to the analysis of interacting, complex processes \cite{lizier_local_2008}.
	
	\begin{equation}
	te(x_t\to y_t) = i(x_{-k:t-1} ; y_t | y_{-j:t-1})
	\end{equation}
	
	For example, imagine we were interested in the relationship between two companies ($C_1,C_2$) traded on the stock market. If we did the full transfer entropy, we might find that $TE(C_1 \to C2) > 0$ bit, however, it is important to remember that this is the \textit{expected information transfer} over the whole dataset and doesn't tell us much about the variability day-to-day. If, instead, we computed the local transfer entropy at every day, we might find that, for the vast majority of days, $te(c_{1t} \to c_{2t}) \approx 0$, and it is only on particular days that the predictive relationship holds. Armed with this information, we could do a more detailed investigation, knowing that $C_1$ is predictive of $C_2$ only at particular moments in time (which may be days that $C_1$ announces a new product, or reports their quarterly earnings, etc). 
	
	While not nearly as frequently used as the general transfer entropy, local transfer entropy has been used to understand the dynamics of cellular automata \cite{lizier_local_2008,goetze_reconstructing_2019}, neural dynamics measured with electrophysiology \cite{martinez-cancino_what_2020}, and animal swarms \cite{crosato_informative_2018}.
	
	\subsubsection{Conditional Transfer Entropy}
	\label{sec:multi_te}
	
	As with any mutual information measure, it is possible to condition the bivariate transfer entropy from $X$ to $Y$ on a larger set of variables to account for common driver effects, redundancies, and synergies in the system \cite{bossomaier_introduction_2016}. The transfer entropy from $X$ to $Y$, conditioned on $Z$ is given by:
	
	\begin{equation}
	TE(X \rightarrow Y | Z) = I(X_{-k:t-1} ; Y_t | Y_{-j:t-1}, Z_{-r:t-1})
	\end{equation}	
	
	The effect of conditioning on $Z$'s past gives us the amount of information $X$'s past gives us about $Y$'s future \textit{above and beyond that which is provided by $Y$'s own past and $Z$'s past}. As with conditional mutual information, conditional transfer entropy is essential when assessing information dynamics in complex systems, which by definition have large numbers of interacting elements. For two variables $X$ and $Y$ belonging to a set of elements $\textbf{V}$, the \textit{complete transfer entropy} between $X$ and $Y$ ($TE(X \to Y | \textbf{V}^{-XY})$) is given by conditioning $TE(X \to Y)$ on the past of every other element in $\textbf{V}$ simultaneously (excluding $X$ and $Y$, of course). The conditional transfer entropy plays a key role in the discovery of effective connectivity networks from time-series data (see Sec. \ref{sec:eff_nets} and \cite{novelli_large-scale_2019,wollstadt_idtxl_2019}). If conditioning $TE(X\rightarrow Y)$ on $\textbf{V}^{-XY}$ \textit{increases} the overall transfer entropy, then $X$ and $Y$ share synergistic dependency that is only ``illuminated" when all the other elements are accounted for. In contrast, if conditioning reduces the transfer entropy, then there are redundancies that are being conditioned out. For more on this, see \cite{stramaglia_disentangling_2024}. Conditional transfer entropy can be localized in the same way as the standard bivariate transfer entropy. 
	
	\subsection{Information Modification}
	\label{sec:info_mod}
	Of the three major elements of computation, information modification has most resisted attempts to develop a satisfactory formalism \cite{lizier_local_2013}. The intuition behind information modification is that the interaction of multiple streams of information causes some non-trivial ``change" in the nature of information being transmitted, although exactly what the nature of that change is can be hard to pin down. 
	
	An early attempt at formalizing information modification was proposed in \cite{lizier_information_2010} called the seperable information. A strictly local measure, seperable information fails as a formal definition of information modification and acts instead as a heuristic based on the interaction between active information storage and bivariate transfer entropy. Subsequently, \cite{lizier_towards_2013} proposed that the synergistic information between a set of sources onto a target could serve as a more robust, formal definition of information modification (for a discussion of synergistic information, see Sec. \ref{sec:pid}). Subsequent work in this vein has used synergistic information as a measure of ``non-trivial" computation in neural systems \cite{timme_synergy_2014,timme_high-degree_2016,faber_computation_2018,sherrill_correlated_2020,sherrill_partial_2021,newman_revealing_2022,varley_information-processing_2023}, where ``computation" is thought to relate closely to the idea of information modification (the output of a process being, in some sense, an irreducible function of both inputs).
	
	\subsection{Excess Entropy}
	\label{sec:excess}
	
	The final measure we will describe here is the excess entropy as, in some sense, it subsumes all of the prior dynamics. The active information storage only tells us how much information from the past is used in computing the immediate next state of our variable. However, not all the information present in the past is used at the immediate next state: some may only become relevant in the distant future, and the total amount can be calculated as the mutual information between the entire past and the entire future \cite{bialek_complexity_2001}:
	
	\begin{equation}
	\label{eq:excess}
	E(X) = \lim_{\substack{k\to\infty \\ j\to\infty}} I(X_{-k:t-1};X_{t:j})
	\end{equation}
	
	When considering the entire lifetime of the variable, this provides something like a measure of the total amount of predictable structure present in the variable. This is an extremely powerful tool with which to characterize a complex system: given the Bayesian interpretation of mutual information, we can intuitively understand the excess entropy as the maximum theoretical limit on our ability to predict the future dynamics of a system given observations of its past \cite{crutchfield_regularities_2003}. The local predictive information has not be explored much, to the best of our knowledge, although it can easily be calculated as a local mutual information. If one can calculate the true excess entropy, then one can construct a so-called ``$\epsilon$-machine": a finite state machine that is provably the minimal, optimal predictive model of the process being analyzed \cite{crutchfield_calculi_1994,shalizi_computational_2001}. These $\epsilon$-machines are information-theoretic objects of study in their own right, and form the foundation of a rich line of work, touching on foundational topics as diverse as irreversibility \cite{crutchfield_times_2009} to emergence \cite{rosas_software_2024}.
	
	In the event that one is dealing with a multivariate system $\textbf{X}=\{X^1,\ldots,X^N\}$, then the excess entropy incorporates all of the previously mentioned dynamics (the information storage, the information transfer, and the information modification), as it counts \textit{all} of the predictive power that the entire past discloses about the entire future. In Sec. \ref{sec:phi_id}, we will discuss how the excess entropy can be decomposed to account for the full space of dynamic dependencies between variables. 
	
	\newpage
	
	\section{Partial Information Decomposition (PID)}
	\label{sec:pid}
	
	We have previously made reference to the idea that information can be synergistically or redundantly shared between multiple sets of elements of a complex system, and that this can complicate our attempts to build models based on empirical data. When discussing conditional mutual information \ref{sec:cond_mi}, we saw how conditioning on multiple variables can reveal when information is present in higher-order configurations: in the case of redundancy, conditioning will reduce the observed quantity of information, while in the case of synergy, conditioning will increase observed quantity of information. Similarly, when discussing multivariate transfer entropy (Sec. \ref{sec:multi_te}) we saw how conditioning on multiple possible parent elements can reveal effective connections that would be invisible when considering only bivariate links in the system.
	
	While conditioning can help account for redundant and synergistic information when building models, conditioning itself doesn't necessarily give us a detailed picture of how information is distributed over all the elements of interest. A set of source elements can provide \textit{both} redundant and synergistic information about a target element simultaneously, and to get a comprehensive breakdown of how information is distributed requires mathematical machinery beyond what is provided by classical Shannon information theory.  
	
	\subsection{PID Intuition}
	\label{sec:pid_intuition}

	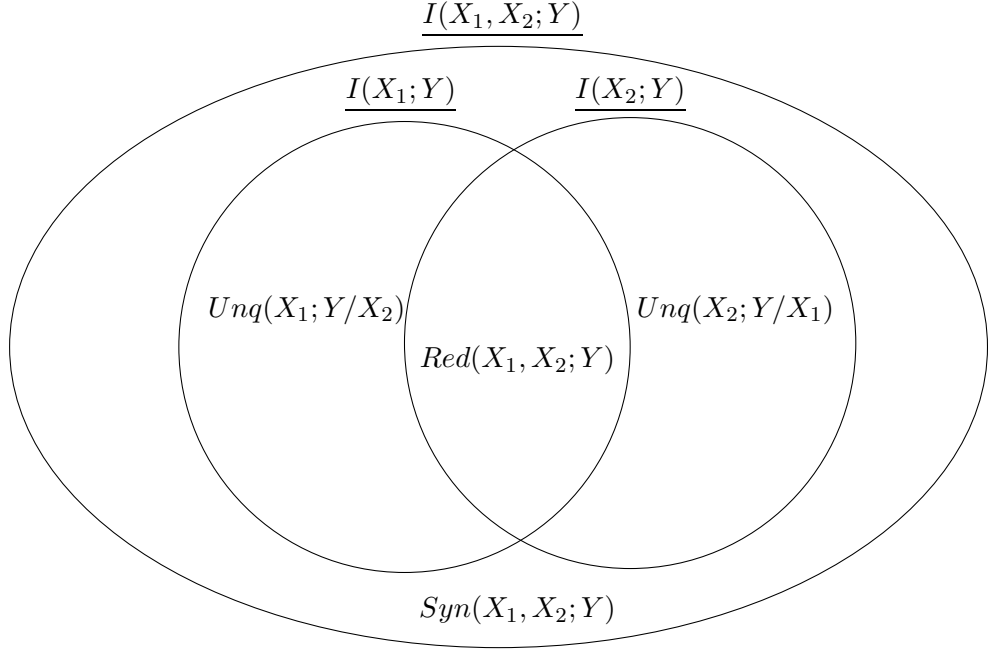
\begin{figure}
		\def\firstcircle{(0,0) circle (3cm)}
		\def\secondcircle{(1:3cm) circle (3cm)}
		\def\elipse{(180:-1.25cm) ellipse (6.5cm and 4cm)}
		\begin{center}
			\begin{tikzpicture}
			
			\draw \firstcircle node[] at (-1.3,0.5) {$Unq(X_1;Y/X_2)$};
			\draw \secondcircle node[] at (4.4,0.5) {$Unq(X_2;Y/X_1)$};
			\draw \elipse node[] at (1.5,-3.5) {$Syn(X_1,X_2;Y)$};
			
			\node[anchor=south] at (1.3,4) {$\underline{ I(X_1,X_2;Y)}$};
			
			\node[anchor=south] at (-0.05,3) {$\underline{ I(X_1;Y)}$};
			\node[anchor=south] at (3,3) {$\underline{ I(X_2;Y)}$};
			
			\node[anchor=south] at (1.5,-0.5) {$Red(X_1,X_2;Y)$};
			
			\end{tikzpicture}
		\end{center}
		\label{fig:pid_venn}
		\caption{A Venn Diagram showing how the various components of partial information (redundant, unique, and synergistic) are related to the joint and marginal mutual information terms for two source variables $X_1$ and $X_2$, and a target variable $Y$. The two circles correspond to the mutual information between each source and the target, while the large ellipse gives the joint mutual information between both sources and the target.
			These diagrams highlight the difference between the marginal mutual information and the unique information: notice that the marginal mutual informations overlap, each one counting the redundant (shared) information towards it's own marginal mutual information. We can also see that $I(X_1,X_2;Y) > I(X_1;Y)\cup I(X_2;Y)$: the difference is the synergistic information which cannot be resolved to either marginal mutual information.
			Finally, the Venn Diagram highlights how the partial information terms relate to mutual information terms: for example: $Syn(X_1,X_2;Y) = I(X_1,X_2;Y) - I(X_1;Y) - I(X_2;Y) + Red(X_1,X_2;Y)$. We have to add a redundancy term back in because it is ``double counted" when subtracting off the marginal mutual informations.}
	\end{figure}

	Given a set of source elements, all of whom synapse onto a single target element, a natural question is how information about the target element is distributed over different combinations of source elements. To develop the intuition in more detail, let's use the simplest possible ``complex system": two parents $X_1$ and $X_2$, both of which synapse onto a child element $Y$. 
	
	\begin{align}
		X_1 \rightarrow Y \leftarrow X_2 \nonumber
	\end{align}
	
	As described above, we know that all the information both parents provide about $Y$ is aggregated by $I(X_1,X_2;Y)$, but this gives us no insights into how that information is distributed over $X_1$ and $X_2$. For example, what information about $Y$ could be learned by observing either $X_1$ alone or $X_2$ alone? What about the information that is only disclosed by $X_1$? To build a complete portrait of how information is distributed, we would like to break the joint mutual information down into all possible ways information can be shared:
	
	\begin{equation}
	\begin{aligned}
	I(X_1,X_2 ; Y) = Red(X_1,X_2 ; Y) \\ + Unq(X_1 ; Y/X_2) \\ + Unq(X_2 ; Y/X_1) \\ + Syn(X_1,X_2 ; Y)
	\end{aligned}
	\label{eq:pid_4_unknowns}
	\end{equation}
	
	Where $Red(X_1,X_2 ; Y)$ corresponds to the information about $Y$ that is redundantly present in both $X_1$ and $X_2$ (i.e. could be learned by observing \textit{either} $X_1$ alone or $X_2$ alone). $Unq(X_{1};Y/X_{2})$ is information about $Y$ that is uniquely present in $X_1$ (i.e. information that could \textit{only} be learned by observing $X_1$), and likewise for $Unq(X_2;Y/X_1)$. Finally, $Syn(X_1,X_2;Y)$ is that information about $Y$ that is synergistically present in the joint states of $X_1$ and $X_2$ (and could not be learned by examining either $X_1$ or $X_2$ individually). A moments reflection should convince the reader that this enumerates all possible combinations. 
	
	Furthermore, given such a breakdown, we would like to be able to decompose each of the marginal mutual informations as well:
	
	\begin{equation}
	\begin{aligned}
	I(X_1 ; Y) = Red(X_1,X_2 ; Y) + Unq(X_1 ; Y / X_2) \\
	I(X_2 ; Y) = Red(X_1,X_2 ; Y) + Unq(X_2 ; Y / X_1)
	\end{aligned}
	\label{eq:pid_pairs}
	\end{equation}
	
	The result is a system of three equations (each equal to $I(X_1,X_2;Y)$, $I(X_1;Y)$ and $I(X_2;Y)$ respectively, see Eqs. \ref{eq:pid_4_unknowns} and \ref{eq:pid_pairs}) and four unknowns ($Red(X_1,X_2;Y)$, $Unq(X_1;Y/X_2)$, $Unq(X_2;Y/X_1)$, and $Syn(X_1,X_2;Y)$). This is an undetermined system: if we were able to calculate any of the four unknowns, we could compute the rest using simple algebra. A number of functions have been proposed targeting the redundancy (see \cite{kolchinsky_novel_2022} for a review, also Sec. \ref{sec:pid_lattice_func}), the unique information (such as \cite{bertschinger_quantifying_2014,james_unique_2018}), and the synergy (such as \cite{rosas_operational_2020}).
	
	In general, however, for $N>2$, it is not the case that the relationship between mutual information terms and partial information ``atoms" is so tightly constrained. For $N = 3$, there are seven known  mutual information terms but eighteen unique partial information atoms. To solve the general care involves developing some mathematical machinery beyond that which is provided by classical Shannon information theory. 
	
	\subsection{Redundancy Lattices \& Functions}
	\label{sec:pid_lattice_func}
	
	Suppose we have a system of $N$ predictor variables, all informing about the state of a single target $Y$. We would like decompose the joint mutual information $I(X_1,\ldots,K_N;Y)$ down into all relevant combinations of sources. Doing this requites two things: the first is an operational definition of what it means for information to be ``redundant" to two or more sources (which we will denote as $I_{\cap}$). The second is an understanding of the \textit{structure} of multivariate information (which is given by the redundancy lattice described below).  Formally, given some (as yet undefined) set of all meaningful sources of information ($\mathfrak{A}$), we want a decomposition such that:
	
	\begin{equation}
	\label{eq:pid-gen}
	I(X_1,\ldots,X_N)=\sum_{\boldsymbol{\alpha}\in\mathfrak{A}}I_{\partial}(\boldsymbol{\alpha};Y)
	\end{equation}
	
	Where $I_{\partial}$ quantifies the information about $Y$ that could only be learned by observing $\boldsymbol{\alpha}$ and no simpler combination of sources. In the case of the two-input example given above, the sources $\boldsymbol{\alpha}$ are the redundant, two unique, and pure synergy terms.
	
	\subsubsection{The Redundancy Lattice}

	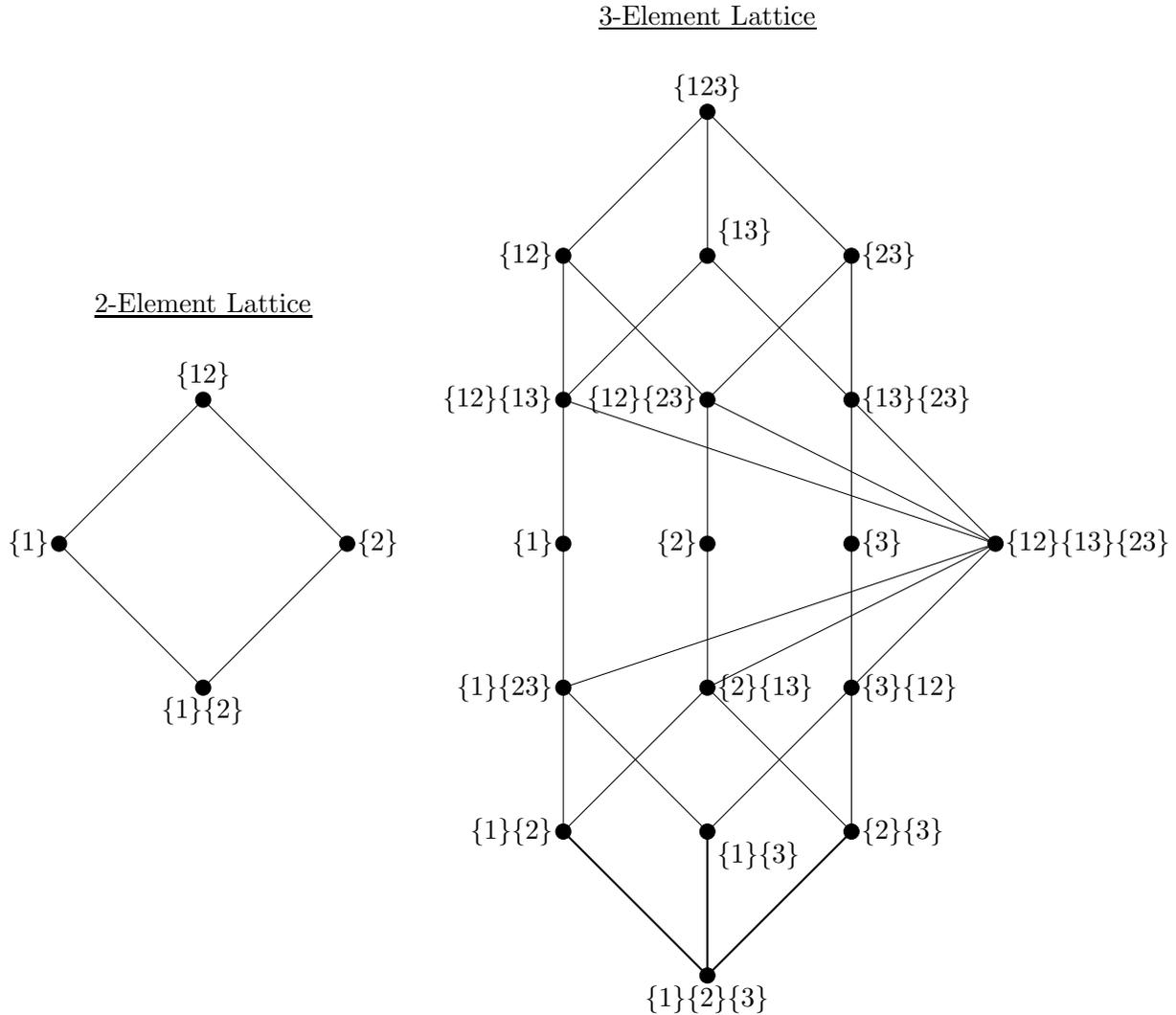
\begin{figure}
		\begin{center}
			\begin{tikzpicture}
			
			
			\node[anchor=south] at (-7,-3) {$\underline{\textnormal{2-Element Lattice}}$};
			
			\filldraw[black] (-7,-4) circle (3pt) node[anchor=south] {$\{12\}$};
			
			\filldraw[black] (-9,-6) circle (3pt) node[anchor=east] {$\{1\}$};
			
			\filldraw[black] (-5,-6) circle (3pt) node[anchor=west] {$\{2\}$};
			
			\filldraw[black] (-7,-8) circle (3pt) node[anchor=north] {$\{1\}\{2\}$};
			
			\draw[] (-7,-4) -- (-9,-6);
			\draw[] (-7,-4) -- (-5,-6);
			\draw[] (-9,-6) -- (-7,-8);
			\draw[] (-5,-6) -- (-7,-8);
			
			
			\node[anchor=south] at (0,1) {$\underline{\textnormal{3-Element Lattice}}$};
			
			\filldraw[black] (0,0) circle (3pt) node[anchor=south] (123) {$\{123\}$};
			
			\filldraw[black] (-2,-2) circle (3pt) node[anchor=east] (12) {$\{12\}$};
			\filldraw[black] (0,-2) circle (3pt) node[anchor=south west] (13) {$\{13\}$};
			\filldraw[black] (2,-2) circle (3pt) node[anchor=west] (23) {$\{23\}$};
			
			\filldraw[black] (-2,-4) circle (3pt) node[anchor=east] (12-13) {$\{12\}\{13\}$};
			\filldraw[black] (-0,-4) circle (3pt) node[anchor=east] (12-23) {$\{12\}\{23\}$};
			\filldraw[black] (2,-4) circle (3pt) node[anchor=west] (13-23) {$\{13\}\{23\}$};
			
			\filldraw[black] (-2,-6) circle (3pt) node[anchor=east] (1) {$\{1\}$};
			\filldraw[black] (-0,-6) circle (3pt) node[anchor=east] (2) {$\{2\}$};
			\filldraw[black] (2,-6) circle (3pt) node[anchor=west] (3) {$\{3\}$};
			\filldraw[black] (4,-6) circle (3pt) node[anchor=west] (12-13-23) {$\{12\}\{13\}\{23\}$};
			
			\filldraw[black] (-2,-8) circle (3pt) node[anchor=east] (1-23) {$\{1\}\{23\}$};
			\filldraw[black] (0,-8) circle (3pt) node[anchor=west] (2-13) {$\{2\}\{13\}$};
			\filldraw[black] (2,-8) circle (3pt) node[anchor=west] (3-12) {$\{3\}\{12\}$};
			
			\filldraw[black] (-2,-10) circle (3pt) node[anchor=east] (1-2) {$\{1\}\{2\}$};
			\filldraw[black] (0,-10) circle (3pt) node[anchor=north west] (1-3) {$\{1\}\{3\}$};
			\filldraw[black] (2,-10) circle (3pt) node[anchor=west] (2-3) {$\{2\}\{3\}$};
			
			\filldraw[black] (0,-12) circle (3pt) node[anchor=north] (1-2-3) {$\{1\}\{2\}\{3\}$};
			
			\draw[] (0,0) -- (-2,-2);
			\draw[] (0,0) -- (-0,-2);
			\draw[] (0,0) -- (2,-2);
			
			\draw[] (-2,-2) -- (-2,-4);
			\draw[] (2,-2) -- (2,-4);
			\draw[] (0,-2) -- (-2,-4);
			\draw[] (0,-2) -- (2,-4);
			\draw[] (-2,-2) -- (0,-4);
			\draw[] (2,-2) -- (0,-4);
			
			\draw[] (-2,-4) -- (4,-6);
			\draw[] (0,-4) -- (4,-6);
			\draw[] (2,-4) -- (4,-6);
			
			\draw[] (-2,-6) -- (-2,-4);
			\draw[] (2,-6) -- (2,-4);
			\draw[] (0,-6) -- (0,-4);
			
			\draw[] (-2,-8) -- (-2,-6);
			\draw[] (2,-8) -- (2,-6);
			\draw[] (0,-8) -- (0,-6);
			
			\draw[] (4,-6) -- (-2,-8);
			\draw[] (4,-6) -- (0,-8);
			\draw[] (4,-6) -- (2,-8);
			
			\draw[] (-2,-10) -- (-2,-8);
			\draw[] (2,-10) -- (2,-8);
			\draw[] (-2,-10) -- (0,-8);
			\draw[] (2,-10) -- (0,-8);
			\draw[] (0,-10) -- (-2,-8);
			\draw[] (0,-10) -- (2,-8);

			\draw[thick] (0,-12) -- (-2,-10);
			\draw[thick] (0,-12) -- (-0,-10);
			\draw[thick] (0,-12) -- (2,-10);
			\end{tikzpicture}
		\end{center}
		\label{fig:pid_lattice}
		\caption{Examples of redundancy lattices for the two simplest possible systems. \textbf{Left:} The redundancy lattice for a set of two sources $X_1$ and $X_2$ synapsing onto a single target. This is a simplified visualization of the Venn Diagram seen in Fig. \ref{fig:pid_venn}: $\{1\}\{2\}$ corresponds to the information redundantly shared between $X_1$ and $X_2$, while $\{12\}$ corresponds to the synergistic information and the single elements $\{1\}$ and $\{2\}$ indicate the unique information in each element. \textbf{Right:} The redundancy lattice for three sources synapsing onto a single target. The three-element lattice makes it clear that, as the number of sources grows, the clean distinctions between ``redundancy", ``unique information" and ``synergy" break down as more complex combinations of sources contribute information about the target. The top and bottom of the lattice can be thought of as ``purely synergistic" and ``purely redundant" respectively, however between the two extremes, the ``PI-atoms" can be thought of as information that is redundantly shared over higher-order combinations of sources: for example $\{1\}\{23\}$ gives the partial information that is redundantly present in both $X_1$ and the joint states of $X_2$ and $X_3$ together. 
		}	
	\end{figure}

	Here, we will provide a sketch of the derivation of the redundancy lattice first proposed by Williams \& Beer \cite{williams_nonnegative_2010}, and explored further by Gutknecht et al., \cite{gutknecht_bits_2020}. Our goal is not a mathematically rigorous derivation, but rather, to provide a basic intuition about where the lattice comes from and what it means. For the interested mathematician, see the above citations.  
	
	When decomposing $I(X_1,\ldots,X_N;Y)$, the first step is to enumerate the set of all possible \textit{sources} of information. For example, information can be disclosed by any $X_i$, but it can also be disclosed by any pair \{$X_i,X_j\}$, every triple, and so on. The set $\textbf{S}$ of all possible (potentially multivriate) sources is given by the power set of the set of $\{X_i\}$: $\textbf{S} = \mathcal{P}_0(\{X_1,\ldots,X_N\})$, where the 0 subscript indicates that the empty set is being excluded from the power set (since a set of no elements discloses no information). 
	
	The seminal insight of Williams and Beer was to realize that the problem of decomposing the multivariate information was equivalent to the problem of understanding how information was redundantly distributed over every element of \textbf{S}. This requires computing another power set: this time the power set of \textbf{S}, to get all possible combinations by which elements of \textbf{S} may redundantly share information:
	
	\begin{equation}
	\mathfrak{A} = \{\boldmath{\alpha}\in\mathcal{P}_0(\textbf{S}) : \forall \textbf{A}_i,\textbf{A}_j\in\boldmath,\textbf{A}_i\not\subset\textbf{A}_j\}
	\end{equation}
	
	Note that not every element of $\mathcal{P}_0(\textbf{S})$ is a valid member of $\mathfrak{A}$: only those collections of sources where no collection is a proper subset of any other (since, if $\textbf{A}_i\subset\textbf{A}_j$, then the information redundantly shared between them would simply be all the information disclosed by $\textbf{A}_i$). The size of $\mathfrak{A}$ grows hyper-exponentially with the number of elements being considered. For two predictors there are four distinct atoms. For three predictors there are eighteen, and for four predictors there are 166. For eight predictor, there are $\approx5.6\times10^{22}$ atoms. Consequently, the current applicability of the partial information decomposition framework to large systems is limited. 
	
	The set $\mathfrak{A}$ has a partially-ordered structure:
	
	\begin{equation}
	\forall \boldsymbol{\alpha}, \boldsymbol{\beta} \in \mathfrak{A},\boldsymbol{\alpha} \preceq \boldsymbol{\beta} \Leftrightarrow \forall \textbf{B} \in \boldsymbol{\beta}, \exists \textbf{A} \in \boldsymbol{\alpha}\,\text{s.t.}\,\textbf{A} \subseteq \textbf{B}
	\end{equation}
	
	This structure allows us to recursively calculate the value of every $\boldsymbol{\alpha}\in\mathfrak{A}$ via Mobius inversion:
	
	\begin{equation}
	I_{\partial}(\boldsymbol{\alpha};Y) = I_\cap(\boldsymbol{\alpha};Y) - \sum_{\boldsymbol{\beta}\prec\boldsymbol{\alpha}}I_{\partial}(\boldsymbol{\beta};Y)
	\end{equation}
	
	From this, we can extract our initial desired Eq. \ref{eq:pid-gen}.
	
	To build intuition, let us revisit our toy example from above. If we have $I(X_1,X_2;Y)$, our first step is to enumerate the set of sources:
	
	\begin{equation}
	\textbf{S} = \big\{\{X_1\}, \{X_2\}, \{X_1,X_2\}\big\}
	\end{equation}
	
	We then need to find the domain of our redundancy function $I_{\cap}$: the set of all sources in \textbf{S} such that no source is a subset of any other:
	
	\begin{equation}
	\mathfrak{A} = \big\{\{X_1\}\{X_2\}, \{X_1\}, \{X_2\}, \{X_1,X_2\}\big\}
	\end{equation}
	
	The four elements of $\mathfrak{A}$ correspond to our intuitive ideas of redundant, unique, and synergistic information described above. The partial ordering shows that $\{X_1\}\{X_2\}\prec\{X_1\}$ and $\{X_2\}$, while both $\{X_1\}$ and $\{X_2\}\prec\{X_1,X_2\}$. This defines the redundancy lattice shown in Fig. \ref{fig:pid_lattice}. 
	
	\subsubsection{The Redundancy Function}
	\label{sec:pid_redundancy}
	
	We have, so far, neglected to define our redundancy function $I_{\cap}$, instead assuming that behaves in a manner intuitive with our notion of redundancy. To formalize these intuitions, the field has converged on a standard set of axioms that any proposed redundancy function \textit{must} satisfy to induce the lattice structure on $\mathfrak{A}$:
	\begin{enumerate}
		\item \textbf{Symmetry:} The function $I_{\cap}(\textbf{A}_1,\ldots,\textbf{A}_k)$ must be invariant under permutation: given a permutation $\sigma$: $I_{\cap}(\textbf{A}_1,\ldots,\textbf{A}_k;Y)=I_{\cap}(\sigma(\textbf{A}_1),\ldots,\sigma(\textbf{A}_k);Y)$
		\item \textbf{Self-Redundancy:} For single sources, the redundancy is equal to the mutual information: $I_{\cap}(\textbf{A};Y)=I(\textbf{A};Y)$.
		\item \textbf{Monotonicity:} As more sources are added, the redundancy must decrease: $I_{\cap}(\textbf{A}_1,\ldots,\textbf{A}_k, \textbf{A}_{k+1};Y) \leq I_{\cap}(\textbf{A}_1,\ldots,\textbf{A}_k;Y)$, with equality if $\textbf{A}_{k}\subset\textbf{A}_{k+1}$. 
	\end{enumerate}
	
	Any function that satisfies these axioms will induce the partial information lattice described above.
	
	When proposing the partial information decomposition for the first time, Williams and Beer provided a ``natural" redundancy function in the form of the \textit{specific information} \cite{williams_nonnegative_2010}:
	
	\begin{equation}
	I_{\cap}^{WB}(\boldsymbol{\alpha};Y)=\sum_{y\in\mathcal{Y}}P(y)\min_{\textbf{a}_i\in\boldsymbol{\alpha}}I(\textbf{A}_i;y)
	\end{equation}
	
	While this function has a number of appealing properties (such as returning exclusively non-negative values for the partial information atoms), it was almost immediately criticized for some counter-intuitive behaviours (see \cite{bertschinger_shared_2013}). For example, it was considered anomalous that it returned non-zero values in the following case:
	
	\begin{equation}
	I^{WB}_{\cap}(X_1,X_2;\{X_1,X_2\}) \ge 0
	\end{equation}
	
	Even when $X_1\bot X_2$. Intuitively, if $X_1$ and $X_2$
	are independent, then the information that either one discloses about their joint state $\{X_1,X_2\}$ should be entirely unique. This is called the two-bit copy problem, and the discovery of the apparent flaw prompted widespread interest in alternative redundancy functions. 
	Unlike the Shannon entropy, where a simple set of intuitive axioms is enough to specify a unique, intuitive function, \textit{many} redundancy functions have been discovered that are consistent with the desired axioms. For a non-exhaustive list, see \cite{bertschinger_shared_2013,harder_bivariate_2013,griffith_intersection_2014,griffith_quantifying_2014,olbrich_information_2015,barrett_exploration_2015,goodwell_temporal_2017,ince_measuring_2017,finn_pointwise_2018,ay_information_2019,kolchinsky_novel_2022,makkeh_introducing_2021}).
	
	These definitions all generally imply slightly different intuitions around the question ``what is redundancy". For example, for Gaussian processes, Barrett showed that a natural redundancy function is based on the minimum mutual information \cite{barrett_exploration_2015}:
	
	\begin{equation}
	I^{MMI}_{\cap}(\boldsymbol{\alpha},Y)=\min_{\textbf{A}_i\in\boldsymbol{\alpha}}I(\textbf{A}_i;Y).
	\end{equation}
	
	Another significant limitation of $I^{WB}_{\cap}$ is that it is not localizable in the way that the mutual information is. This has prompted a search for functions that both satisfy the initial axioms, as well as localizability criteria. One of the first was the pointwise common change in surprisal, proposed by Ince \cite{ince_measuring_2017} and based off the inclusion-exclusion criteria that defines the co-information:
	
	\begin{equation}
	i^{ccs}_{\cap}(\boldsymbol{\alpha};Y)=co_Q(\textbf{a}_1,\ldots,\textbf{a}_k,y)
	\end{equation}
	
	Where $co_Q$ refers to the local co-information computed with respect to a maximum-entropy distribution that preserves pairwise marginals. Another, less computationally intensive proposal from Finn \& Lizier \cite{finn_probability_2018} was based on the logic of informative and misinformative local mutual information. They decomposed the local mutual information into two components corresponding to informative and misinformative probability mass exclusions \cite{finn_probability_2018}:
	
	\begin{equation}
	i^{\pm}_{\cap}(\boldsymbol{\alpha};y) = \min_{\textbf{a}_i\in\boldsymbol{\alpha}} h(\textbf{a}_i) -\min_{\textbf{a}_i\in\boldsymbol{\alpha}} h(\textbf{a}_i|y)
	\end{equation}
	
	They showed that, while $i^{\pm}_{\cap}$ was not entirely well-behaved (for example, returning negative partial information atoms), the two components, when decomposed individually, satisfied all the original desiderata. This approach was then expanded by Makkeh et al., \cite{makkeh_introducing_2021} to a form that is differentiable: 
	
	\begin{equation}
	i^{sx}_{\cap}(\textbf{a}_1,\ldots,\textbf{a}_k;y) = \log_2\frac{P(y) - P(y \cap (\bar{\textbf{a}}_1\cap\ldots\cap\bar{\textbf{a}}_k))}{1 - P(\bar{\textbf{a}}_1\cap\ldots\cap\bar{\textbf{a}}_k)} - \log_2P(y).
	\end{equation}
	
	Where $P(\bar{\textbf{a}})$ indicates the probability of seeing \textbf{A} in any state other than \textbf{a}. While originally formulated for discrete random variables, the $i^{sx}_{\cap}$ function also has a non-parametric form for continuous random variables based on K-nearest neighbors estimators \cite{ehrlich_partial_2024} (for further discussion see Sec. \ref{sec:knn}).
	
	This is an extremely abbreviated survey of redundancy functions. Other, more complex algorithms have been proposed, including ones based on advanced concepts such as information geometry \cite{harder_bivariate_2013}, probabilistic graphical models \cite{sigtermans_path-based_2020}, or the Gacs-Korner common information \cite{griffith_intersection_2014}. When attempting to implement a PID analysis, care should be taken when choosing a redundancy function, as different functions can return very different results \cite{kay_comparison_2022,kolchinsky_novel_2022}. 
	
	\subsection{Information Modification, Revisited}
	
	In Sec. \ref{sec:info_mod} we discussed how the third information dynamic, information modification, has been discussed in a synergy-based context. We now have the techniques to make this more explicit: for a single element that receives two inputs from parent elements, the information modification can be equated with the synergy \cite{lizier_towards_2013}. In the context of a dynamic process, synergy can be thought of as the ``novel" information generated from the interaction of two streams of information coming together on a single element \cite{newman_revealing_2022}.
	
	For higher order interactions involving more that two variables interacting with a single target, the question of the correct decomposition can be tricky. For example, given three upstream elements all synapsing onto a single target, what should the information modification be? One natural answer would be $I_{\partial}(\{X_1,X_2,X_3\}$, which is the triple synergy and the most ``irreducible" information in the system. However, there are other combinations of sources that use all three elements. For example $I_{\partial}(\{X_1,X_2\}\{X_1,X_3\}\{X_2,X_3\}$ is the information redundantly distributed over all pairs of joint states. To what extent does this also count as synergistic information modification? As it stands, there is not a single best answer to this question. It may be the case that different atoms communicate different aspects of information processing in complex systems and the idea of a single notion of ``information modification" will prove to be overly simplistic. 
	
	The PID can also prompt a reinterpretation of the transfer entropy \cite{williams_generalized_2011,james_information_2016}. Recall that:
	
	\begin{equation}
	TE(X \rightarrow Y) = I(X_{-k:t-1} ; Y_t | Y_{-j:t-1})
	\end{equation} 
	
	Using the PID framework, we can decompose the conditional mutual information into:
	
	\begin{equation}
	TE(X\rightarrow Y) = Unq(X_{-k:t-1};Y_{t}) + Syn(X_{-k:t-1},Y_{-j:t-1};Y_t)
	\end{equation}
	
	This shows that the transfer entropy consists of two, non-overlapping informational components. The first is referred to as the \textit{state-independent information transfer} (SITE): $Unq(X_{-k:t-1};Y_{t})$ is the information that passes from $X$'s past of $Y$'s future and is independent of the state of $Y$'s own past. In contrast, the \textit{state-dependent information transfer} (SDTE), $Syn(X_{-k:t-1},Y_{-j:t-1};Y_t)$, which is the novel information generated at time $t$ in $Y$ by the interaction of $X$'s past and $Y$'s past. 
	
	This complicates transfer entropy's status as a measure of information transfer, as it appears to conflate the true transfer of information with synergistic information modification \cite{james_information_2016}. As it stands, the significance of this has largely remained under-explored. Daube, Gross, and Ince explored the contribution of synergy to sensory EEG data analysis, finding that the apparent transfer entropy was largely reflecting the synergistic component (with the unique ``true flow" component being negative) \cite{daube_whitening_2022}. More recently \cite{varley_synergistic_2024} argued that the transfer entropy could be understood as conflating information transfer and information modification (or computation), with the SITE reflecting information transfer and the SDTE reflecting a kind of computation that $Y$ performs based on its own past and the past of $X$. This is one area in particular with a lot of potential to support future work. 
	
	\subsection{Extensions of the Partial Information Decomposition}
	
	Since the initial proposal by Williams and Beer, there has been work on generalizing the partial information decomposition in various ways. Here we will discuss several: the partial entropy decomposition \cite{ince_partial_2017,finn_generalised_2020} (PED, which decomposes the joint entropy in a manner analogous to the PID), the generalized information decomposition \cite{varley_generalized_2024} (GID, which provides a decomposition of arbitrary Kullback-Leibler divergences), the integrated information decomposition \cite{mediano_beyond_2019} ($\Phi$ID, which extends the PID to multiple targets). Other derivatives that we will not discuss in detail include the synergistic disclosure framework \cite{rosas_operational_2020} (which takes a synergy-first approach rather than a redundancy first approach), a cooperative-game theoretic approach \cite{ay_information_2019} (which uses an alternative lattice and has no notion of redundancy), and an approach based on secret-key cryptography and dependency lattices \cite{james_unique_2018,james_unique_2019,kay_exact_2018}. These approaches are, at present, not as developed or widely applied, although each has considerable promise and is worthy for future exploration. Finally, there is a weaker decomposition, the $\alpha$-synergy decomposition \cite{varley_scalable_2024}, which does not provide a ``complete" picture of the structure of multivariate information, but instead provides a univariate ``backbone" decomposition. This decomposition has the benefit of being much more scalable than the PID or any of its complete derivatives. 
	
	\subsubsection{Partial Entropy Decomposition}
	
	The most straightforward generalization the PID is the partial entropy decomposition (PED). First proposed by Ince \cite{ince_partial_2017}, instead of decomposing the mutual information that some set of predictors discloses about a target, the PED decomposes the joint entropy of a set of variables without requiring categorizing some as sources and others as targets. The general logic is largely the same as the PID: given a redundant entropy function that satisfies all of the axioms described in Sec. \ref{sec:pid_redundancy}, it is possible to define a redundancy lattice that can be solved via Mobius inversion in the same way as in PID. In a sense, it is related to an early proposed extension of the PID from Bertschinger et al., \cite{bertschinger_shared_2013}, who proposed a ``strong symmetry" axiom that required the redundancy to be insensitive to permutations of sources and targets together, although in the PED framework, the bivariate redundancy $H_{\cap}(X_1,X_2)$ is not necessarily equal to the mutual information $I(X_1;X_2)$, as in Bertschinger's proposed framework. 
	
	The PED reveals interesting, and sometimes counter-intuitive properties of information. For example, consider the decomposition of $H(X_1,X_2)$. As before, we get the decomposition:
	
	\begin{align}
	H(X_1,X_2) = H_{\partial}(\{X_1\}\{X_2\}) + H_{\partial}(\{X_1\}) + H_{\partial}(\{X_2\}) + H_{\partial}(\{X_1,X_2\})
	\end{align}
	
	which again corresponds to a redundant, two unique, and one synergistic atoms (although these are partial entropy atoms, which are conceptually quite different from partial information atoms). Furthermore, the marginal entropies can be decomposed in the same way:
	
	\begin{align}
	H(X_1) = H_{\partial}(\{X_1\}\{X_2\}) + H_{\partial}(\{X_1\}) \\
	H(X_2) = H_{\partial}(\{X_1\}\{X_2\}) + H_{\partial}(\{X_2\}) 
	\end{align}
	
	Based on this decomposition, it is possible to re-write the bivariate mutual information in terms of partial entropy atoms:
	
	\begin{align}
	I(X_1;X_2) =  H_{\partial}(\{X_1\}\{X_2\}) - H_{\partial}(\{X_1,X_2\})
	\end{align}
	
	This formulation reveals that the Shannon mutual information can be thought of as the difference between the redundant and synergistic entropies. The exact interpretation of this is difficult, and hinges on what ``redundant" and ``synergistic" entropy is taken to mean. Ince argued that this showed that mutual information obscured the ``true" redundancy ($H_{\partial}(\{X_1\}\{X_2\})$) with the higher-order synergistic entropy and prompted a ``corrected" mutual information $I^C(X;Y) = I(X;Y) + H_{\partial}(\{X,Y\})$, although it remains unclear how to interpret this value. 
	
	The structure of even bivariate mutual information becomes even more complex as the context in which is exists expands. For example, consider the case of three variables $X_1$, $X_2$, and $X_3$. Following the logic above, it is possible to decompose the bivariate mutual information between two ($I(X_1,X_2)$) into a set of partial entropy atoms that, crucially, can involve redundant and synergistic interactions with the third variable.
	
	\begin{align}
	I(X_1;X_2) &= 
	H_{\partial}(\{X_1\}\{X_2\}\{X_3\}) + H_{\partial}(\{X_1\}\{X_2\}) \\
	&- H_{\partial}(\{X_3\}\{X_1,X_2\}) - H_{\partial}(\{X_1,X_2\}\{X_1,X_3\}\{X_2,X_3\}) \nonumber \\ 
	&- H_{\partial}(\{X_1,X_2\}\{X_1,X_3\}) - H_{\partial}(\{X_1,X_2\}\{X_2,X_3\}) \nonumber \\
	&- H_{\partial}(\{X_1,X_2\})  \nonumber
	\label{eq:biv_from_tri}
	\end{align}
	
	Here, we can see that the bivariate mutual information also accounts for information redundantly copied over all three elements ($H_{\partial}(\{X_1\}\{X_2\}\{X_3\})$), as well as penalizing any information in the joint state of $X_1$ and $X_2$ combined. This shows that, depending on the context, the bivariate mutual information may not be ``specific": a dependency between two variables does not necessarily contain only information shared by only those two, but that non-local, ``higher-order" information can be present as well. In this case, higher-order information can be counted multiple times, globally inflating the overall amount of apparent dependency in ths system.
	
	The partial entropy decomposition inherent the same unusual property that it requires deciding on one of a number of possible redundant entropy functions. As of this writing, there are three well-explored ones, all of which are defined for local realizations. The first, from Ince \cite{ince_partial_2017}, is based on the co-information:
	
	\begin{equation}
	h^{cs}_{\cap}(\boldsymbol{\alpha})=\max(co_{Q}(\textbf{a}_1,\ldots,\textbf{a}_k), 0)
	\end{equation}
	
	With $co_Q$ once again referring to the local co-information computed over a maximum entropy distribution $Q(\textbf{X})$. This redundancy function can return negative partial entropy values, which is difficult to interpret, although Ince et al., argue that a negative partial entropy atom is an instance of misinformation, similar to the negative local mutual information. The $h^{cs}_{\cap}$ measure has been critiqued for being somewhat recursive \cite{varley_partial_2023}: it is possible to apply the PED to the redundancy function itself, which reveals that the co-information conflates redundant and synergistic entropy:
	
	\begin{align}
	h^{cs}_{\cap}(x_1;x_2;x_3) &= h_{\partial}(\{x_1\}\{x_2\}\{x_3\}) \\ &- \nonumber h_{\partial}(\{x_1\}\{x_2,x_3\}) - h_{\partial}(\{x_2\}\{x_1,x_3\}) -  \nonumber h_{\partial}(\{x_3\}\{x_1,x_2\}) \\ &- 2\times h_{\partial}(\{x_1,x_2\}\{x_1,x_3\}\{x_2,x_3\})  \nonumber\\ 
	&-  h_{\partial}(\{x_1,x_2\}\{x_1,x_3\}) - h_{\partial}(\{x_2,x_3\}\{x_1,x_3\}) - h_{\partial}(\{x_1,x_2\}\{x_2,x_3\})  \nonumber\\ &+ h_{\partial}(\{x_1,x_2,x_3\})
	\end{align}
	
	Based on this decomposition, we can see that, despite being based on inclusion-exclusion criteria analogous to the intersection of sets, that the co-information does not return only the expected triple-intersection, but rather, conflates the true redundancy with various synergistic terms. 
	
	Finn and Lizier independently derived the same framework \cite{finn_generalised_2020}, where they define the redundant entropy of an ensemble as the minimum entropy of any element:
	
	\begin{equation}
	h^{\min}_{\cap}(\boldsymbol{\alpha})=\min_{\textbf{a}_i\in\boldsymbol{\alpha}}h(\textbf{a}_i)
	\end{equation}
	
	This function returns provably non-negative partial entropy atoms, and can be applied to discrete and continuous data. Finally, Varley et al., \cite{varley_partial_2023} proposed a redundancy function inspired by Makkeh et al.'s $i_{sx}$ function \cite{makkeh_introducing_2021}:
	
	\begin{equation}
	h^{sx}_{\cap}(\boldsymbol{\alpha})=-\log\frac{1}{P(\textbf{a}_1\cup\ldots\cup\textbf{a}_k)}
	\end{equation}
	
	Like $h^{min}_{\cap}$, $h^{sx}_{\cap}$ returns provably non-negative partial entropy atoms, although it is only well-defined for discrete random variables. As it stands, no existing $h_{\cap}$ function is completely satisfying: all have their limitations and care should be taken to ensure that the strengths and weaknesses of a chosen function are appropriate for a given application.
	
	One of the benefits of $h^{sx}_{\cap}$ and $h^{min}_{\cap}$ is that they can be readily interpreted in the more well-understood context of the PID. Given the identity that: $i(\textbf{x};\textbf{x})=h(\textbf{x})$, it can be shown that: 
	
	\begin{equation}
	h^{sx}_{\cap}(\textbf{x})=i^{sx}_{\cap}(x_1,\ldots,x_k;\textbf{x})
	\end{equation}
	
	For example, $h^{sx}_{\cap}(\{x_1\}\{x_2\}$ is equivalent to asking ``what information about the joint state of $\{x_1,x_2\}$ could be learned by observing either just $x_1$ or just $x_2$?" and likewise for $h^{min}_{\cap}$. Given its comparative novelty, the PED is much less explored than the PID. While the formal mathematics is well-developed, deep questions of interpretation remain and the application of the framework to real-world complex systems is an area of ongoing research. 
	
	\subsubsection{Generalized Information Decomposition}
	\label{sec:gid}
	
	While the PED successfully relaxes the requirement that a system be partitioned into multiple ``inputs" and a single ``target", it does so at the expense of being a true information decomposition, instead decomposing the entropy. To construct a general information decomposition that relaxes the ``input"/``target" distinction while still retaining the intuitions about what information is, \cite{varley_generalized_2024} introduced the \textit{generalized information decomposition} (GID) based on the Kullback-Leibler divergence. 
	
	Recall from Sec. \ref{sec:dkl} that the Kullback-Leibler divergence can be written as the expected difference in local entropies:
	
	\begin{align}
		D_{KL}(P||Q) = \mathbb{E}_{P(X)}[h^Q(x) - h^P(x)]
	\end{align}

	Given a localizable PED, it is possible to decompose each of those local entropy terms, with the differences between each partial entropy atom defining a partial Kullback-Leibler divergence:
	
	\begin{equation}
		D_{KL}^{P||Q}(\boldsymbol{\alpha}) = \mathbb{E}_{P(X)}[h^Q_\partial(\boldsymbol{\alpha}) - h^P_\partial(\boldsymbol{\alpha})]
	\end{equation}

	This notation is admittedly a little \textit{ad hoc}. The result, is a decomposition of the Kullback-Leibler divergence, such that:
	
	\begin{equation}
		D_{KL}(P||Q) = \sum_{\boldsymbol{\alpha}\in\mathfrak{A}}D_{KL}^{P||Q}(\boldsymbol{\alpha})
	\end{equation}

	Importantly, since the GID decomposes the information gained when updating from an arbitrary prior to an arbitrary posterior, it the PID as a special case: by decomposing the single-target mutual information when it is understood as:
	
	\begin{equation}
		\label{eq:single_target_gid}
		I(\textbf{X};Y) = D_{KL}(P(\textbf{X}|Y)||P(\textbf{X}))
	\end{equation}

	The link between the PID and the PED illuminated by the GID also applies to redundancy functions as well. For example, if one decomposes Eq. \ref{eq:single_target_gid} using the $h_\cap^{\min}$ function, the result is same as the PID computed using the $i_\cap^{\pm}$ function. Similarly, a GID using the $h^{sx}_\cap$ function is equivalent to a PID using the $i^{sx}_\cap$ function. The relationship between the GID, PID, and PED is deep: the PID is a special case of the GID, while the GID is constructed from the (local) PED. The PED itself is just a special case of the PID, when decomposing the information that the ``parts" disclose about the ``whole". All are in some sense fundamental, while all are also constructable from each-other. 
	
	\subsubsection{Integrated Information Decomposition}
	\label{sec:phi_id}

	\begin{figure}
		\begin{center}
			\begin{tikzpicture}
			
			\filldraw[black] (0,1) circle (3pt) node[anchor=south] {$\{12\}\rightarrow\{12\}$};
			
			\filldraw[black] (-4.25,-1.5) circle (3pt) node[anchor=east] {$\{1\}\rightarrow\{12\}$};
			
			\filldraw[black] (-1.5,-1.5) circle (3pt) node[anchor=east] {$\{2\}\rightarrow\{12\}$};
			
			\filldraw[black] (1.5,-1.5) circle (3pt) node[anchor=west] {$\{12\}\rightarrow\{1\}$};	
			
			\filldraw[black] (4.25,-1.5) circle (3pt) node[anchor=west] {$\{12\}\rightarrow\{2\}$};	
			
			\filldraw[black] (0,-4) circle (3pt) node[anchor=east] {$\{2\}\rightarrow\{1\}$};	
			
			\filldraw[black] (-6,-5) circle (3pt) node[anchor=west] {$\{1\}\{2\}\rightarrow\{12\}$};
			
			\filldraw[black] (-3.25,-5) circle (3pt) node[anchor=west] {$\{1\}\rightarrow\{1\}$};	
			
			\filldraw[black] (3.25,-5) circle (3pt) node[anchor=east] {$\{2\}\rightarrow\{2\}$};	
			
			\filldraw[black] (6,-5) circle (3pt) node[anchor=east] {$\{12\}\rightarrow\{1\}\{2\}$};
			
			\filldraw[black] (0,-6) circle (3pt) node[anchor=west] {$\{1\}\rightarrow\{2\}$};
			
			\filldraw[black] (-4.25,-8.5) circle (3pt) node[anchor=east] {$\{1\}\{2\}\rightarrow\{1\}$};
			
			\filldraw[black] (-1.5,-8.5) circle (3pt) node[anchor=east] {$\{1\}\{2\}\rightarrow\{2\}$};
			
			\filldraw[black] (1.5,-8.5) circle (3pt) node[anchor=west] {$\{1\}\rightarrow\{1\}\{2\}$};	
			
			\filldraw[black] (4.25,-8.5) circle (3pt) node[anchor=west] {$\{2\}\rightarrow\{1\}\{2\}$};	
			
			\filldraw[black] (0,-11) circle (3pt) node[anchor=north] {$\{1\}\{2\}\rightarrow\{1\}\{2\}$};		
			
			\draw[] (0,1) -- (-4.25,-1.5); 
			\draw[] (0,1) -- (-1.5,-1.5);
			\draw[] (0,1) -- (4.25,-1.5);
			\draw[] (0,1) -- (1.5,-1.5);
			
			\draw[] (-4.25,-1.5) -- (-3.25, -5);
			\draw[] (4.25,-1.5) -- (3.25, -5);
			\draw[] (-4.25,-1.5) -- (0,-6);
			\draw[] (4.25,-1.5) -- (0,-6);
			
			\draw[] (0,-4) -- (-4.25,-8.5);
			\draw[] (0,-4) -- (4.25,-8.5);
			
			\draw[] (-1.5,-1.5) -- (-6,-5);
			\draw[] (1.5,-1.5) -- (6,-5);
			\draw[] (-4.25,-1.5) -- (-6,-5);
			\draw[] (4.25,-1.5) -- (6,-5);
			
			\draw[] (-1.5,-1.5) -- (3.25,-5);
			\draw[] (1.5,-1.5) -- (-3.25,-5);
			\draw[] (1.5,-1.5) -- (0,-4);
			\draw[] (-1.5,-1.5) -- (0,-4);
			
			\draw[] (-3.25,-5) -- (1.5,-8.5);
			\draw[] (3.25,-5) -- (-1.5,-8.5);
			
			\draw[] (0,-6) -- (1.5,-8.5); 
			\draw[] (0,-6) -- (-1.5,-8.5); 
			
			\draw[] (-6,-5) -- (-4.25,-8.5);
			\draw[] (6,-5) -- (4.25,-8.5);
			\draw[] (-6,-5) -- (-1.5,-8.5);
			\draw[] (6,-5) -- (1.5,-8.5);
			
			\draw[] (-3.25,-5) -- (-4.25,-8.5);
			\draw[] (3.25,-5) -- (4.25,-8.5);
			
			\draw[] (-4.25,-8.5) -- (0,-11);
			\draw[] (-1.5,-8.5) -- (0,-11);
			\draw[] (4.25,-8.5) -- (0,-11);
			\draw[] (1.5,-8.5) -- (0,-11);
			
			\end{tikzpicture}
		\end{center}
		\label{fig:double_lattice}
	\end{figure}

	The third generalization of the PID is the \textit{integrated} information decomposition ($\Phi$ID)  \cite{mediano_beyond_2019}, which expands the PID to multiple sources and multiple targets. Doing so requires proposing a ``double redundancy" function that captures some notion of the information that is doubly-redundant across multiple sources and targets, and a double-redundancy lattice that captures the structure of multivariate information. The double redundancy lattice is a \textit{product lattice} $\mathcal{A}^2 = \mathcal{A}\times\mathcal{A}$ (where $\mathcal{A}$ is the single-target redundancy lattice derived above). Each vertex represents an ordered pair $\boldsymbol{\alpha}\to\boldsymbol{\beta}$, with $\boldsymbol{\alpha}, \boldsymbol{\beta}\in\mathcal{A}$.
	
	The product lattice is a partially ordered set, with:
	
	\begin{equation}
	\boldsymbol{\alpha} \to \boldsymbol{\beta} \preceq \boldsymbol{\alpha}' \to \boldsymbol{\beta}' \Leftrightarrow \boldsymbol{\alpha} \preceq \boldsymbol{\alpha}', \boldsymbol{\beta} \preceq \boldsymbol{\beta}'
	\end{equation}
	
	In the $\Phi$ID framework, the double-redundancy lattice is not derived from the properties of redundancy function  $I_{\phi\cap}(\boldsymbol{\alpha}\to\boldsymbol{\beta})$ (as is the case in the single-target PID) but rather, it emerges from the product of the ``marginal" lattices. This is a non-trivial difference. For the PID, while there has been much contention over the correct definition of ``redundancy", essentially everyone agrees that the logic of the lattice is correct (although for an alternative, see \cite{kolchinsky_novel_2022}) and that the idealized redundancy function ``works". Not so for the $\Phi$ID, though, which puts the framework on more mathematical shaky ground than the PID/PED/GID (although given its novelty, future work shoring up the foundations is certainly possible). 
	
	This does not mean that any redundancy function is valid: there are constraints on what kind of redundancy function will work with the product lattice. For example, Mediano et al., \cite{mediano_beyond_2019} give a ``compatibility constraint" that requires that, in the case of multiple sources and a single target, the $\Phi$ID should revert to a standard PID. Formally, given two variables \textbf{X}, \textbf{Y} and two double-redundancy atoms $\boldmath{\alpha}, \boldmath{\beta} \in \mathcal{A}^{2}$
	\begin{equation}
	I_{\phi\cap}({\boldsymbol{\alpha}\to\boldsymbol{\beta}}) = 
	\begin{cases}
	I_{\cap}(\textbf{X}^{\alpha_1},\ldots,\textbf{X}^{\alpha_k} ; \textbf{Y}^{\beta_1}) & \iff |\textbf{Y}|=1 \nonumber \\
	I_{\cap}(\textbf{Y}^{\beta_1},\ldots,\textbf{Y}^{\beta_j} ; \textbf{X}^{\alpha_1}) & \iff |\textbf{X}|=1 \nonumber \\	
	I(\textbf{X};\textbf{Y}) & \iff |\textbf{X}| = |\textbf{Y}| = 1
	\end{cases}
	\end{equation}
	
	The compatibility axiom requires that, if one of the variables (\textbf{X} or \textbf{Y}) is univariate, then the double redundancy function reduces to a classic, single-target redundancy function, and the $\Phi$ID reduces to the classic PID. This sets up the PID as essentially a ``special case" of the $\Phi$ID (a nice property for a proposed generalization). 
	
	As it stands, there has been less empirical work using the $\Phi$ID than the PID, and almost all of it has been done in areas of computational neuroscience, using modalities such as fMRI \cite{luppi_synergistic_2022,luppi_reduced_2023,luppi_synergistic_2024,gatica_transcranial_2024}, cell electrophysiology \cite{varley_decomposing_2022}, and \textit{in silico} dynamic models \cite{menesse_integrated_2024,varley_evolving_2024}. Given the paucity of work on $\Phi$ID generally, the question of what a natural redundancy function might be has received limited attention. The most commonly used redundancy function (used in all the above-cited works but one) is the minimum mutual information function:
	
	\begin{equation}
	I^{MMI}_{\phi\cap}(\boldsymbol{\alpha}\to\boldsymbol{\beta}) = \min_{i,j}I(\textbf{A}_i;\textbf{B}_j)
	\end{equation}
	
	which can be computed for continuous variables under Gaussian assumptions, but does not readily localize to individual realizations. Unfortunately, while the $I^{MMI}_{\Phi\cap}$ function has practical appeal, it has counter-intuitive behaviors that have led some to question its appropriateness \cite{varley_considering_2024}. In particular, it appears to conflate first-order autocorrelation and higher-order temporal synergy in potentially fatal ways. Other double redundancy functions have been proposed, including one generalizing the $i^{sx}_\cap$ function \cite{varley_decomposing_2022}, one based on the $i^{ccs}_\cap$ function \cite{mediano_towards_2021}, and one based on the dependency lattice approach of James et al., \cite{james_unique_2018,mediano_beyond_2019}.
	
	\subsubsection{Information Dynamics, Revisited}
	\label{sec:info_dynamics_revisted}
	The $\Phi$ID framework, like the PID framework, is agnostic to the type of data to which it is applied. However (at the time of this writing), all published applications of the $\Phi$ID have been to temporal data: pairs of elements that are evolving through time. In this case, the $\Phi$ID can be understood as decomposing the excess entropy (see Eq. \ref{eq:excess}). The most common approach is lag-1 excess entropy for two variables: $I(X^1_{t-1},X^2_{t-1};X^1_t, X^2_t)$. In this case, the various integrated information atoms take on particular interpretations in terms of information dynamics \cite{mediano_beyond_2019}:
	
	\begin{itemize}
		\item[] \textbf{Information storage:} Information in one source at time $t-1$ that \textit{stays} in that source at time $t$. E.g. $\{X_i\}\to\{X_i\}$, $\{X_1\}\{X_2\}\to\{X_1\}\{X_2\}$, and $\{X_1,X_2\}\to\{X_1,X_2\}$.
		\item[] \textbf{Information transfer:} Information in one element at time $t-1$ that moves to another element at time $t$. E.g. $\{X_i\}\to\{X_j\}$.
		\item[] \textbf{Information copying:} Information that is in just one element at time $t$ that is duplicated to be redundantly present in both elements at time $t+1$. E.g. $\{X_i\}\to\{X_1\}\{X_2\}$.
		\item[] \textbf{Information erasure:} information that is redundantly present in both elements at time $t$ that is ``pruned" from one, and left in the other. E.g. $\{X_1\}\{X_2\}\to\{X_i\}$.
		\item[] \textbf{Upward causation:} When information in a ``lower order" atom is moved ``upward" into a synergistic atom. E.g. $\{X_i\}\to\{X_1,X_2\}$, or $\{X_1\}\{X_2\}\to\{X_1,X_2\}$.
		\item[] \textbf{Downward causation:} When information in a ``higher order" atom moves ``down" to influence lower-order atoms. E.g. $\{X_1,X_2\}\to\{X_i\}$, or $\{X_1,X_2\}\to\{X_1\}\{X_2\}$. This one is philosophically contentious: the question of whether ``downward causation" is a real phenomenon is a topic of active discussion \cite{bedau_downward_2002,gibb_routledge_2019}.
		\item[] \textbf{Causal decoupling:} A particularly unusual form of information storage that gets it's own category: irreducible information in the ``whole" that persists through time: $\{X_1,X_2\}\to\{X_1,X_2\}$. This has also been discussed in the context of ``emergence" \cite{rosas_reconciling_2020,mediano_beyond_2019}.
	\end{itemize}
	
	In theory, for temporal data, each $\Phi$I atom corresponds to a unique information dynamic. As the number of elements gets larger, however, the intuitive interpretations get harder. For example, for a three-element system, we might propose a dynamic:
	
	\begin{itemize}
		\item[] \textbf{Modification:} When information in the joint state of two sources influences the output of a third. E.g. $\{X_1,X_2\}\to\{X_3\}$.
	\end{itemize}
	
	This definition is analogous to the one proposed by Timme in the context of the single-target PID \cite{timme_synergy_2014}, and later explored by Sherill \cite{faber_computation_2018,sherrill_correlated_2020,sherrill_partial_2021,newman_revealing_2022}. Other atoms are more difficult to classify. For a three element system, how might one understand $\{X_1\}\{X_2\}\{X_3\}\to\{X_1,X_2\}\{X_1,X_3\}\{X_2,X_3\}$? This is the information redundantly copied over all three nodes at time $t$ that is converted into being copied over all pairs of bivariate joint states at time $t+1$?
	
	Like the PED, the comparative lack of investigation into the $\Phi$ID can be seen as a blessing and a curse. As an under-developed set of tools, off-the-shelf applications are probably pretty few and far between at this point. The flip side of this, however, is that there is much fertile ground to be explored asking fundamental questions and exploring basic applications of these frameworks to complex systems. 

	\newpage
	
	\section{Network Inference}
	
	One of the foundational methods in complex systems science is the construction of network models from data \cite{barabasi_network_2016,menczer_first_2020}. It is hard to overstate how fundamental networks are to the modern study of complex systems, appearing in everything from brain science \cite{sporns_networks_2010} to sociology \cite{granovetter_strength_1973,kossinets_empirical_2006}, from ecology \cite{dunne_network_2004,thompson_food_2012} to economics \cite{vie_how_2020}. In applications, networks generally break down into two broad classes: there are physical networks (where the nodes and edge are unambiguous, physical objects in the real world), and statistical networks (where the links between nodes reflects some statistical dependency between them). Physical networks are generally reasonably easy to infer the structure of. For example, the airline network is conceptually simple: nodes may be airports (or cities), and a directed edge exists between two nodes if there is a regular flight from one to another \cite{du_analysis_2016}. Another example might be a structural connectivity network map of the brain: using diffusion tensor imaging (DTI) it is possible to get an estimate of the white matter tracts in the brain that physically connect disparate brain regions \cite{garyfallidis_dipy_2014}.  
	
	In contrast, statistical networks are more abstract: rather than referring directly to some physical structure in the real world (like a white-matter tract, or airline route), the edges in a statistical network refer to some degree of statistical dependency between elements. In an information-theoretic context, these dependencies almost always take the form of some kind of mutual information: nodes are usually said to be connected if knowing \textit{something} about the state of one node reduces our uncertainty about the state of another node. There are other ``kinds" of statistical networks not based on mutual information (for example, dynamic causal modeling builds statistical networks using Bayesian model comparison \cite{friston_dynamic_2003,friston_dynamic_2019}), however for our purposes, we will focus on information-theoretic approaches. Within the category of statistical networks, we will describe two general classes: the ``functional" connectivity network, which has no notion of temporal order and produces undirected networks, and the ``effective" connectivity network, which does take time into account (if time is a relevant feature) and produces directed networks. Finally, we will discuss how the problem of synergies in data can complicate the very idea of a bivariate network model of a complex system and briefly touch on higher-order models such as hypergraphs and simplicial complexes (although a rigorous mathematical treatment of these topics is beyond the scope of this tutorial).   
	
	\subsection{Functional Connectivity Networks}
	\label{sec:fc_nets}
	
	The simplest form of network construction is that of a ``functional connectivity" (FC) network. In an FC network, the edges between nodes are \textit{undirected} and describe some symmetric statistical relationship between the two elements of a system. Functional connectivity analysis is generally most well known as a tool in computational neuroscience \cite{sporns_networks_2010,fornito_fundamentals_2016} and an Internet search for ``functional connectivity" will return overwhelmingly neuroscience-focused results, however the notion of using time-series statistics to infer a network is much more broadly applicable and has been used when constructing genomic networks \cite{butte_mutual_1999,liang_gene_2008}, financial networks \cite{fiedor_networks_2014,guo_development_2018}, climate networks \cite{donges_complex_2009}, and in developmental biology \cite{blackiston_revealing_2024}.
	
	The basic logic of functional connectivity is the equivalence between a correlation matrix and the adjacency matrix of a network. For a system with $N$ features, the correlation matrix is an $N\times N$ matrix that is symmetric about the diagonal (since the correlation is \textit{usually} symmetric in its arguments). If we imagine a network with $N$ nodes, we can draw a link between every pair of nodes, where the weight of edge $i,j$ is equivalent to the corresponding correlation in cell $i,j$. This link between square, symmetric matrices and networks forms the basis of a large field of mathematics known as spectral graph theory (for reference, see \cite{brouwer_spectra_2012}), although for our purposes, the isomorphism is sufficient. For a data set with $N$ features (such as $N$ time series, each corresponding to a dynamic element), the basic algorithm for network inference is very simple. In Python, a basic implementation using \texttt{numpy} arrays might be:
	
	\begin{python}
		N0, N1 = data.shape #Elements are rows, frames are columns. 
		mi_mat = np.zeros((N0, N0))
		for i in range(N0):
			for j in range(i): #Only need the triangle, since MI is symmetric.
				mi = mutual_information(data[i,:], data[j,:])
				mi_mat[i,j] = mi
				mi_mat[j,i] = mi
	\end{python}

	The result is a dense matrix, corresponding to a fully-connected network (there is a non-zero edge between every pair of nodes). 
	
	Given an efficient estimator of mutual information, this appears to be the end of the line. However, despite the apparently simplicity, there are two, surprisingly subtle issues that must be accounted for: the question of how to determine if an edge (dependency) is ``real", and the inherent bias in the naive estimation of mutual information.
	 
	\subsubsection{Bias in Naive Entropy Estimators}
	
	The most common way to estimate entropies and mutual information from discrete data is through a naive estimator, where $P(X=x)$ is a function of how many times $x$ appears in a sample, divided by the total number of samples. Due to finite size effects, however, the naive entropy $\hat{H}(X)$ systematically \textit{under-estimates} the true entropy, and by extension, \textit{over-estimates} the true mutual information. Being strictly positive, the naive mutual information estimator is known to be upwardly-biased, reliably over-estimating the true mutual information \cite{bossomaier_introduction_2016}. This means that, unless large amounts of data are included in the estimates that define the edge weights, the overall information-density of the system is being over-estimated. 
	
	Depending on the specific problem being studied, this may not be a problem. Often, as scientists, we don't actually care about the exact value of the mutual information in bits or nats, but rather, the \textit{relative} values of some set of correlations compared to another. For example, if one is interested in comparing brain data recorded from a conscious person versus data from the same individual after being anesthetized (as in \cite{luppi_synergistic_2024}), the fact that all the inferred mutual information values are upwardly inflated may not be a critical issue (so long as the upward bias is uniform across conditions). 
	
	If accurate estimates of the mutual information are critical, there are two possible solutions. The first is analytic corrections. There is a large body of mathematical research dedicated to correcting the inherent bias in naive entropy and mutual information estimators \cite{schurmann_bias_2004} (for discussion, see Sec. \ref{sec:discrete_signals}), and analytic corrections can typically be implemented with little-to-no computational overhead. The second approach is to construct a ``null" distribution that quantifies the expected null mutual information if there was no correlation between variables. Due to the upward bias of naive mutual information estimators, the expected value of this distribution will be greater than zero, and can be subtracted off from the empirical mutual information to improve the estimate \cite{miller_note_1955,ince_statistical_2016}. While constructing null distributions is often computationally expensive, it gives the analyst a high degree of control over exactly how to formulate null models: for example, perhaps one might want to preserve certain features of the variables (such as autocorrelation) while randomizing the dependency between them. This can be built into a null, but cannot usually be addressed using analytic corrections. 
	
	\subsubsection{Significance Testing \& Thresholding}
	
	The other concern when inferring functional connectivity networks is whether an edge is ``real", and deciding on a criteria for whether to include an edge in the network model. The most common way to do this is to compute an analytic or bootstraped p-value for a null-hypothesis significance test. For Gaussian and discrete variables, it is possible to compute analytic p-values based on Chi-squared distributions \cite{brillinger_data_2004,cheng_data_2006}. The estimators are infrequently used, although are implemented in the \texttt{JIDT} package \cite{lizier_jidt_2014}. In the absence of an analytic p-value, null hypothesis significance testing can be done by constructing a null distribution (as discussed above and described in detail below). Given some suitably large number of nulls, it is easy to compute an empirical p-value by quantifying the number of null mutual information values greater than or equal to the empirical value. 
	
	A final point of discussion is whether a functional connectivity network should be sparsified beyond excluding non-significant edges. This is particularly popular in the human neuroimaging literature, where functional connectivity networks are routinely thresholded at upwards of ninety percent (keeping only the top one to ten percent of edges) \cite{van_den_heuvel_proportional_2017}. This is often done to justify the use of graph-theoretic measures, which typically are most well-defined on sparse networks. However, recent work has shown that this kind of arbitrary thresholding can produce the appearance of non-trivial topology even in unstructured networks: for example, Cantwell et al., found that thresholding random networks at high values can create plausibly heavy-tailed degree distributions, even when the generating data was Gaussian noise \cite{cantwell_thresholding_2020}. Other systematic studies of thresholding have also found that it can induce systemic biases in the topology of the resulting network (e.g. \cite{schwarz_negative_2011,garrison_stability_2015,adamovich_thresholding_2022}), and as such, this author recommends \textit{against} sparsifying functional connectivity networks beyond significance testing edges. At the risk of being polemical, radically sparsifying a network to make it receptive to a particular graph-theoretic algorithm is probably indicative that one is putting the cart before the horse: analyses should be constructed to suit the nature of the data, rather than torturing the data to make one's preferred analysis possible. We should not paint with too broad a brush, however, and there may be cases where such choices are appropriate. In general, rather than providing a prescriptive checklist of DOs and DON'Ts, we hope that researchers critically consider the pros and cons of every choice made in a pipeline, rather than simply acting on autopilot or because it was done in a particular way by a previously highly-cited paper. 
	
	\subsubsection{Null Models for Testing Empirical Mutual Information}
	\label{sec:fc_nulls}
	
	We have made reference to the idea that one can use null distributions to significance test empirical mutual information values, as well as for correcting the inherent bias associated with naive entropy estimators. The question of building strong null models is a surprisingly deep one, and we will not be able to do it justice in this tutorial (for an accessible review, see \cite{vasa_null_2022}). In general, null distributions are built by randomizing the existing data in some way, and re-computing the mutual information some large number of times. The question of how exactly to randomize, however, is where a lot of the nuance comes in. There is a fairly simple maxim that can be of practical use here:\newline
	
	\textit{A null model should ideally preserve every possible feature of your data \textbf{except} the one that you are testing.}\newline
	
	What does this mean? In many complex systems, patterns or statistics can be consequences of constraints imposed by other features of the system or data (in evolutionary biology, these patterns are called ``spandrels" \cite{gould_spandrels_1997,rubinov_constraints_2016}). When generating a null model to significance-test something like the mutual information, we want to ensure that we include all features of the data that might inflate the apparent correlation \textit{other than} the actual statistical dependency of the data. 
	
	A commonly-encountered example of this is in the analysis of time series data. It is well-known that autocorrelation in time series data reduces the effective degrees of freedom of the variable and consequently inflates the correlation between two variables \cite{afyouni_effective_2019,cliff_exact_2020}. This means that, if the data being studied is time series data, whatever randomization is used to disrupt the dependence between $X$ and $Y$ should preserve the autocorrelation of both variables (for example, using a circular shift null, rather than shuffling the data). By preserving the autocorrelation but randomizing the joint statistics of the two variables, we ensure that the null mutual informations benefit from the same ``bump" caused by the autocorrelation that the empirical mutual information did. This is an apples-to-apples comparison, whereas building a null distribution that destroyed autocorrelation in the surrogate data would be apples-to-oranges. 
	
	There are many different kinds of randomization that have been explored in the literature (the \texttt{IDTxl} packages conveniently implements a number of options and describes them in the documentation), and choosing the best null for a given dataset is as much an art as it is a technical concern. Depending on exactly what features of the data must be preserved (such as autocorrelation, marginal entropies, etc), the choice of how best to randomize can become quite nuanced. It is not uncommon for the actual analysis of empirical data to take much less time (and be much easier) than the process of designing and generating the correct null model against which to test that empirical data. 
	
	\subsection{Effective Connectivity Networks}
	\label{sec:eff_nets}
	
	The functional connectivity network has no notion of ``time" - it describes the instantaneous correlation between elements and would be the same if one were to permute frames (preserving the joint state). To account for temporal dynamics requires an alternative approach, one that can infer directed edges. Historically, this was done using the temporal mutual information, quantifying how much knowing the \textit{past} of $X$ reduces our uncertainty about the \textit{future} of $Y$. The temporal mutual information has a significant limitation, however: it will conflate true information transfer with active information storage if the signals are autocorrelated. Consider the case of two time series: $X=\sin(t)$ and $Y=\cos(t)$. Since these signals are periodic and offset by $\pi/2$, knowing the state of $X$ at time $t$ uniquely specifies the state of $Y$ at time $t+\tau$, which is indicated by the non-zero temporal information. However, the \textit{same} information about $Y_{t+\tau}$ could also be learned by observing $Y$'s own past. There is no \textit{additional} information disclosed by $X$'s past about $Y$'s future that isn't disclosed by $Y$'s own past.
	
	A more natural measure of temporal information flow is the transfer entropy (see Sec. \ref{sec:te} for the introduction to transfer entropy). Here we repeat Eq. \ref{eq:te}) for convenience:
	
	\begin{equation}
	TE(X \rightarrow Y) = I(X_{-k:t-1}; Y_t | Y_{-l:t-1}). \nonumber
	\end{equation}
	
	As described above, the transfer entropy quantifies the information that $X$'s past discloses about $Y$'s future after $Y$'s own past has been taken into account. Importantly, this can include the unique information that flows from past to future, as well as the synergistic interaction between both pasts (\cite{williams_generalized_2011,james_information_2016}, see Sec. \ref{sec:info_dynamics_revisted} for further discussion).
	
	In contrast to the functional connectivity, which is strictly undirected, effective connectivity networks are directed and model how information \textit{flows} from one element of the system to another. The effective network is \textit{not} usually equivalent to the causal network and it is known that multiple different causal structures can give rise to the same effective structure \cite{ay_information_2008, lizier_differentiating_2010, razak_quantifying_2014}, but is nevertheless a useful tool that can provide insights beyond a symmetric functional connectivity analysis. 
	
	\subsubsection{Optimizing Parameters}
	
	Unlike the functional connectivity, which is parameter-free, the transfer entropy requires deciding on how much of the history from the source and target to consider when computing the conditional mutual information. In Eq. \ref{eq:te}, these are the two parameters $k$ (for the source past) and $l$ (for the target past). The question of how best to optimize them is a complex topic, and I will not be able to do it justice here. For readers interested in the gory details, I recommend Bossomaier et al.'s \textit{Introduction to Transfer Entropy} \cite{bossomaier_introduction_2016}. Here, I will briefly review the general concerns (over/underestimating the true value), as well as describing a simple, serial conditioning algorithm (based on \cite{novelli_large-scale_2019,novelli_inferring_2020}) and briefly discussing null-models (again). 
	
	If $k$ is underestimated and not enough of the source's history is taken into account, then the estimate of the effective connectivity will be erroneously low, as information that takes too long to propagate will be excluded (for example, in the case of axon conduction delays \cite{lemarechal_brain_2021}). On the other hand, if $k$ is overestimated, then the effective connectivity will be inflated by finite size effects. For a discrete time series that can adopt $N$ states for a given history dimension of $k$, there will be $N^k$ possible unique past states. For large $N$, $k$, or in the case of small data sets, most configurations will only be seen a small number of times, creating the illusion of deterministic transitions and inflating the apparent transfer entropy. Similarly, if the target time series history embedding dimension $l$ is too short, then long-range autocorrelations might be missed, potentially inflating the apparent transfer entropy by incomplete conditioning. On the other hand, if $l$ is too large, the same finite size effects can create illusory determinism in the target's own information storage dynamics. 
	
	Picking the correct embedding then, requires balancing a series of trade-offs. Too little history accounted for will underestimate the true value, while too much history accounted for can compromise it through finite size effects. One possible solution to this problem is a serial conditioning approach (described in detail by \cite{novelli_inferring_2020}). Briefly, the algorithm works by testing whether each additional bin of history produces a statistically significant increase in the transfer entropy when added. For example, if optimizing the target $Y$'s past, one begins with the active information storage with one bin of history: $I(Y_{t-1};Y_{t})$. Then, one bin of history is added in the form of a conditional mutual information: $I(Y_{t-2};Y_{t}|Y_{t-1})$, and this value is significance-tested against a null distribution. If the increase in the active information storage is significantly \textit{greater} than would be expected by chance, then the increase is likely to be real and not simply a finite-size effect. This process can be repeated with increasing values of $k$ (conditioning on all previous $k-1$ values each time) until either subsequent values are no longer significant, or a pre-defined $k_{\max}$ is reached. Once the optimal value $k^*$ is selected, the same algorithm can be used to optimize the source dimensionality ($l^*$) using the optimal $k^*$. It is also possible to use the same logic to infer a non-uniform embedding, given some alternative stopping criterion (such as a pre-selected $k_{\max})$ \cite{faes_information-based_2011}. 
	
	This approach, while powerful, comes with a nontrivial computational cost. For every successive possible value of $k$ and $l$, there must be a null-hypothesis significance test, typically involving the construction of a large distribution of null values. For large systems with many source-target pairs (each requiring a different value of $l$), this can become prohibitively expensive. The sheer number of significance tests also requires stringent correction for false discoveries, necessitating strict corrections. 
	
	\subsubsection{Null Models}
	In Sec. \ref{sec:fc_nulls}, we discussed the importance of null models that preserve as many potentially salient features of the data as possible, while randomizing only the relationship of interest. In the case of transfer entropy, this generally necessitates that only the \textit{source} time series is randomized and that the \textit{target} time series remains unchanged. This is based on the logic that the transfer entropy is the information flow that is revealed only when the past of the target is taken into account. A null model then, must preserve the active information storage of the target. A null that involved randomizing both source and target series would be too liberal. Similarly, the randomization of the source time series should preserve the autocorrelation in the the data, disrupting only the interaction between the variables. A circular-shift null is generally optimal for this purpose. 
	
	In some cases, such as in spiking neural recordings, bursty activity or other global signals can artificially inflate the number of significant directed edges (e.g. in \cite{ito_extending_2011,ito_large-scale_2014,shimono_functional_2015}), which can necessitate further processing of the edges using additional tests, such as the ``coincidence index" \cite{ito_extending_2011} or the ``sharpness index" \cite{kajiwara_inhibitory_2021}. In general, whatever network comes out of the effective connectivity inference algorithm should be ``sanity checked" against known features of the system (i.e. is it implausibly dense)?
	
	\subsubsection{Redundancy \& Synergy in Effective Connectivity Networks}
	
	The bivariate transfer entropy, like the bivariate mutual information, has a significant limitation in the context of network inference: it will double-count redundancies while missing higher-order, synergistic relationships. For example, if $Z_{t}=X_{t-k}\bigoplus Y_{t-l}$, then $TE(X\rightarrow Z)=0$ bit and $TE(Y\rightarrow Z)=0$ bit, while the joint transfer entropy $TE(X,Y\rightarrow Z)=1$ bit. In this case, there is a synergistic interaction between both parents and the target produced by the logical-XOR computation performed by the target. In the case of a system with synergies, a bivariate transfer entropy network is not a ``complete" description of the system. The opposite can happen when there is redundant information: if $X$ and $Y$ are both being driven by a common driver $W$ (which is a ``grandparent" element to $Z$, then the influence of $W$ can show up multiple times in the interactions between $X$ or $Y$ and $Z$, inflating both transfer entropies and increasing the ``apparent" integration of the whole system. 
	
	One possible solution to this is with the \textit{global transfer entropy}, which conditions the bivariate transfer entropy on the state of every other element in the system. If $X$ and $Y$ are both elements of \textbf{V}:
	
	\begin{equation}
	TE_G(X\rightarrow Y|\textbf{V}) = I(X_{-k:t-1}; Y_t | Y_{-l:t-1}, \textbf{V}^{-\{X,Y\}}_{-q:t-1}). \nonumber
	\end{equation}
	
	Where $\textbf{V}^{-\{X,Y\}}_{-q:t-1}$ indicates the past state of every element of \textbf{W} \textit{excluding} $X$ and $Y$. The global transfer entropy is conceptually very powerful, as all synergiestic dependencies that inform on the relationship $X\to Y$ will be illuminated, and redundant information will be pruned. This can make it useful for the study of synergies in complex systems (as in \cite{marinazzo_synergy_2019}). However, for large systems, the amount of data required to estimate the multidimensional probability distribution can be prohibitively large. 
	
	To address this, \cite{novelli_large-scale_2019} propose a multivariate transfer entropy algorithm that uses robust, multilevel statistical tests and a greedy optimization procedure to identify the minimal set of parent elements $\textbf{V}^{*}$ that optimizes the conditional transfer entropy $TE(X\to Y | \textbf{V}^{*})$. In doing so, the mutlivarite transfer entropy algorithm builds a bivariate network, but where the weights of the edge between source $X$ and target $Y$ is ``informed by" the interaction between every other parent of $Y$ as well. Analysis with simulated Gaussian and mean-field field models has found that the multivariate transfer entropy does a much better job at extracting the ground truth generating model of the simulated data then the bivariate transfer entropy \cite{novelli_inferring_2020,ursino_transfer_2020}. The multivariate transfer entropy network inference algorithm has been applied to inferring interaction networks between cryptocurrencies \cite{garcia-medina_network_2020}, biological neuronal networks \cite{varley_information-processing_2023,varley_serotonergic_2024}, and human electrophysiological data \cite{harmah_measuring_2019}.
	
	It is my recommendation that the multivariate transfer entropy analysis should be used whenever plausible (the only existing implementation is provided by the \texttt{IDTxl} package \cite{wollstadt_idtxl_2019}), but its limitations should also be made clear. While it is less data-intensive than the global transfer entropy, the optimization of $\textbf{V}^{*}$, like the optimization of the embedding dimensions and lags, requires considerable computational effort in the form of multiple sequences of large null distributions. This can produce extremely long runtimes as multiple optimizations are run in sequence. 
	
	\subsection{Hypergraphs, Simplicial Complexes, \& Other Higher-Order Frameworks}
	
	While the multivariate transfer entropy is sensitive to synergies (and prunes redundancies) in a way that the bivariate transfer entropy is not, it could still be argued that networks are simply too restrictive a model for complex systems rich in higher-order redundancies and synergies. For example, consider our toy logical-XOR system (for simplicity, we will exclude the lags from consideration): $Z = X\bigoplus Y$. If we take a conditioning approach, we will find that $I(X;Z|Y)=1$ bit and and $I(Y;Z|X)=1$ bit, which will appear as two, bivariate links in the final network. Unfortunately, looking at the network, it is impossible to differentiate between the case where two links represent a \textit{synergistic} interaction between two parents and a target versus the case where both parents have independent, specific relationships with the target. 
	
	To fully represent the possible, higher-order dependencies, it is necessary to leave the space of strictly bivariate graphs behind and use other frameworks that are more general \cite{battiston_networks_2020,torres_why_2021,bick_what_2022}. There are two commonly explored generalizations of the bivariate network: the \textit{hypergraph} and the \textit{simplicial complex}. 
	
	Hypergraphs allow edges to be incident on more than two nodes. A standard network is described by two sets: $\textbf{V}=\{V_1,\ldots,V_k\}$ which defines the nodes of the network, and $\textbf{E}=\{(V_i,V_j)|i,j\in k\times k\}$, which defines the edges are pairs of nodes. Importantly, every $E_i\in\textbf{E}$ is a (potentially ordered) pair consisting only of two entries. If we relax the requirement that the elements of $\textbf{E}$ can only be pairs, but instead tuples of arbitrary length, we have a hypergraph. In the hypergraph, edges can be incident on arbitrary numbers of nodes, and can even be subsets of each-other (i.e. it is possible to have $(X_i,X_j,X_k)$ and $(X_i,X_j)$ defining distinct edges with distinct weights. This makes hypergraphs natural frameworks for assessing higher-order synergies and redundancies in data. For example, Varley et al., used hypergraph community detection \cite{kumar_new_2020} to look for patterns in sets of three redundantly or synergistically interacting brain regions \cite{varley_partial_2023}, while Marinazzo et al., used information theory to infer hypergraphical relationships in psychometric data \cite{marinazzo_information-theoretic_2024}. Finally, Medina-Mardones et al., propose an analytical framework to asses hypergraphical relationships in a computationally tractable way \cite{medina-mardones_hyperharmonic_2021}. 
	
	The other generalization of networks that we will touch on is the simplicial complex. A simplicial complex can be thought of as analogous to a clique, or a fully-connected subgraph. A simplicial complex is made of up simplexes, with are higher-order generalizations of points (0-simplex), lines (1-simplex), triangles (2-simplex), and tetrahedra (3-simplex). A set of simplexes $\textbf{K}$ is called a simplicial complex if:
	\begin{enumerate}
		\item The face of every simplex in $\textbf{K}$ is itself a member of $\textbf{K}$ and,
		\item The intersection of any two faces $\textbf{K}_1$,$\textbf{K}_2\in \textbf{K}$ is a face of $\textbf{K}_1$ and $\textbf{K}_2$.
	\end{enumerate}
	
	For example, if we have a set of three elements $X_1$, $X_2$, and $X_3$, and the triangle $(X_1,X_2,X_3)$ is a member of our simplical complex \textbf{K}, then so too must every edge and point also belong to $\textbf{X}$ (i.e. $(X_1,X_2)$, $(X_1,X_3)$, etc). This recursive structure makes simplcial complexes tractable with the mathematical machinery of algebraic topology and there is already a very well-developed toolkit for analyzing these structures \cite{battiston_networks_2020}. Despite this, to the best of our knowledge, at the time of this writing, there have not been any \textit{direct} applications of simplicial complexes to representing higher-order redundancies and synergies, although a closely related approach comes from Santoro et al., \cite{santoro_higher-order_2023}, who apply simplicial complexes to the multivariate structure of so-called ``edge time series" in human brain data \cite{faskowitz_edge-centric_2020}. Given the nature of the relationship between local correlation and local redundancy \cite{varley_partial_2023}, we conjecture that Santoro et al's method is likely sensitive to higher order redundancies moreso than higher-order synergies, although this remains an area of active research.
	
	Higher-order network representations of complex systems is still in the early stages of development as a field. Many fundamental questions remain, and while this may mean that there aren't as many ``off-the-shelf" tools that can be applied (as there are in network science), the opportunities for foundational insights is still very much open. 
	
	\newpage
	
	\section{Integration, Segregation, \& Complexity}
	\label{sec:complexity}
	
	A core feature of complex systems is that they incorporate elements of ``integration" and ``segregation" \cite{tononi_measure_1994}. Integration refers to a dynamic where all of the elements of the system are interacting and affecting each-other, while segregation refers to a dynamic where subsets of elements are involved in their own processes that are not shared with other elements. As an example, consider the brain: it is known that particular brain areas are involved in some processes and not others (functionality is segregated between different regions), however, at the same time, the brain is integrated enough that all the different local processes can be combined into an integrated, unified whole: a unitary organism with an (apparently) single consciousness. It has been hypothesized that this balance of integration and segregation is key for healthy brain function \cite{tononi_information_2004,sporns_networks_2010}. Similarly, in economics, successful firms maintain a healthy balance of segregation (each branch working on its own mission), while all the work is overseen and broadly directed by a centralized executive office. In global politics, the internal dynamics of individual nations are segregated from each-other by national borders, languages, and cultures, while between-nation integration exists in the form of treaties, trade, and historical entanglements. 
	
	This mixture of integration and segregation is an inherently multi-scale phenomenon, with different scales often displaying different biases one way or another. Consider a modular network: within each module, there is high integration (which could be assessed using total correlation \cite{puxeddu_leveraging_2024}), but each module may only be sparely connected to other modules, indicating system-wide higher-scale segregation. Depending on how the network is wired together (for instance, having a small-world \cite{watts_collective_1998} or scale-free \cite{barabasi_emergence_1999} topology), the structure of the overall dynamics being run on the network will change. As mentioned in Sec. \ref{sec:complex_systems}, multi-scale or scale-free structure is one of the hallmarks of complex systems, and is deeply related to the issue of balancing integration and segregation: a system that is fully integrated at the micro-scale cannot support a ``higher" scale (a fully connected network has no higher-order communities), nor can a fully segregated system (a set of non-interacting elements barely qualifies as a system). It is a balance of the two that allows non-trivial, multi-scale structure to emerge. 
	
	This has been the foundation of several proposed scalar measures of ``how complex" a given system is \cite{ay_unifying_2006,tononi_measure_1994}. The notion of ``complexity" as reflecting a balance of integration and segregation recalls the notion discussed in Sec. \ref{sec:complex_systems} that complexity balances predictable order and randomness (an idea that can be formalized in the context of critical phase transitions \cite{timme_criticality_2016,popiel_emergence_2020}).
	
	\subsection{TSE-Complexity}
	
	One canonical, information-theoretic attempt to formalize ``complexity" as a balance between integration and segregation is given by the Tononi-Sporns-Edelan (TSE) Complexity \cite{tononi_measure_1994}, which iterates over all bipartitions of a given system to quantify how integration and segregation are distributed.
	
	\begin{equation}
	\label{eq:tse_1}
	TSE(\textbf{X}) = \sum_{k=1}^{ \lfloor N/2 \rfloor }\langle I(\textbf{X}_j^k ; 
	\textbf{X}_j^{-k}) \rangle
	\end{equation}
	
	Where $\textbf{X}_j^k$ indicates the $j^{th}$ set of $k$ elements in \textbf{X} and $\textbf{X}_j^{-k}$ indicates the complement of $\textbf{X}_j^k$ (all elements in \textbf{X} not in $\textbf{X}_j^k$). Eq. \ref{eq:tse_1} shows us that the TSE complexity of \textbf{X} can be thought of as the sum of the average mutual informations across all bipartitions of size $k$, and consequently, the measure is high when there is non-trivial information structure observable at every scale. Every element, or combination of elements, on average, discloses some information about the rest of the system, whether we're talking about single units, or large collections.
	
	The TSE complexity can also be formalized in terms of the total correlation of the whole system \textbf{X} and the expected total correlation of all subsets of \textbf{X}:
	
	\begin{equation}
	TSE(\textbf{X}) = \sum_{k=2}^{N} \bigg[ \biggl(\frac{k}{N}\biggr) TC(\textbf{X}) - \langle TC(\textbf{X}^{k}_{j}) \rangle_{j}\bigg] 
	\label{eq:tse_2}
	\end{equation}
	
	This makes it clear that TSE complexity is high when the average integration of all subsets of size $k$ is \textit{less than} what we would expect if integration increased linearly with subset size, \textit{and} the overall integration of the whole system is high \cite{tononi_measure_1994,rosas_quantifying_2019}. This describes a dynamic where sets of small elements are generally independent from each-other, but nevertheless, the entire system behaves as an integrated whole. The TSE complexity can be used as an objective function when attempting to build systems that strike an optimal balance between integration and segregation \cite{sporns_evolving_2006}. 
	
	Since the TSE-complexity requires brute-forcing all possible bipartitions of the set of elements, it is impossible to fully compute for large systems. Random sampling of subsets can be used to heuristically estimate the complexity \cite{marshall_analysis_2016,varley_multivariate_2022}, although for large systems, even this approach can be computationally intensive. An alternative approach is to consider only the top ``layer" of the system: the difference between the total correlation of the whole, and the expected decrease in total correlation when each element is removed \cite{tononi_complexity_1998,sporns_theoretical_2002}:
	
	\begin{equation}
	C(\textbf{X}) = TC(\textbf{X})-\frac{TC(\textbf{X})}{N} - \mathbb{E}[TC(\textbf{X}^{-i})] \label{eq:cd}
	\end{equation}
	
	While a fairly crude approximation for systems with many scales, the \textit{description complexity} is a tractable heuristic that can be applied to large data sets. The conceit of removing each element once gives it some conceptual similarities to the dual total correlation (Eq. \ref{eq:dtc}), and in fact the relationship is direct:
	
	\begin{equation}
	DTC(\textbf{X}) = N\times C(\textbf{X})
	\end{equation}
	
	This is consistent with the intuition discussed in Sec. \ref{sec:dtc} that the dual total correlation, unlike the total correlation is highest in a critical zone between highly redundant structure, and total independence. Interestingly, there appears to be a relationship between the TSE complexity and the synergy: the generalized information decomposition, as applied to the total correlation, reveals that redundant information generally negatively contributes to the value of the TSE, while synergistic information positively contributes \cite{varley_generalized_2024}, although the exact nature of this relationship requires further study. 
	
	\subsection{O-Information \& S-Information}
	
	The TSE-complexity can be thought of as a measure of how the dependencies that characterize a system are distributed over the elements at very scale. However, it does not give any direct insight into the \textit{nature} of those dependencies. As an alternative measure, Rosas et al., proposed a measure they called the O-information: 
	
	\begin{equation}
	\Omega(\textbf{X}) = TC(\textbf{X}) - DTC(\textbf{X})
	\label{eq:o-info}
	\end{equation}
	
	The O-information (which was originally named the ``enigmatic information" by James and Crutchfield \cite{james_anatomy_2011}) has the useful property that if \textbf{X} is dominated by synergistic interactions, then $\Omega({\textbf{X}}) < 0$ bit and conversely, if \textbf{X} is dominated by redundant interactions, $\Omega({\textbf{X}}) > 0$ bit. It is 0 bit if \textbf{X} is completely disintegrated (i.e. all $X_i$ are independent), and it is \textit{only} sensitive to beyond-pairwise dependencies (i.e. the bivariate mutual information doesn't contribute towards $+\Omega$ or $-\Omega$). A negative value of $\Omega(\textbf{X})$ only indicates that \textbf{X} is \textit{dominated} by synergistic dependencies - redundant dependencies can also exist, which makes it more of a heuristic than the \textit{complete} decompositions provided by the PID and PED. The O-information has generated considerable interest in diverse fields, including neuroscience \cite{gatica_high-order_2021,varley_multivariate_2022}, psychology \cite{marinazzo_information-theoretic_2024}, and music theory \cite{scagliarini_quantifying_2022}. Unlike the PID, which grows hyperexponentially with the number of elements in the system, the O-information scales much more gracefully and can be computed for several hundred interacting elements (assuming sufficient data). The measure has also been extended in a number of directions, including a modified transfer entropy \cite{stramaglia_quantifying_2021}, a local perspective to capture time-resolved redundancy/synergy dynamics \cite{scagliarini_quantifying_2022,pope_time-varying_2024}, and a time-lagged, frequency-specific version \cite{faes_new_2022}.
	
	Varley et al., \cite{varley_multivariate_2022} showed that the O-information can be re-written in terms of sums and differences of just total correlations:
	
	\begin{equation}
	\Omega(\textbf{X}) = (2-N)TC(\textbf{X}) + \sum_{i=1}^NTC(\textbf{X}^{-i})
	\end{equation}
	
	This can help provide some intuition into what synergy means. The left-side term $(2-N)TC(\textbf{X})$ is the total integration of the ``whole", duplicated $N-2$ times (and made negative). The right-hand term ``adds back" the total correlation associated with the $N$ possible ways that single elements can be removed. If $\sum_{i=1}^NTC(\textbf{X}^{-i})>(2-N)TC(\textbf{X})$, then there must be correlations in the $\textbf{X}^{-i}$ that are redundant, and so get double counted. Conversely, if $(2-N)TC(\textbf{X})$ is greater, then there is information in the structure of the ``whole" that can't be found in the parts when all added together. 
	
	The O-information can be analyzed locally, in the same way that the bivariate mutual information can be \cite{scagliarini_quantifying_2022, pope_time-varying_2024}. For a particular configuration \textbf{x} of \textbf{X}:
	
	\begin{equation}
	\omega(\textbf{x}) = tc(\textbf{x})-dtc(\textbf{x})
	\end{equation}
	
	Where $tc$ and $dtc$ are the local total correlation and dual total correlation (which can be computed from local entropies). Generally, the sign of $\omega$ is interpreted in the same way as the in the expected case: a negative sign indicates greater synergy while a positive sign indicates greater redundancy. The local total correlation and local dual total correlation can both be negative, however (unlike their strictly non-negative expected forms). This can create more complex situations: suppose that $tc(\textbf{x})=-3.0$ bit and $dtc(\textbf{x})=-8.0$ bit. In this case, $\omega(\textbf{x})=+5.0$ bit and we must ask: if $dtc$ is \textit{more misinformative} than the $tc$, is that the same thing as the overall system being ``redundancy dominated"? The exploration of the local O-information at this point is very limited (although see \cite{pope_time-varying_2024} for recent work) and so there remain fundamental questions waiting to be answered. 
	
	A closely related measure to the O-information is the \textit{S-information} (again, coined by Rosas et al., \cite{rosas_quantifying_2019}). Formally:
	
	\begin{equation}
	\Sigma(\textbf{X})=TC(\textbf{X})+DTC(\textbf{X})
	\end{equation}
	
	Originally studied by James and Crutchfield \cite{james_anatomy_2011}, it was initially referred to by the somewhat tongue-in-cheek term ``very mutual information", which reflects the fact that it quantifies all of the dependencies between each neuron and the rest of the system:
	
	\begin{equation}
	\Sigma(\textbf{X})=\sum_{i=1}^{N}I(X_i;\textbf{X}_{-i})
	\end{equation}
	
	The S-information is relatively less explored than the O-information, but analysis of discrete and Gaussian random variables has found that it is extremely tightly correlated ($r\approx1$, \cite{rosas_quantifying_2019,varley_multivariate_2022}). In fact, $\Sigma(\textbf{X})$ is a much more effective proxy measure for $TSE(\textbf{X})$ than the initial description complexity/dual total correlation proposed in \cite{tononi_complexity_1998}. Like the O-information, the S-information can be localized:
	
	\begin{align}
	\sigma(\textbf{x}) &= \sum_{i=1}^{N}i(x_i;\textbf{x}_{-i}) \\
	&= tc(\textbf{x}) + dtc(\textbf{x}) \nonumber
	\end{align}
	
	As before, the local total correlation and local dual total correlation can be negative, although since the relationship is addative, it is conceptually less difficult: informative dependencies contribute to $\sigma(\textbf{x})$ while misinformative dependencies reduce it. The S-information (both local and average) has potential as a kind of "total statistical density" measure: quantifying how tightly individual nodes are collectively ``embedded" in the system under study. 
	
	\subsection{Whole-Minus-Sum Complexity \& Integrated Information}
	
	The above measures are all based on sums and differences of multivariate generalizations of mutual information (total correlation, dual total correlation), which means that they are not dynamic: they operate on static probabilities distributions. A temporal approach (conceptually very similar to some of the information dynamics explored in Sec. \ref{sec:info_dynamics}) is with the measure of \textit{integrated information} proposed by Balduzzi and Tononi \cite{balduzzi_integrated_2008}. Consider a dynamic system comprised of two elements: \textbf{X}$=\{X_1,X_2\}$. We have previously defined the excess entropy as the mutual information that \textbf{X}'s past discloses about it's own future (Sec. \ref{sec:excess}), and shown how the excess entropy can be decomposed with multitarget information decomposition (Sec. \ref{sec:phi_id}). A simpler heuristic measure is the difference between the excess entropy and the sum of the two ``marginal" temporal mutual informations:
	
	\begin{equation}
	\Phi^{WMS}(\textbf{X}) = E(\textbf{X}) - \sum_{i=1}^2E(X^i)
	\end{equation}

	Typically, rather than estimating the whole excess entropy (which is computationally non-trivial), it is typical to just use single steps:

	\begin{align}
		I(\textbf{X}_t;\textbf{X}_{t+1}) - \sum_{i=1}^2 I(X^i_{t};X^i_{t+1})
	\end{align}
	
	If $\Phi(\textbf{X})>0$, then there is ``integrated" information about the future of the whole system that is only accessible when the past of the whole system is known. In contrast, if $\Phi(\textbf{X})<0$, then the system is dominated by redundant information duplicated over both elements that swamps any synergistic signal \cite{mediano_beyond_2019}. Like the O-information, the sign in this case is a measure of \textit{dominance}, rather than absolute quantity (for a more detailed discussion of the relationship between O-information and whole-minus-sum integration, see \cite{rosas_characterising_2024}).
	
	Using the $\Phi$ID framework, Mediano et al., were able to propose a modification to $\Phi(\textbf{X})$, named $\Phi^R$ (pronounced ``fire"):
	
	\begin{equation}
	\Phi^R(\textbf{X}) = \Phi^{WMS}(\textbf{X}) - I_{\partial}(\{1\}\{2\}\to\{1\}\{2\})
	\end{equation}
	
	where $I_{\partial}(\{1\}\{2\}\to\{1\}\{2\})$ is the double redundancy described by the integrated information decomposition (Sec: \ref{sec:phi_id}). $\Phi^R$ is known to be strictly non-negative (unlike the classic $\Phi$), and satisfies the initial desire for a measure that describes how much information as in the joint state but not the marginals (for an in-depth discussion, see \cite{mediano_beyond_2019,rosas_reconciling_2020}). 
	
	$\Phi$ was originally posed in the context of \textit{integrated information theory} \cite{tononi_consciousness_2008}, which make very strong claims linking information quantities and phenomonological consciousness. We will remain agnostic on the c-word here, however, a weaker form of integrated information theory has recently emerged as an object of interest in complex systems more generally \cite{mediano_strength_2022,mediano_integrated_2022}. Mediano et al., found that $\Phi^R$ generally tracked intuitive ideas of relative ``complexity" in complex systems (elementary cellular automata, flocking models, etc) \cite{mediano_integrated_2022} and proposed $\Phi^R$ as a principled, interpretable measure of dynamical complexity. While $\Phi^R$ is only well-defined for the simplest case of a two-element dynamical system, Mediano et al., leverage an old idea first proposed by Tononi and Sporns: the minimum information bipartiton (the MIB) \cite{tononi_measuring_2003}. For a system of many interacting variables, the MIB is the cut that minimizes the information lost by disintegrated the system. For a system of $N$ elements then, which may be too large to do a full $\Phi$ID analysis of, if one can find the MIB, then it is possible to do a bivairate analysis of both partitions (using either $\Phi$ID or $\Phi^R$) to get a heuristic sense to which the system has synergistic dynamics. Finding the minimum information bipartion is easier said than done, however: it is often an intractable problem, and several different heuristics have been proposed to solve it (see Toker and Sommar for review \cite{toker_information_2019}). 
	
	Despite these limitations however, I am of the opinion that ``weak integrated information theory" holds tremendous promise that has only begun to be realized. It provides an appealing link between the worlds of information theory and computation and dynamical systems approaches on the other. It also is built on strong axiomatic foundations (Shannon information theory, and extensions), which make measures interpretable in ways that other complexity measures may not be. And finally, being purely based on information theory, it is extremely general and can be applied to a variety of domains. 
	
	\subsection{Do We Even Need Complexity Measures?}
	
	The question of a scalar measure of complexity is a naturally interesting one, and, as mentioned, considerable work has gone into attempting to derive the most ``natural" one. It is worth taking a step back and, borrowing from the title of \cite{feldman_measures_1998} ask: "measures of statistical complexity: why?" As a mathematical question, the project of finding a scalar measure that captures the nuances of what it means to be complex is an interesting one, but ultimately it is often unclear what has been learned by announcing that ``this system has \textit{x} units of complexity". While there are some cases where relative changes in complexity can be informative (for example, see \cite{schartner_global_2017,schartner_increased_2017,varley_differential_2020,varley_topological_2021,luppi_reduced_2023,luppi_synergistic_2024} for examples), in general it is more useful to begin by defining what particular process or dynamic one is trying to quantify and going from there. Is it synergy/redundancy dominance? Memory? Integration? Predictability? As complex systems science continues to mature, the ability to answer \textit{specific} questions about \textit{specific} aspects of system dynamics will become increasingly important, while vaguer, more abstract notions about ``complexity" may have to fall by the wayside. 
	
	\newpage 
	
	\section{Information Estimators}
	\label{sec:estimators}
	
	As a branch of pure mathematics, information theory is largely unconcerned with the practical, day-to-day realities of actually \textit{using} information theoretic measures. Pure mathematicians are generally quite happy to live in the land of beautiful abstraction, where measures like entropy, mutual information, and synergy can be formally defined and studied as objects in their own right. For scientists interested in information-theoretic analyses of real world systems, however, there is an immediate issue that presents itself: given a set of data, how does one go about actually estimating the required probabilities, entropies, and related quantities? This is an issue that many aspiring scientists rarely encounter: in most cases, the problems of estimating probabilities are handled ``under the hood" of various closed-form statistical tests. When attempting information-theoretic analysis, however, the questions of accurately estimating probabilities is much close to the surface, and care must be taken to ensure that the estimators being used are valid and their various limitations taken into account. 
	
	In this section, we will discuss how to estimate entropies and mutual informations for discrete and continuous data. The discrete case is reasonably straightforward (although, as discussed in Sec. \ref{sec:fc_nets}, bias associated with finite size effects is a concern). The situation becomes more complex for continuous data. We will discuss parametric (Gaussian) estimators for continuous data, as well as briefly touching on non-parametric estimators using K-nearest neighbor graphs. Each one of these has its own strengths and weaknesses, and critical considerations should be made at the outset of an analysis about what approach is best. This section will not cover all possible families of information estimators: for example, there is a small, but growing field concerned with the problem of estimating information-theoretic dependencies between point processes in continuous time \cite{spinney_transfer_2017,shorten_estimating_2021,mijatovic_information-theoretic_2021,mijatovic_measuring_2022}, with application to processes such as spiking neurons and heart-rate. Similarly, the problem of estimating mutual information between mixtures of discrete and continuous variables is an area of research under active development \cite{ross_mutual_2014,gao_estimating_2017}. Finally, Bayesian estimators have been developed with many appealing properties \cite{wolpert_estimating_1995}. To those interested in these more niche applications, we refer to the aforementioned literature. 
	
	\subsection{Discrete Signals}
	\label{sec:discrete_signals}
	
	Historically, the standard practice to infer the relevant probabilities has been to estimate them from the counts of events, so that $\hat{P}(x_i) = n_i / N$, where $n_i$ is the number of occurrences of the $i^{th}$ event and $N$ is the total number of samples. These estimated probabilities can be plugged into Eq. \ref{eq:entropy} to give the maximum-likelihood estimate of the entropy given the data:
	
	\begin{equation}
	\hat{H}(X) = -\sum_{x\in\mathcal{X}}\hat{P}(x)\log(\hat{P}(x))
	\end{equation}
	
	This is variously known as the \textit{naive entropy estimator}, the \textit{plugin estimator}, as well as the \textit{maximum likelihood estimator}, and is probably the most commonly used method for estimating entropies from data. Despite its frequency of use, however, the plugin estimator has fundamental limitations: the most significant of which is that it \textit{systematically under-estimates the true entropy of the process.} This is known as \textit{bias} and is given by:
	
	\begin{equation}
	\mathbb{E}[\hat{H}(X)] \le H(X)
	\end{equation} 
	
	The severity of the bias is a function of both the number of states of the system and the number of samples used to construct the probability distributions \cite{miller_note_1955}, and a correction term can be introduced for a bias-corrected estimator:
	
	\begin{equation}
	\hat{H}_{c}(X) = \hat{H}(X) + \frac{M-1}{2S}
	\end{equation}
	
	where $M$ is the size of the state-space (eg. $|\mathcal{X}|$) and $S$ is the number of samples. In cases where the number of samples is much larger than the number of states, the bias term becomes negligible, however as the number of states grows (for example when considering the joint entropy of multiple variables), or when the dataset is small and the number of samples limited, bias can quickly overwhelm the plugin estimate, resulting in incorrect estimates unless bias correction is included. 
	
	The mutual information suffers from a similar bias, although unlike the entropy, the plugin estimator of mutual information will generally \textit{over-estimate} the true information quantity. Since the mutual information is often calculated as a difference in multiple entropy measures (eg. Eqs. \ref{eq:mi_1}), if the bias in MLE estimates of the entropy is greater than the difference, significant errors can be introduced. A number of bias-corrected estimators of mutual information exist (see: \cite{kraskov_estimating_2004}), although the majority of work has been done on estimating the mutual information from continuous signals, although the risk of bias is also of concern even when looking at naturally discrete data. Finally, the bias in entropy can also impact discrete estimators of higher-order measures such as the O-information, biasing it towards synergy and these measures can require correction when the number of states is much larger than the number of samples \cite{gehlen_bias_2024}.
	
	\subsection{Continuous Signals}
	
	So far, everything that we have discussed assumes that we are analyzing data that is naturally discrete: variables can take on one of a finite number of mutually exclusive states (faces of a die, coin flips, etc) that they switch between according to some probability distribution over all states. While this can be very useful when examining naturally digital processes like spiking neurons or a system like a cell that has discrete states (apoptotic, mitotic, etc), the vast majority of empirical data is continuous, from electro-physiological records to climate data, and often does not naturally fall into a set number of obvious ``states" that can be assigned to a discrete probability distribution. 
	
	This may seem like a significant blow to information theory's utility when analyzing the natural world, however considerable work has been done addressing this issue, and several possible solutions are available. They generally fall into two categories: coarse-graining approaches (binning, point-processing, ordinal partition embedding), and density-based approaches (such as the KSG-estimator for mutual information). In general, density-based estimators are considered to be the gold-standard and if possible, should be the first line of attack on a set continuous data. The coarse-graining estimators are presented mostly due to the frequency that these have been used in the past so that the aspiring scientist understands what has been done, as well as providing an opportunity to discuss the limitations inherent to these kinds of analysis. 
	
	\subsubsection{Coarse Graining}
	
	Coarse graining, or discretizing, is the process by which every observation of a continuous random variable is mapped to one of a finite number of discrete states. It is generally the easiest approach: once a signal is coarse grained, the naive entropy estimators can be used and difficult questions associated with the differential entropy can be avoided. Because of this ease-of-use, coarse graining is far and away the most commonly seen approach to information-theoretic analysis of continuous signals. Unfortunately, this strategy has some fundamental limitations that, when not critically considered at the outset, can seriously compromise the integrity of an analysis. 
	
	\begin{flushleft}
		\small{\textit{Histogram Binning}}
	\end{flushleft}	
	
	The simplest method of coarse graining continuous data is to bin it, constructing probabilities from the histogram of binned values and feeding those values into the naive entropy estimator described above. The apparent simplicity of this method is misleading however, as a number of problems become apparent after looking more closely. The most significant is the trade-off between bias and information loss. We have already seen that the plug-in estimator for Shannon entropy systematically under-estimates the true entropy and, importantly, that the bias gets worse as the number of bins increases. Consequently, we are left attempting a tricky optimization problem: enough bins that we capture the important information in the signal, but not so many that bias renders our numerical estimate worthless. While histogram-based binning approaches have been widely used, at this point they are largely obsolete and should \textbf{not} be used unless no other option is available. More advanced approaches can provide the same insights without the inherent biases. In cases where histogram binning must be used, considerable care should be taken to minimize the risk of bias and ensure that the loss of information from the signal is not too great.
	
	\begin{flushleft}
		\small{\textit{Point Processing}}
	\end{flushleft}
	
	An alternative to binning the data that applies naturally to time series data is creating a ``point-process", which maps the continuous signal to a binary time-series by marking ``events"  at the most extreme points of the signal, usually above some threshold. Point processing assumes that only a subset of moments (typically unusually high-amplitude samples) are significant or informative about the underlying dynamics, and that the majority of the signal is essentially noise. Assuming the assumptions are valid, point-processing is a much more natural method of compressing a complex, continuous signal, as it only records significant events and does not conflate signal and noise in the same way that histogram binning does. When analyzing point-processed data, one can use a standard discrete entropy estimator, or more recent and sophisticated measures of continuous-time entropy that operate on inter-even intervals rather than treating each frame as a draw from a fixed distribution \cite{shorten_estimating_2021,mijatovic_information-theoretic_2021,mijatovic_measuring_2022}.
	
	In fMRI data, point-processing has been found to preserve a surprising amount of information, allowing for the reconstruction of higher-order structures such as canonical functional connectivity networks \cite{tagliazucchi_criticality_2012,sporns_dynamic_2021}. When considering a point-process analysis, care must be taken to set the threshold correctly: too low and noisy, low-amplitude fluctuations will confuse the data, while if it is too high, so few moments will be retained that no information will be preserved. In the absence of an \textit{a priori} reason to select a particular threshold, it would be best to run the analysis using several thresholds, to ensure that the results are reasonably robust to small changes in threshold. An interesting method for choosing a threshold was used in \cite{meisel_fading_2013,shriki_neuronal_2013,varley_differential_2020}: the authors compared the z-scored empirical distribution of instantaneous amplitudes to the best-fit Gaussian, and set the threshold as the standard deviation at which the empirical data and the best-fit Guassian began to diverge. This ensures that events greater than the threshold are unlikely to have been generated by noise. This assumes that the events of interest are being drawn from some heavier-than Gaussian distribution (e.g. a power-law or lognormal) contaminated by Gaussian noise. Many processes studied in complex systems science are thought to follow heavy-tailed distributions, so this is often not a bad starting place. 
	
	As with the histogram-based binning, the point-process throws out a considerable amount of data: a binary process is inherently less able to encode information than a process with more states. If the assumption that all the important information is in these rare ``events" is wrong, huge amounts of signal are irretrievably lost. Furthermore, like histogram binning it assumes that every moment in time is something like an independent draw from an underlying distribution: temporally extended fluctuations are ignored, as only the local maxima are retained. While more principled than histogram-based binning, point processes should be treated with similar caution: in fields like neuroscience where point processes are well-explored, it may be a useful technique, but care should be taken to ensure that all assumptions are satisfied.
	
	\begin{flushleft}
		\small{\textit{Ordinal Partition Embedding}}
	\end{flushleft}
	
	for time series data, the histogram-based binning transformation assumes that every observation of a time-series is a random draw from some underlying distribution, ignoring the possibility that temporally-extended patterns are meaningful as well. An alternative coarse-graining approach is the \textit{permutation embedding} or \textit{ordinal partition embedding}, which transforms continuous data into a sequence based on temporally-extended amplitude patterns \cite{bandt_permutation_2002}.
	
	Constructing ordinal partitions requires embedding the time-series $X_t$ in a $d$-dimensional space (with an optional lag of $\tau$), as is commonly done with Taken's embeddings \cite{rand_detecting_1981}. Each embedded vector $v_t = [x_t,x_{\tau+t},x_{2\tau+t}...x_{(d-1)\tau+t}]$ is part of a finite-sized set of possible vectors of size $d!$. Once the associated discrete timeseries has been constructed, it is possible to use all the normal information-theoretic measures, such as entropy \cite{bandt_permutation_2002}, transfer entropy \cite{ruan_ordinal_2019}, and a variety of other measures of dynamical complexity \cite{varley_topological_2021}.
	
	As with the point-processing, the ordinal partition embedding as parameters that need to be set: the embedding dimension $d$ and the temporal lag $\tau$. For a discussion of the practical considerations in selecting these parameters, see: \cite{riedl_practical_2013,mccullough_measuring_2015,mccullough_time_2015}. If $d$ is too large, then the size of the state-space balloons and under-sampling is a serious concern, while if it is too small, meaningful temporal dependencies can get lost. Similarly, $\tau$ must be optimized to align with the natural temporal dynamics of the system. 
	
	\subsubsection{Differential Entropy}
	
	In the original proposal by Shannon \cite{shannon_mathematical_1948}, a generalization of the discrete entropy to continuous random variables was presented in the form of the \textit{differential entropy}:
	
	\begin{equation}
	H^{dx}(X) = -\int_x P(x)\log (P(x))dx 
	\end{equation}
	
	This satisfies the definition that $H^{dx}(X)=\mathbb{E}[-\log(P(x))]$, however, it fails to preserve several other properties that make the discrete entropy a useful measure \cite{cover_elements_2012}. For instance, it can be negative (as in the case of very tight Gaussian distributions), and it is not invariant under under rescaling due to the additive nature of the logarithm:
	
	\begin{equation}
	H^{dx}(\alpha X) = H^{dx}(X)+\log(|\alpha|)
	\end{equation}
	
	While the differential entropy is not entirely well-behaved, the differential Kullback-Leibler divergence maintains the desired properties of the discrete divergence:
	
	\begin{equation}
	D_{KL}^{dx}(P||Q) = \int_xP(x)\log\bigg(\frac{P(x)}{Q(x)}\bigg)dx
	\end{equation}
	
	Unlike the differential entropy, the differential relative entropy is strictly non-negative (which implies that the differential mutual information is non-negative as well), and preserved under scaling. 
	
	If one has a strong reason to believe that continuous data is drawn from a closed-form, parametric distribution, then all of these differential entropies and derivatives can be calculated analytically, and exact estimators exist for a number of distributions. By far, the most common, and well-developed estimators are for univariate and multivariate Gaussian distributions, which are described below. 
	
	\subsubsection{Gaussian Estimators}
	
	One of the biggest benefits of information theoretic analysis of is that it is sensitive to nonlinear relationships between interacting variables. In complex systems, which are dominated by nonlinear interactions, this can be significant. However, if a researcher knows that their data is already multivariate-normally distributed (or only cares about linear interactions), analytic relationships between information theoretic and classical parametric measures exist. For the marginal and joint entropies:
	
	\begin{equation}
	H^{dx}(X) = \ln(\sigma\sqrt{2\pi e})
	\end{equation}
	
	\begin{equation}
	H^{dx}(X_1, X_2 ... X_N) = \frac{\ln[(2 e)^N|\Sigma|]}{2}
	\end{equation}
	
	Where $\sigma$ is the variance and $|\Sigma|$ is the determinate of the $N$-dimensional covariance matrix. For a continuous, univariate distribution on the interval $[-\infty,\infty]$ and with fixed mean $\mu$ and variance $\sigma$, the Gaussian distribution is the maximum entropy distribution. For a multivariate distribution with infinite support, fixed means, and fixed, pairwise covariances, the multivariate Gaussian is the maximum entropy distribution possible \cite{cover_elements_2012}. 
	
	For largely historical reasons, Guassian estimates of information-theoretic quantities are given in nats, rather bits. For bivariate mutual information, the following equality holds:
	
	\begin{align}
	I^{dx}(X;Y) &= \frac{-\ln(1-r^2)}{2}
	\end{align}
	
	Where $r$ is the Pearson correlation coefficient between $X$ and $Y$. If \textbf{X} and \textbf{Y} are multivariate, the generalized Gaussian mutual information is of the form:
	
	\begin{align}
	I^{dx}(\textbf{X};\textbf{Y}) &= \frac{1}{2}\ln\bigg(\frac{|\Sigma_X||\Sigma_Y|}{|\Sigma_{XY}|}\bigg) \\
	&= \frac{1}{2}\times\big(\ln(|\Sigma_X|) + \ln(|\Sigma_Y|) - \ln(\Sigma_{XY})\big)
	\end{align}
	
	Which has obvious symmetry with the classing formulation of discrete mutual information. It is also possible to construct Gaussian estimators of local information values. Recall that, for a univariate Guassian:
	
	\begin{equation}
	P(x) = \frac{1}{\sigma \sqrt{2\sigma^2} } e^{-\frac{1}{2}\left(\frac{x-\mu}{\sigma^2}\right)^2}
	\end{equation}
	
	And for a multivariate Gaussian:
	
	\begin{equation}
	P(x_1, x_2...x_n) = \frac{1}{(2\pi)^\frac{n}{2}|\Sigma|^\frac{1}{2}}e^\frac{-(\boldsymbol{x-\mu})^T\Sigma^{-1}(\boldsymbol{x-\mu})}{2}
	\end{equation}
	
	Where $\boldsymbol{x}$ is the vector of $x_1, x_2...x_n$ and $\boldsymbol{\mu}$ is the mean vector. We can then substitute these values in any standard information theoretic function (e.g. Eqs. \ref{eq:mi_main}) and extract expected or local values. This substitution trick will work for any information theoretic value, allowing more complex estimators, such as the Gaussian transfer entropy, Guassian total correlation, or Gaussian predictive information. 
	
	\subsubsection{Density-Based Estimators Estimators}
	\label{sec:knn}
	
	While the parametric Gaussian estimator can be used to efficiently compute information-theoretic measures for continuous signals, it comes with the significant limitation of being blind to nonlinearities in the data. For time-series from complex systems such as financial markets, neural data, or climatic measurements, this can be a severe limitation. To address this, there are non-parametric estimators of information-theoretic measures based on K-nearest neighbors (KNN) inferences of underlying probability manifolds. The field of manifold learning from sparse data is a complex and difficult one so the gory details are beyond the scope of this review. Here we will provide the formulae, and the underlying mathematical intuitions behind the derivations can be found in the cited literature. 
	
	With any KNN-based analysis, it is crucial to pre-specify the particular distance function to be used and assess its appropriateness. While in principle any valid metric could be used for the following estimators, in practice, concerns such as the curse of dimensionality often limit what measures will be useful. Here, we are assuming that the reader is planning to use the $L_{\infty}$ (Chebyshev) distance, which is robust to high dimensional data and simplifies some of the equations. Readers interested in alternatives such as the $L_1$ (Manhattan distance) or the $L_2$ (Euclidean distance) should review the cited literature. 
	
	\begin{flushleft}
		\textbf{Kozachenko-Leonenko Entropy Estimator}
	\end{flushleft}
	
	The earliest K-nearest neighbors estimator of the entropy of a single or multivariate time-series is given by the Kozachenko-Leonenko estimator \cite{kozachenko_sample_1987,delattre_kozachenkoleonenko_2017}. For a potentially multivariate time-series \textbf{X}:
	
	\begin{equation}
	\hat{H}(\textbf{X}) = -\psi(k) + \psi(N) + \frac{1}{N}\sum_{i=1}^N\log(d_i)
	\end{equation}
	
	where $\psi$ is the Digamma function, $d_i$ is twice the distance from the $i^{th}$ point to it's $k^{th}$ nearest neighbor, and $N$ is the dimensionality of \textbf{X}. The Kozachenko-Leonenko estimator can also be used to calculate the local entropy:
	
	\begin{equation}
	\hat{h}(\textbf{x}) = -\psi(k) + \psi(N) + \log(d_i)
	\end{equation} 
	
	Since this estimator requires calculating the K-nearest neighbors for every point in the series, the runtime gets large fast. A naive estimator will run in $O(N^2))$ time, although efficient KNN algorithms, such as a KD-Tree can bring the runtime down to $O(N\log(N))$ time. 
	
	\begin{flushleft}
		\textbf{Kraskov, Stogbauer, and Grassberge mutual information Estimator}
	\end{flushleft}
	
	Kraskov, Sogbauer, and Grassberge provided a generalizable nonparametric estimator of mutual information for continuous signals by building off of the Kozachenko-Leonenko KNN-based entropy estimator \cite{kraskov_estimating_2004}. 
	
	There are two different estimators, the first is
	
	\begin{equation}
	\hat{I}^1(X;Y) = \psi(K) - \langle\psi(n_x + 1) + \psi(n_y + 1) \rangle + \psi(N)	
	\end{equation}
	
	which is more appropriate for smaller samples. Here $n_x$ and $n_y$ correspond to the number of points in the marginal spaces that fall strictly between $\pm$ the maximum distance from the $i^{th}$ point to it's $K^{th}$ nearest neighbour. The second implementation is:
	
	\begin{equation}
	\hat{I}^2(X;Y) = \psi(K) - K^{-1} - \langle\psi(n_x) + \psi(n_y) \rangle + \psi(N)
	\end{equation}
	
	and is more appropriate for larger sample sizes. Here $n_x$ and $n_y$ correspond to the number of points in the marginal space that fall between $\pm$ the distance from the $i^{th}$ point to it's $K^{th}$ nearest neighbour \textit{on their respective axes}. This estimator can also readily generalize to the total correlation by incorporating more dimensions, although it does not generalize to the dual total correlation. 
	
	As with the Kozachenko-Leonenko estimator, the KSG estimator can be localized by looking only at the local neighborhood of a particular point:
	
	\begin{equation}
	\hat{i}^1(x;y) = \psi(K) - \psi(n_x + 1) + \psi(n_y + 1) + \psi(N)	
	\end{equation}
	
	\begin{equation}
	\hat{i}^2(x;y) = \psi(K) - K^{-1} - \psi(n_x) + \psi(n_y) + \psi(N)
	\end{equation}
	
	\begin{flushleft}
		\textbf{KSG Conditional mutual information}
	\end{flushleft}
	
	As with the Kozachenko-Leonenko entropy estimator and the KSG mutual information estimator, it is not appropriate to simply substitute the various estimators in to established relationships to construct more complex measure: a new estimator of conditional mutual information must be constructed \cite{frenzel_partial_2007,wibral_transfer_2014,gomez-herrero_assessing_2015}. The two bivariate KSG mutual information estimators can be generalized as follows:
	
	\begin{equation}
	\hat{I}^1(X;Y|Z) = \psi(K) - \langle \psi(n_z + 1) - \psi(n_{xz} + 1) + \psi(n_{yz} + 1) \rangle + \psi(N)	
	\end{equation}
	
	\begin{equation}
	\hat{I}^2(X;Y|Z) = \psi(K) -\frac{2}{K} - \big\langle \psi(n_z) - \psi(n_{xz}) + n_{xz}^{-1} - \psi(n_{yz}) + n_{yz}^{-1} \big\rangle 
	\end{equation}
	
	The local measures can be calculated as described previously.
	
	From the bivariate and conditional mutual informations, non-parametric estimators of almost all the information dynamics measures described above can be constructed, including active information storage, transfer entropy, multivariate transfer entropy, and predictive information. A recently introduced KNN-based estimator of the $i^{sx}_\cap$ redundancy function opens the door to non-parametric PID \cite{ehrlich_partial_2024}, and the JIDT package includes KNN-based estimators of the total correlation. 
	
	\section{Packages for Information Theory Analysis of Data}
	A large number of packages exist for information theory analysis, far more than we can exhaustively catalogue and review. Many of these are not maintained and guarantees of quality are impossible. Here, we provide references and brief descriptions of four, well-established toolboxes written and maintained by experts in the field. We have tried to ensure that the variety of programming languages used in modern scientific research are accounted for, including Java, Python, and MATLAB. 
	
	\subsection{Java Information Dynamics Toollkit (JIDT)}
	
	The Java Information Dynamics Toolkit (JIDT) is one of the most versatile of the packages listed here, and also has a shallower learning curve than either IDTxl or DIT \cite{lizier_jidt_2014}. Written in Java, JIDT is a compiled program, and can be called from a variety of programming environments, including MATLAB, Octave, Python, and Java. Like IDTxl and the Neuroscience Information Toolbox, JIDT is focused primarily on analyzing time-series data, and users can read in timeseries for quick analysis of a variety of information dynamics phenomena, including transfer entropy (local and global), active information storage (local and global), mutual information, and more. 
	
	Unlike all the other listed packages, JIDT also has a point-and-click GUI, allowing users to manually load in datasets, select parameters, and run the analysis without needing to write the script themselves. The package will also provide the script that replicates the given analysis, which can be copy/pasted for reuse in the future. JIDT will provide the scripts in MATLAB, Octave, Python, and Java. This makes JIDT an excellent ``starter package" aspiring complex systems scientists who may lack the more advanced Python scripting skills required to dive directly into something like IDTxl or DIT. 
	
	Available at: \url{https://github.com/jlizier/jidt}
	
	\subsection{Information Dynamics Toolkit xl (IDTxl)}
	\label{sec:idtxl}
	IDTxl is a package built on JIDT specifically aimed at using information theory to infer complex effective networks from time-series data in Python \cite{wollstadt_idtxl_2019,novelli_large-scale_2019}. The package includes several different effective network inference methodologies, including multivariate and bivariate transfer entropy, multivariate and bivariate mutual information, as well as tools to analyse active information storage and basic PID. IDTxl gives users a high degree of control over the inference pipeline, including choosing what information estimator to use (discrete naive, KSG, etc), as well as the number and type of surrogates used during the intensive null-hypothesis significance testing for significant edges. IDTxl's network inference capability integrates with Networkx, a standard Python library for network science analysis \cite{hagberg_exploring_2008}.
	
	Available at: \url{https://github.com/pwollstadt/IDTxl}
	
	\subsection{Discrete Information Theory Toolbox (DIT)}
	
	The Discrete Information Theory Toolbox (DIT) \cite{james_dit_2018} provides Python implementations of a large number of information theoretic functions, including both standards like entropy and mutual information, as well as a suite of more exotic measures, such as the TSE complexity. Unlike JIDT and IDTxl, DIT is not optimized for time-series analysis: the basic object DIT operates on is a multivariate probability distribution, from which the various measures are calculated. 
	
	Of the all the packages mentioned here, DIT is the most well-equipped to handle PID analysis: it comes with a large number of redundancy functions built-in, including $I_{min}$ \cite{williams_nonnegative_2010}, $I_{mmi}$ \cite{bertschinger_shared_2013}, $I_{ccs}$ \cite{ince_measuring_2017}, and more, as well as a redundant entropy function for PED \cite{ince_partial_2017}. 
	
	Available at: \url{https://github.com/dit/dit}
	
	\subsection{Neuroscience Information Theory Toolbox}
	
	Written in MATLAB, the Neuroscience Information Theory toolbox provides functions for those interested in analyzing neural spiking data (typically represented as a binary raster) \cite{timme_tutorial_2018}. The toolbox will compute basic Shannon entropies, mutual information, as well as bivariate transfer entropy for building transfer entropy networks. It also implements PID estimators for two sources and a single target using the \textit{$I_{min}$} redundancy function \cite{williams_nonnegative_2010}. This toolbox has been used extensively for analysis of multielectrode array recordings: see \cite{timme_high-degree_2016,timme_tutorial_2018,faber_computation_2018}. While other packages presented here are capable of more advanced analyses (multivariate transfer entropy, more complex PID, etc), the Neuroscience Information Theory Toolbox has a shallower learning curve, and the available MATLAB scripts are useful references for efficient implementation of many information theoretic measures. 
	
	Available at: \url{https://github.com/nmtimme/Neuroscience-Information-Theory-Toolbox}
	
	\newpage
	
	\section{Limitations of Information Theory}	
	
	Throughout this review, we have discussed why information theory is a natural \textit{lingua franca} for the field of complex systems, discussed (in some detail) the kinds of insights information theory can give us, and even what packages might be used to do the analysis on empirical data. We should end, however, with a discussion of when, and why, you should \textbf{not} use information theory. Claude Shannon himself bemoaned the fact that, following his introduction of information theory, it became something of a ``bandwagon" \cite{shannon_bandwagon_1956}, with researchers racing to apply the framework in ways that he saw as wholly inappropriate.
	
	We have stressed the fact that information theory is largely model-free (certainly compared to standard parametric statistical analyses commonly used in psychology and sociology), however this can be a double-edged sword: without the ability to leverage existing models, information-theoretic analyses generally require considerably more data (sometimes orders of magnitude more) to reliably infer a relationship. In small-N studies, such as those commonly done in psychology, cognitive science, or sociology, hypothesis testing with information theory will likely be more trouble than it is worth (although Gaussian estimators may help alleviate this issue, at the cost of making some constraining assumptions). In addition to big-data requirements, most information theory measures must be calculated by brute force, manually counting the occurrence and co-occurrence rates of different states, often many thousands of times in a row (for significance testing). This can extend the run-times of information theoretic inferences significantly, leaving the researcher in an unfortunate catch-22: having enough data to be confident of a result often means that the runtime will be considerable. 
	
	Furthermore, it is difficult to use information theory to build true ``causal models" of a process. Measures like active information storage, transfer entropy, or predictive information aggregate the whole set of statistical relationships into single scalar measures without providing much insight into how, or why, certain information is conserved or transmitted. The standard mantra of ``correlation does not imply causation" applies just as much to mutual information and transfer entropy as it does to the Pearson product-moment correlation. This is part of the reason that experts in transfer entropy have stressed that effective connectivity is \textit{not} the same thing as causal connectivity \cite{bossomaier_introduction_2016}. Different causal models can have identical effective connectivity structures, and information theory, at least in it's current formulation, will not provide the means to distinguish between them \cite{ay_information_2008,lizier_differentiating_2010,razak_quantifying_2014}. 
	
	Finally, it is vitally important to remember what information \textit{is} and \textit{is not}. We talk naturally about information as if it were a substance: quantifying ``how much" information available, how it ``flows" through systems and how it can be decomposed and separated into distinct ``pools," as in PID. This is a case where our choice of language may be subtlety introducing certain assumptions and biases that are not entirely appropriate. There is, in some corners of complex systems science, a tendency to talk about ``information" as if it were a kind of \textit{elan vital}, some kind of aether-like element from which certain mysterious qualities (such as consciousness) might spring. This is, in my opinion, a mistake, and a misunderstanding of what information is and is not. As we discussed in the Introduction, information theory is fundamentally about how to resolve uncertainty in an often-noisy world and inherently requires some notion of an observer who can be more or less certain about the state of the observed. It does not make much sense to ask ``what is a single neuron's perspective on integration of information", as the information being integrated only in reference to the observer who is modeling the system (this is a oft-discussed topic in the field of autopoesis, for a fascinating reference, see \cite{varela_autopoiesis_1991}). Two observers (perhaps with different priors, or different apparatuses) might observe the same system and infer different information structures. These models, while distinct from each-other, both accurately describe how observable statistics reduce the specific uncertainties of both modellers. 
	
	\section{Conclusions}
	
	In this piece, I have attempted to provide a robust (although far from comprehensive) introduction to the theory and applications of multivariate information theory to the study of complex systems. In doing so, I hope to make information theory a more widely known and commonly used tool in the toolkit of modern science. I feel that this document should not be the \textit{only} exposure to information theory that an interested reader explores: different authors present the same material in different ways, and there is value in viewing things from different perspectives. Textbooks such as Bossomaier et al., \cite{bossomaier_introduction_2016}, Cover and Thomas \cite{cover_elements_2012}, and McKay \cite{mackay_information_2003} were all consulted during the writing of this review and are themselves invaluable resources. There are also topics that this review does not even touch on, including maximum entropy models, source coding, information bottlenecks, the links between information entropy and physical entropy, and the role of information theory in machine learning. While these are all deep topics worth exploration, they are either beyond the scope of this review (such as information theory in machine learning) or require concepts from other domains that may not be relevant to complex systems scientists as a whole (such as links to thermodynamics or maximum-entropy models). 
	
	To summarize, information theory is a mathematical framework that describes the process of making inferences under conditions of uncertainty. By beginning with a measure of uncertainty (the entropy), it then defines information as the reduction in uncertainty that occurs after making some observation. From this basic idea, that information is the reduction in uncertainty, we can build a rich ecosystem of measures, from multivariate measures of information, to temporal measures of information-processing dynamics. These different measures are natural tools for understanding the structure and dynamics of complex systems, which can be rich in higher-order redundancies, synergies, and computational processes. After decades of development, the interest in information theory and the number of teams working on fundamental topics appears to be accelerating. There is undoubtedly much waiting to be revealed. 
	
	\bibliographystyle{apsrev4-1}
	
	%

	\section*{Appendix}
	
	\subsection*{Basic Probability Theory}
	
	Probability theory is a branch of mathematics that deals with the question "how likely events are to occur?" 
	
	The core structure of study in probability theory is a \textit{random variable}, which can be thought of as any kind of thing that can take on various states. In our case, we will restrict ourselves to \textit{discrete} random variables, where there are only a finite number of possible states that the variable can adopt, and it will only adopt one at a time (states are mutually exclusive). Examples of discrete random variables include coin flips (which can be Heads or Tails), die (which can roll one of six faces), or a deck of cards (from which 52 unique cards can be drawn). For a random variable $X$ (``big-ecks"), we denote specific outcomes as $X=x$ or just $x$ (``little-ecks"). 
	
	The \textit{support set} of the a random variable is the set of all possible outcomes. For a variable $X$, it's support set is typically denoted as $\mathcal{X}$. The cardinality of the support set ($|\mathcal{X}|$) gives the number of unique possible states our variable can take on. If $X$ is a fair, six-sided die, than $\mathcal{X} = \{1,2,3,4,5,6\}$ and $|\mathcal{X}| = 6$. The support set of a variable is sometimes also referred to as the ``image" of that variable. 
	
	Every element in the support set has an associated numerical value, between 0 and 1, called the \textit{probability}, denoted as $P(X=x)$. The question of ``what is probability" is a deep philosophical one, with ongoing battles between Frequentist and Bayesian philosophers of mathematics and we won't dig into the gory details here. For now, we will take a Bayesian approach and say that the probability of a specific outcome is a measure of our belief about how likely that particular outcome is. If we say that there is a 90\% chance of rain tomorrow, we are indicating that we are very confident that it will rain tomorrow. Similarly, if we say there is only a 1\% chance of snow, we are indicating that we are very confident that it will not snow. A 50\% chance of snow suggests that we are not very confident and it could go either between both binary outcomes (snow and no snow).
	
	By definition, the probabilities of every outcome in $\mathcal{X}$ must sum to 1:
	
	\begin{equation}
	\sum_{x \in \mathcal{X}} P(X = x) = 1
	\end{equation} 
	
	\subsubsection*{Properties of Probabilities}
	
	For two random variables $X$ and $Y$, we can look at the probabilities of specific pairs of events co-occurring. This is the joint probability and is given by $P(X = x, Y = y)$. As with the individual probabiliteis (called ``marginal probabilities"), the joint probabilities must all sum to 1:
	
	\begin{equation}
	\sum_{\substack{x \in \mathcal{X} \\ y \in \mathcal{Y}}} P(X = x, Y = y) = 1
	\end{equation}
	
	We can also calculate how the probabilities of specific outcomes of $X$ changes depending on the state of $Y$ using the conditional probability $P(X=x | Y=y)$. Two examples can help make the intuition behind conditional probability clear: first, imagine two fair independent coins. The outcome of the first coin flip was Heads and we're interested in whether than affects the probability that the second coin flip will be heads. Since the coins are fair and independent, the outcome of the first flip has no effect on the second, so $P(X = Heads | Y = Heads) = 1/2$. For a second example, imagine a deck of 52 shuffled playing cards. We draw two cards and are interested in whether either one of them is the Ace of Spades. For the first card, $P(Card_1=A\spadesuit) = 1/52$. Let's say that, instead of drawing the Ace of Spades on the first draw, we drew the 3 of Hearts instead. What, then, is the probability that the second card is the Ace of Spades, given that we have already removed the 3 of Hearts from the deck? Clearly: $P(Card_2 = A\spadesuit) | Card_1 = 3\heartsuit) = 1/51$.
	
	Joint and conditional probabilities are related to each-other:
	
	\begin{equation}
	P(X,Y) = P(X|Y) \times P(Y) = P(Y|X) \times P(X)
	\end{equation}  
	
	And by extension:
	
	\begin{equation}
	P(X|Y) = \frac{P(X,Y)}{P(Y)}
	\end{equation}  
	
	From these relationships we can derive Bayes Rule:
	
	\begin{equation}
	P(X|Y) = \frac{P(Y|X) \times P(X)}{P(Y)}
	\end{equation}
	
	Bayes Rule is a profound fact about probabilities that deserves (and has received) far more attention than we can give it here. 
	
	\subsubsection*{Independent Events}
	If two variable events $X$ and $Y$ are independent, then the outcome of one event has no effect on, or discloses no information about, the state of the other. Two coin flips from the same coin are generally assumed to be independent: flipping heads on the first flip has no effect on the probability of getting heads on the next flip.  
	
	This is formalized by the conditional probability:
	
	\begin{equation}
	P(X | Y) = P(X) \iff X \bot Y
	\end{equation}
	
	The probability of a particular event $X$ is unaffected by the outcome of $Y$. 
	
	Assuming independence, the joint probability of two events is equal to the product of the marginals:
	
	\begin{equation}
	P(X,Y) = P(X) \times P(Y) \iff X \bot Y
	\end{equation}	
	
	\subsection*{Expected Values}
	
	For some probability $P(X)$ and an associated value function $f : \mathcal{X} \mapsto \mathbb{R}$, then the \textit{expected value} of $X$ is given by:
	
	\begin{equation}
	\mathbb{E}[X] = \sum_{x\in\mathcal{X}} P(x) \times f(x)
	\end{equation}
	
	For example, consider a fair coin \textit{C}, with a support set $\{H, T\}$, and that the probability of both is 1/2. We will also say that if $C$ flips $H$, then you, the player, get \$5.00 and if it flips $T$, you loose \$3.00. The long-term expected value of this game can be quantified with:
	
	\begin{equation}
	\mathbb{E}[C] = \frac{1}{2}\times \$5.00 + \frac{1}{2}\times -\$3.00 = \$1.00
	\end{equation}
	
	So, on average, over many trials, you would make \$1.00 each turn. On the other hand, if the coin were weighted so that $T$ came up 80\% of the time, the outcome would be quite different:
	
	\begin{equation}
	\mathbb{E}[C] = \frac{1}{5}\times \$5.00 + \frac{4}{5}\times -\$3.00 = -\$1.40
	\end{equation}
	
	So in this case you would \textit{lose} money, on average, each term. 
	
	\subsection*{Common Logical Gates}
	\subsubsection*{Logical AND}
	\begin{center}
		
		\begin{table}[!h]
			\begin{tabular}{@{}cccccc@{}}
				\toprule
				$P(x_1, x_2, y)$   & $X_1$&$\land$ & $X_2$ & = & $Y$ \\ \midrule
				1/4 & 0    & & 0     &  & 0   \\
				1/4 & 0    & & 1     &  & 0   \\
				1/4 & 1    & & 0     &  & 0   \\
				1/4 & 1    & & 1     &  & 1   \\ \bottomrule
			\end{tabular}
		\end{table}
	\end{center}	
	
	\subsubsection*{Logical OR}
	\begin{center}
		
		\begin{table}[!h]
			\begin{tabular}{@{}cccccc@{}}
				\toprule
				$P(x_1, x_2, y)$   & $X_1$&$\lor$ & $X_2$ & = & $Y$ \\ \midrule
				1/4 & 0    & & 0     &  & 0   \\
				1/4 & 0    & & 1     &  & 1   \\
				1/4 & 1    & & 0     &  & 1   \\
				1/4 & 1    & & 1     &  & 1   \\ \bottomrule
			\end{tabular}
		\end{table}
	\end{center}
	
	\subsubsection*{Logical Exclusive-OR (XOR)}
	\begin{center}
		\begin{table}[!h]
			\begin{tabular}{@{}cccccc@{}}
				\toprule
				$P(x_1, x_2, y)$   & $X_1$&$\oplus$ & $X_2$ & = & $Y$ \\ \midrule
				1/4 & 0    & & 0     &  & 0   \\
				1/4 & 0    & & 1     &  & 1   \\
				1/4 & 1    & & 0     &  & 1   \\
				1/4 & 1    & & 1     &  & 0   \\ \bottomrule
			\end{tabular}
		\end{table}
	\end{center}

\end{document}